\documentclass[12pt]{article}
\pdfoutput=1
\usepackage{amsmath,amssymb,amsthm,amscd}
\usepackage{tikz}
\usepackage{jheppub}
\usepackage{amsmath}
\usepackage{enumerate}
\usepackage{placeins}
\usepackage{array}

\theoremstyle{definition}

\def\P{{\sf P}}

\def\e{\textrm{e}}

\newcommand{\bea}{\begin{eqnarray}}
\newcommand{\eea}{\end{eqnarray}}
\newcommand{\be}{\begin{eqnarray}}
\newcommand{\ee}{\end{eqnarray}}
\newcommand{\nn}{\nonumber}
\newcommand{\Tr}{\mathrm{Tr}}

\newcommand{\udl}[1]{\mathrm{d} #1 \,}

\newcommand{\xfac}[1]{\left( #1; x^2 \right)_\infty}

\newcommand{\sbfunc}[1]{s_b\left( #1\right)}
\newcommand{\Sfunc}[1]{S_2\left( #1\right)}
\newcommand{\SFunc}[1]{S_3\left( #1\right)}

\def\ga{\alpha}
\def\gb{\beta}
\def\gc{\gamma}

\def\Gd{\Delta}
\def\gd{\delta}

\def\Gl{\Lambda}

\def\Gp{\Phi}

\def\mun{FM[SU(N)]}
\def\gun{G[U(N)]}

\title{\boldmath$3d$ dualities from \boldmath$2d$ free field correlators: recombination and rank stabilization
}

\abstract{We propose various new $3d$ $\mathcal{N}=2$ dualities exploiting their recently discovered connection to  the duality relations  for  $2d$ free field CFT correlators.
Most of the dualities  involve, as the main building block, a quiver theory with monopole superpotential which enjoys various interesting properties such as being self-dual and reducing, in a suitable real mass deformation, to the familiar  $T[SU(N)]$ theory.
In particular we propose a duality for the  $U(N)$ theory with one adjoint and $k+1$ fundamental flavors.
By iterating some  basic dualities we can bring the theory to a stable form which, in turns, allows us to find a dual frame where the rank of the original theory appears as a parameter.
}

\author[1]{Sara Pasquetti}
\author[1]{Matteo Sacchi}

\affiliation[1]{Dipartimento di Fisica, Universit\`a di Milano-Bicocca \& INFN, Sezione di Milano-Bicocca, \\
I-20126 Milano, Italy
}

\emailAdd{sara.pasquetti@gmail.com}
\emailAdd{m.sacchi13@campus.unimib.it}

\begin{document}

\maketitle

\section{Introduction and Outlook}

Recently in \cite{US1} it was  observed that the $3d$ $\mathcal{N}=2$ duality relating the $U(N)$ theory with one adjoint and one flavor to a Wess-Zumino (WZ) model with $3N$ chiral fields proposed in \cite{Benvenuti:2018bav} enjoys  an intriguing  relation to the 3-point correlator in Liouville CFT in the free field realization. More precisely, the observation is that the integral identity  encoding the equality of the $S^2\times S^1$ partition functions of the  $U(N)$ theory and of the  WZ model, reduces in a suitable {\it Coulomb} limit, to a complex integral identity 
providing the evaluation formula for the  Liouville 3-point correlator in the free field representation.

$3d$ $\mathcal{N}=2$ theories can be formulated on compact 3-manifolds and by applying supersymmetric localization the path integral of certain protected observables can be reduced to a matrix integral capturing the contribution of the Gaussian fluctations 
above the BPS vacua (see for example Contribution 6 in \cite{Pestun:2016zxk} and references therein). $3d$ localized partition functions are independent on the gauge coupling so they are ideal tools to test IR dualities relating pairs of theories conjectured  to flow to the same strongly coupled fixed point.

Here we will be mostly interested in the squashed three-sphere $S^3_b$ and $S^2\times S^1$ partition functions.
In particular for the  latter the BPS vacua are labelled by a set of continuous zero modes and by a set of discrete parameters, the magnetic flux through the sphere so the partition function consists of  an integral for each factor in the Cartan subalgebra of the gauge group $G$, whose domain of integration is $\mathrm{rank}\,G$ copies of the unit circle, and a sum over the quantized magnetic fluxes:
\be
\label{base}
Z^{S^2\times S^1}=\sum_{{\bf m}}\oint \prod_{j=1}^{\mathrm{rank}\,G}\frac{\udl{u_j}}{2\pi i\, u_j}Z_{int}\, ,
\ee
where $Z_{int}$ is the contribution of the classical action evaluated on the BPS vacua and of the quadratic fluctations.
In  \cite{US1} it was shown how, in the  {\it Coulomb} limit, where  we shrink the ratio $\gb$ between the radius of the $S^1$ of the $S^2$ and suitably rescale the mass parameters, the sum over magnetic fluxes can be approximated by an integration and  \eqref{base} reduces to a complex integral:
\be
\sum_{{\bf m}}\oint \prod_{j=1}^{\mathrm{rank}\,G} \frac{\udl{u_j}}{2\pi i\,u_j}\underset{\beta\rightarrow0}\longrightarrow\int_{\mathbb{C}}\prod_{j=1}^{\mathrm{rank}\,G}\frac{\udl{^2z_j}}{\pi\beta|z_j|^2}\, \,.
\label{measure}
\ee

Complex integrals of this type appear in the study of CFT correlators as follows. Correlators of $k$ primary operators in Liouville theory exhibit poles when the momenta satisfy the screening quantization condition \cite{GL}: 
\begin{equation}
\ga\equiv\alpha_1+\cdots +\alpha_k=Q-Nb, \qquad N \in \mathbb{N}\, ,
\label{onshell}
\end{equation}
where $Q=b+b^{-1}$ and $b$ is the coupling constant appearing in the central charge $c= 1+6Q^2$.
The residue in turn takes the form of a free field Dotsenko-Fateev (DF) correlator with $N$ screening charges:
\begin{eqnarray}
\label{genres}
   \underset{\alpha=Q-Nb}{\text{res}}\langle V_{\alpha_1}(z_1) V_{\alpha_2}(z_2)\cdots V_{\alpha_k}(z_k) \rangle&=&
   (-\pi\mu)^N \prod_{i<j}^k |z_i-z_j|^{-4\alpha_i\alpha_j}\\
\!\!\!\!\!\!\nonumber&&   \times\int
\prod_{i=1}^N \prod_{j=1}^k |x_i-z_j|^{-4b \alpha_k}  \prod_{i<j}^N|x_i-x_j|^{-4b^2} \udl{^2\vec{x}_N}\,.
\end{eqnarray}
where
\be
 \udl{^2\vec{x}_N}=\frac{1}{\pi^N N!} \prod_{j=1}^N d^2x_j\,.
\ee

In the case of the 3-point correlator,  it is possible to find an evaluation formula for the complex  integral and, as mentioned above,  this formula coincides with the {\it Coulomb} limit of  the identity of
the $S^2\times S^1$ partition functions of the $U(N)$ theory and of the  WZ model. In this sense we can say that the IR $3d$ duality is an {\it uplift} of the  evaluation formula for the 3-point correlator in Liouville CFT in the free field realization \cite{Fateev:2007qn}.

Even more interestingly we can establish a complete parallel  between the manipulations done
to the screening integral to prove the evaluation formula and the field theory steps leading to prove the IR duality. Indeed in \cite{Fateev:2007qn} the evaluation formula for the 3-point correlator was derived
by an  iterative procedure based on the application two-specializations of a fundamental duality identity. At each step the number of integrations in the screening integral is lowered by one unit until all the integrations are performed and the evaluation formula is obtained.

\begin{figure}[t]
	\centering
		\makebox[\linewidth][c]{
	\includegraphics[scale=0.34]{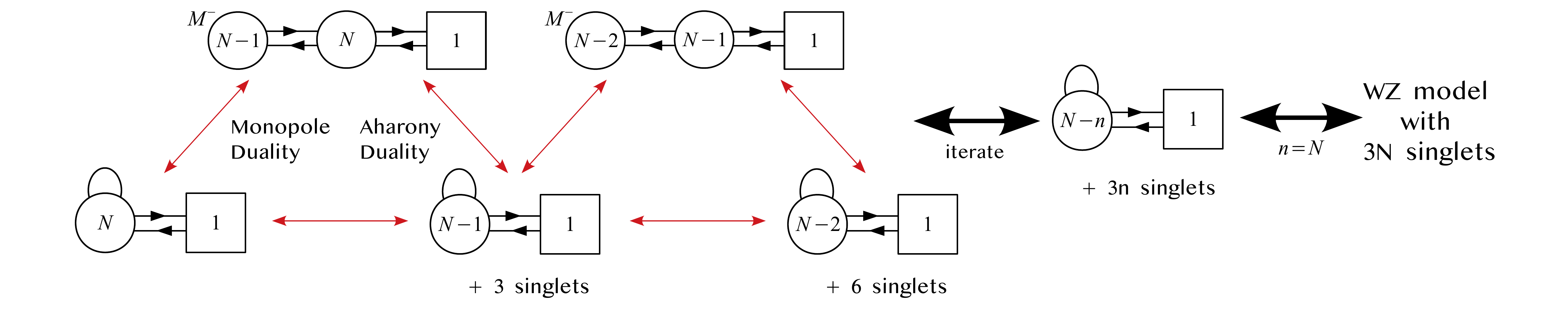}}
	\caption{Derivation of the duality between the $U(N)$ theory with one adjoint and one flavor and the WZ model.
	We  start from an auxiliary  quiver theory with gauge group $U(N-1)\times U(N)$, where at the $U(N-1)$ node we have the negative fundamental monopole turned on in the superpotential.  Applying the  monopole duality on the first node confines it and yields our original theory. On the other hand, if  we apply Aharony duality on the $U(N)$ node, we confine it and we go back to the original theory  with rank lowered by one unit and 3 extra singlets.	Iterating this procedure $N$ times, we get the WZ dual frame.
	}
	\label{d1}
\end{figure}

In \cite{US1} it has been shown that also the fundamental duality identity of \cite{Fateev:2007qn}  can  be obtained by taking the {\it Coulomb} limit of an identity expressing the equality of the $S^2\times S^1$ partition functions of pair of dual theories related by a $3d$ Seiberg-like duality with monopole superpotential  derived in \cite{Benini:2017dud}. The two  specializations relevant for the recursion in particular correspond to the Aharony duality \cite{Aharony:1997gp} for the $U(N_c)$ SQCD theory with $N_f=N_c$ flavors and to the one-monopole duality for the $U(N_c)$ SQCD with $N_f=N_c+1$ flavors and the negative fundamental monopole operator turned on in the superpotential \cite{Benini:2017dud}.  In both cases the electric theories confine and the dual theories are WZ models.
As sketched in  Figure \ref{d1} these two dualities can be used iteratively to obtain a sequence of dual theories
where at each step, we still have SQCD with one flavor and one adjoint - in this sense we say that the theory is stable under the combination of the two basic dualities - but the rank of the SQCD is decreased by one unit and 3 extra singlets are produced.  After $n$ iterations the rank is decreased to  $N-n$ and there are $3n$ singlets.  
The sequence ends after $n=N$ iterations when we reach the WZ frame with $3N$ chiral fields.

Coming back to the CFT side, we notice that obtaining an evaluation formula for the screening integral is actually only the first step towards the evaluation of the correlation function. Indeed ideally we would like 
to reconstruct correlators for generic values of the momenta  lifting the screening condition \eqref{onshell}.
To do so  one would need to perform analytic continuation in $N$.
 In the 3-point case  this  can be done by using various special function properties allowing us to
 recast the evaluation formula in a form where $N$ enters  as parameter which can be analytically continued to non-integer values.
In \cite{US1}, we interpreted this analytic continuation in the gauge theory context as a geometric transition to the $5d$ $T_2$ geometry, 
obtained by interesecting two $(1,0)$, two $(0,1)$ and two $(1,1)$ 5-branes  \cite{Benini:2009gi}.

For more general correlators, performing the analytic continuation becomes quickly quite tricky. Only in few cases
an evaluation formula can be obtained.
 Nevertheless as shown in \cite{Fateev:2007qn,Fateev:2007ab,Fateev:2008bm} in some special cases it is still possible to recast the screening integral in a form suitable for analytic continuation by some quite non-trivial applications of the duality relations for complex integrals.

In this paper we show that it is possible to {\it uplift} also these more sophisticated  duality relations between higher point free field correlators in $2d$ CFT to new genuine $3d$  IR dualities.

The duality relations we focus on are those obtained in \cite{Fateev:2007qn} for the study of correlators with 3 primaries and $k$ degenerate operators in Liouville theory. For this special choice of the momenta, the $N$-dimensional integral \eqref{genres} can be massaged in a form suitable for analytic continuation in $N$ involving  the  \emph{kernel} function $K_k^\Delta(m_1,..,m_k|t_1,..,t_k)$:
\begin{multline}\label{m3point}
\langle V_{-\frac{b}{2}}(z_1)\dots
  V_{-\frac{b}{2}}(z_k)V_{\alpha_1}(0)V_{\alpha_2}(1)V_{\alpha_3}(\infty)\rangle=
  \Omega^N_k(\alpha_1,\alpha_2,\alpha_3)\prod_{a=1}^k\vert z_a\vert^{2b\alpha_1}
  \vert z_a-1\vert^{2b\alpha_2}\times\\\times
  \prod_{a<b}^k|z_a-z_b|^{-b^2}
  \int\prod_{a=1}^k\vert x_a \vert^{2A}\vert x_a-1\vert^{2B}K_k^C(x_1,..,x_k|z_1,..,z_k)
  \prod_{a<b}^k|x_a-x_b|^{-4b^2} \,\udl{^2\vec{x}_k}\, ,
\end{multline}
where
\be
       &&A=b\left(\alpha-2\alpha_1-Q+kb/2\right),\;\;\nn\\
       &&B=b\left(\alpha-2\alpha_2-Q+kb/2\right),\;\;\nn\\
       &&C=b\left(Q+(2-k)b/2-\alpha\right)\,.
\ee
The kernel $K_k^\Delta(m_1,..,m_k|t_1,..,t_k)$ is represented by a complex rank $k(k-1)/2$ integral.

In the integral on the r.h.s of eq.~\eqref{m3point} $N$ enters just as a parameter in the  sum of the momenta $\ga$, fixed by the screening condition \eqref{onshell},  appearing in $A$, $B$, $C$.
The  prefactor $\Omega^N_k(\alpha_1,\alpha_2,\alpha_3)$ instead is the product of $4N-3k$ factors of the function
$\gamma(x)=\Gamma(x)/\Gamma(1-x)$. However 
by using the periodicity property of the $\Upsilon$-function
\begin{equation}
 \begin{aligned}
   &\Upsilon(x+b)=\gamma(bx)b^{1-2bx}\Upsilon(x)\\
   &\Upsilon(x+b^{-1})=\gamma(b^{-1}x)b^{2b^{-1}x-1}\Upsilon(x)\, ,
 \end{aligned}
\end{equation}
we can re-express the contribution of $N-k$ $\gc$-functions in terms of a single $\Upsilon$-function moving the dependence on 
 $N$ inside the argument of the $\Upsilon$, so that also   $\Omega^N_k(\alpha_1,\alpha_2,\alpha_3)$ depends parametrically on $N$. 
This equivalent form of the free field correlator is then suitable for analytic continuation in $N$.

\begin{figure}[t]
	\centering
	\includegraphics[scale=0.5]{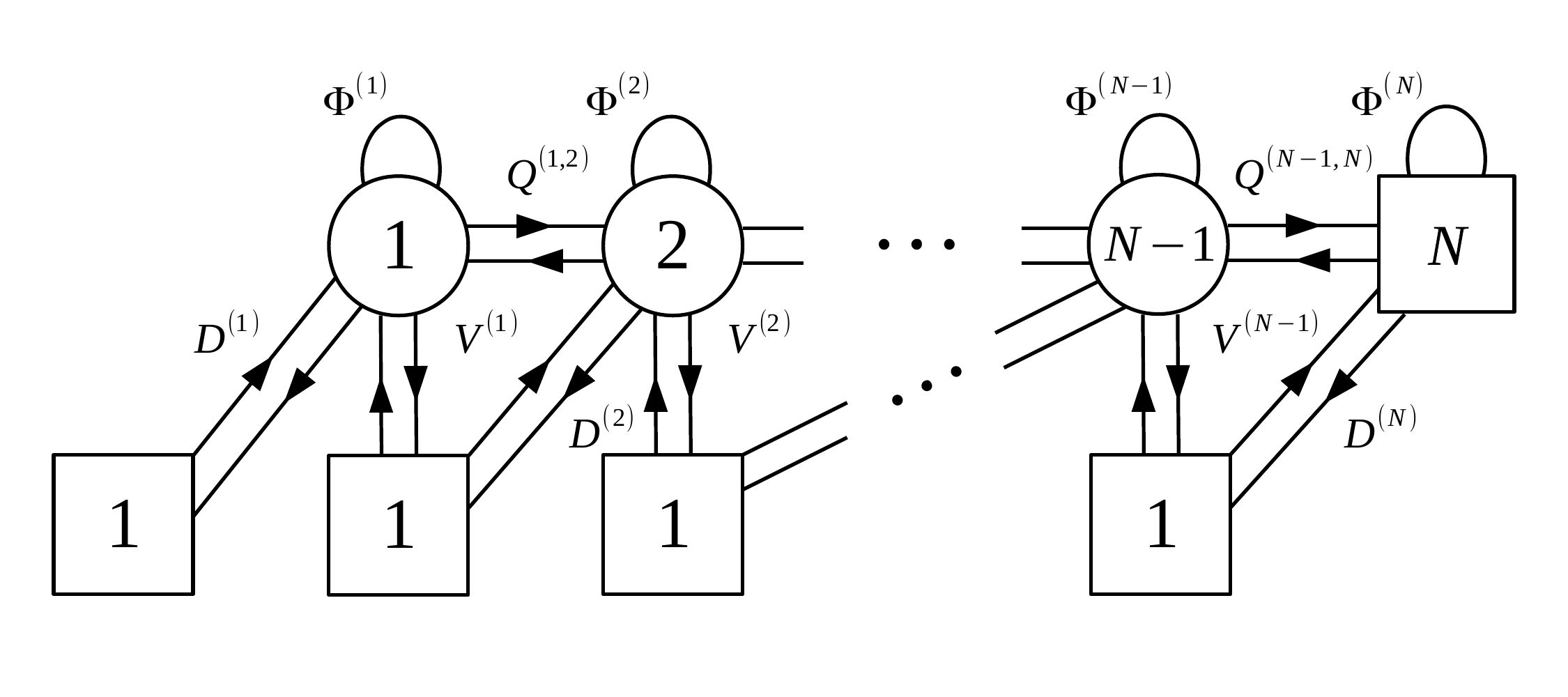}
	\caption{Quiver diagram of the $FM[SU(N)]$ theory. Double-lines connecting two nodes represent pairs of bifundamental chirals in conjugate representations with respect to the corresponding symmetries. Lines that start and end on the same node correspond to chirals in the adjoint representation.}
	\label{mtsun}
\end{figure}

The kernel function $K_k^\Delta(m_1,..,m_k|t_1,..,t_k)$ satisfies various remarkable properties, such as being symmetric under the exchange $m_a\leftrightarrow t_a$
\be
K_N^\Delta(m_1,..,m_N|t_1,..,t_N)=K_N^\Delta(t_1,..,t_N|m_1,..,m_N)\, .
\label{kernelsymm}
\ee
As in the case of the 3-point correlator, for which we were able to find a $3d$ {\it uplift} given by the
$\mathcal{N}=2$ $U(N)$ theory with one flavor and one adjoint, we claim that also the kernel function has a $3d$ avatar, which is the $\mathcal{N}=2$ quiver theory depicted in Figure \ref{mtsun} and that we name as the $\mun$ theory.  In $\mun$ each $U(m)$ gauge node has $2m+2$ flavors and the two fundamental monopoles turned on in the superpotential. There is also the standard $\mathcal{N}=4$ cubic superpotential coupling adjoint and bifundamental fields and a cubic superpotential involving vertical, diagonal and bifundamental chirals in each triangle of the quiver. The global symmetry group of the theory is 
\be
SU(N)_M\times SU(N)_T\times U(1)_{m_A}\times U(1)_\Delta\, .
\ee
The $SU(N)_T$ symmetry is actually emergent at low energies and in the UV only the $U(1)^{N-1}$ rotating the flavors in the saw is visible.

We show that $\mun$ is left invariant by the action of a (self)-duality which swaps the operators transforming  under  $SU(N)_M$ with those transforming under $SU(N)_T$, leaving the charges under the two $U(1)$ invariant. This is basically the $3d$ uplift of the symmetry property \eqref{kernelsymm} of the kernel function. 
Indeed, it is a straightforward  exercise to show that by taking the  Coulomb limit, defined in \cite{US1},
 of the $S^2\times S^1$ partition function of $\mun$ we recover the kernel function and the $3d$ self-duality reduces to \eqref{kernelsymm}.

For $N=2$ the theory is abelian and we are able to prove the self-duality {\it piecewise} by iterating a fundamental duality with two monopoles in the superpotential \cite{Benvenuti:2016wet}. For larger $N$ we provide evidences of the self-duality by mapping the operators in the chiral ring and matching various orders of the perturbative expansion of the superconformal index. In particular, we are able to explicitly construct for arbitrary $N$ a gauge invariant chiral operator in the adjoint representation of $SU(N)_T$, which contains as its bottom component the moment map for this enhanced symmetry.

The self-duality of $\mun$ is reminiscent of the self-duality of the $T[SU(N)]$ theory \cite{Gaiotto:2008ak} under Mirror Symmetry, which swaps the $SU(N)$ rotating the Higgs branch operators and the $SU(N)$ rotating the Coulomb branch operators. Moreover, if one considers an axial mass deformation for the $U(1)_A$ symmetry which is the anti-diagonal combination of $U(1)_R\in SU(2)_R$ and $U(1)_L\in SU(2)_L$ (where $SU(2)_{R}\times SU(2)_L$ is the non-abelian $\mathcal{N}=4$ R-symmetry), then Mirror Symmetry also changes the sign of this mass. The self-duality of $\mun$ in this respect is even closer to that of the $FT[SU(N)]$ theory  (the $T[SU(N)]$ with an extra set of singlets flipping the Higgs branch moment map studied in \cite{Zenkevich:2017ylb}, \cite{Aprile:2018oau}) under spectral duality, which swaps the two $SU(N)$ symmetries leaving $U(1)_A$ invariant. Indeed we show that $\mun$ reduces to $FT[SU(N)]$ when a suitable real mass deformation associated to the  $U(1)_\Delta$ axial symmetry is taken.

\begin{figure}[t]
	\centering
	\includegraphics[scale=0.5]{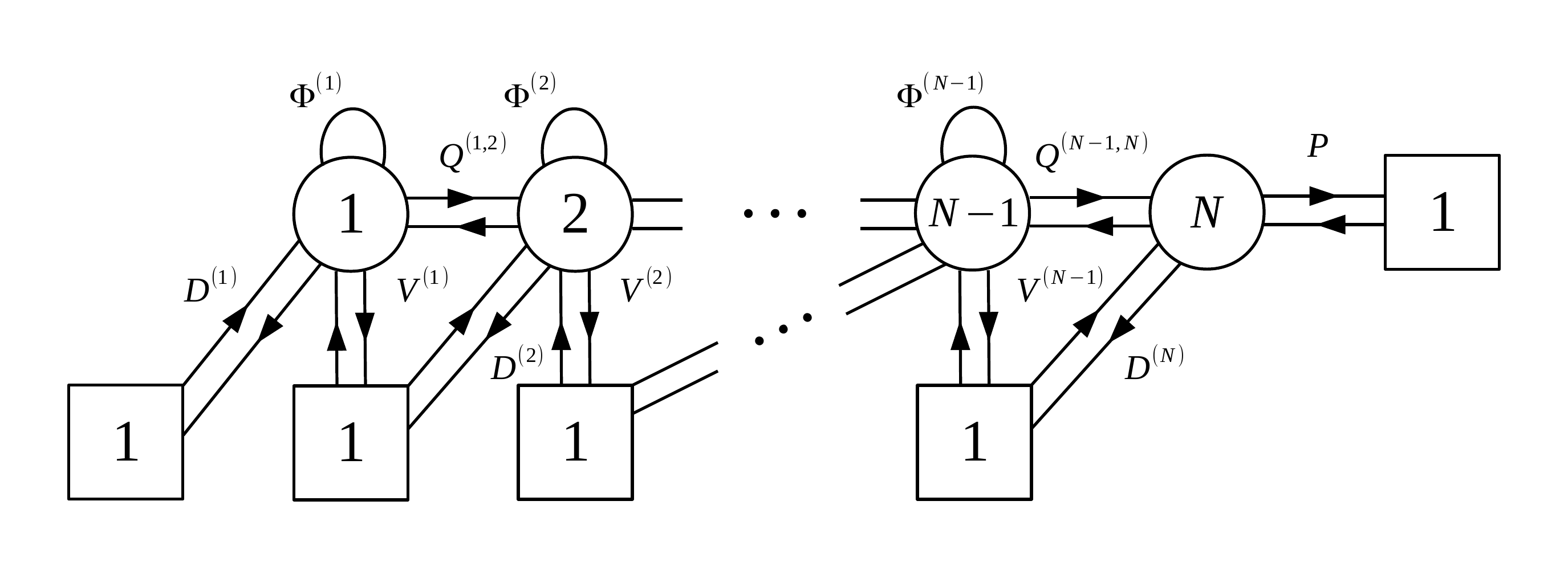}
	\caption{Quiver diagram of the $G[U(N)]$ theory.}
	\label{gunfig}
\end{figure}

The $\mun$ theory with the last node gauged (with no adjoint) and one extra flavor attached to it, depicted in Figure \ref{gunfig}, will also play a central role in many of the dualities we are going to present, so it deserves its own name $\gun$. We show that $\gun$ satisfies a very curious recombination property, which is the $3d$ avatar of the factorization property of the kernel function discussed in \cite{Fateev:2007qn}.

Namely $\gun$  has a family of dual frames obtained by joining two smaller theories $G[U(N-k)]$ and $G[U(k)]$ by a bifundamental and various cubic, quartic and monopole superpotential terms (see Figure \ref{gluedmun}). The recombination property follows from the sequential application of Aharony duality, starting from the last $U(N)$ node of $\gun$. This node has no monopoles turned on in the superpotential nor an adjoint so we can use Aharony duality to turn it into a $U(1)$ node. This operation has also the effect of removing the adjoint chiral from the adjacent $U(N-1)$ node and to modify the charges of its monopoles, so that they are actually removed from the superpotential (following the same argument used in \cite{US1}). This allows us to apply again Aharony duality on the second node. This procedure can be repeated for an arbitrary number $k$ of iterations, giving exactly the claimed duality. 

\begin{figure}[t]
	\centering
	\makebox[\linewidth][c]{
	\includegraphics[scale=0.4]{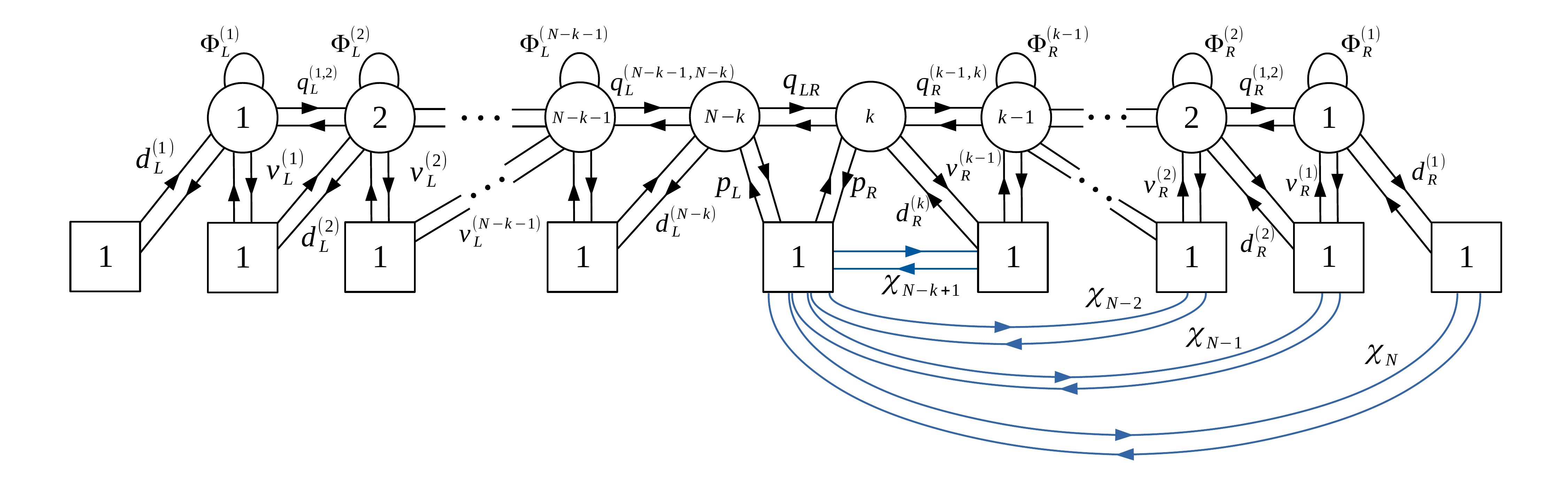}
	}
	\caption{Recombination dual frame of $\gun$.}
	\label{gluedmun}
\end{figure}

It is also noteworthy that the recombination property of $\gun$ allows us to find a dual frame where the rank of the gauge group is minimal. Indeed, Aharony duality reduces the rank of the gauge node to which we apply it until we arrive at the middle of the tail. From this point, the following applications of Aharony duality start to increase the rank back and when we arrive at the end of the tail we recover $\gun$, but ``reversed".

\begin{figure}[b]
	\centering
		\includegraphics[scale=0.38]{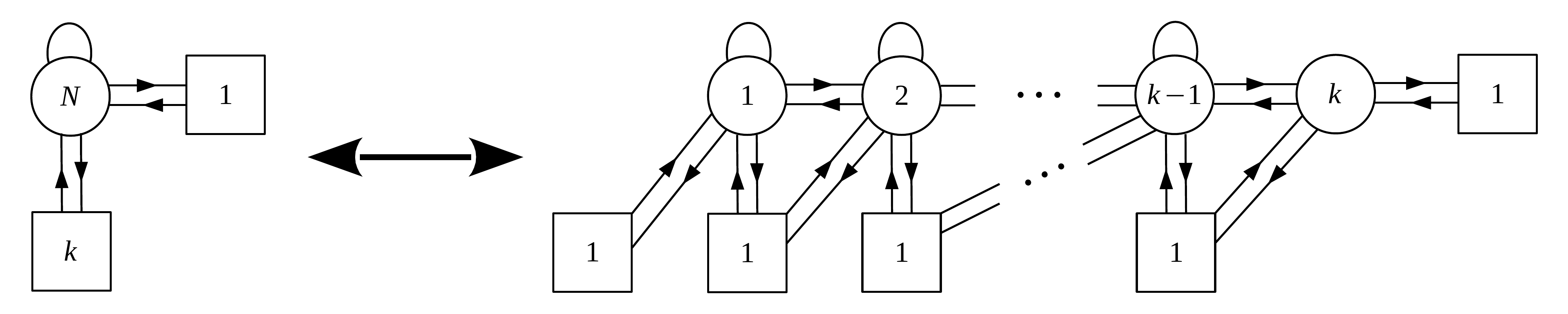}
	\caption{Rank stabilization uality for generic $k$.}
	\label{generalk}
\end{figure}

Finally, we {\it uplift} the duality relation \eqref{m3point} to a $3d$ IR duality between the $U(N)$ theory with one adjoint and $k+1$ flavors\footnote{More dualities for adjoint SQCD have been recently discussed in \cite{Giacomelli:2019blm}.}, with the adjoint coupled to $k$ of the flavors only, and the $G[U(k)]$ theory plus some gauge singlets shown in Figure \ref{generalk}. We call this \emph{rank stabilization duality} as the rank $N$ of the gauge group of the original theory only appears as a parameter in the $U(1)$ charges of the chiral fields and as the number of singlet fields in the dual theory. The strategy we follow to derive it retraces again the steps done in CFT in \cite{Fateev:2007qn}. We apply a sequence of basic dualities trying to reduce the theory to a frame which is stable under the application of the sequence of basic dualities discussed above for the $k=0$ case.

\begin{figure}[t]
	\centering
		\includegraphics[scale=0.32]{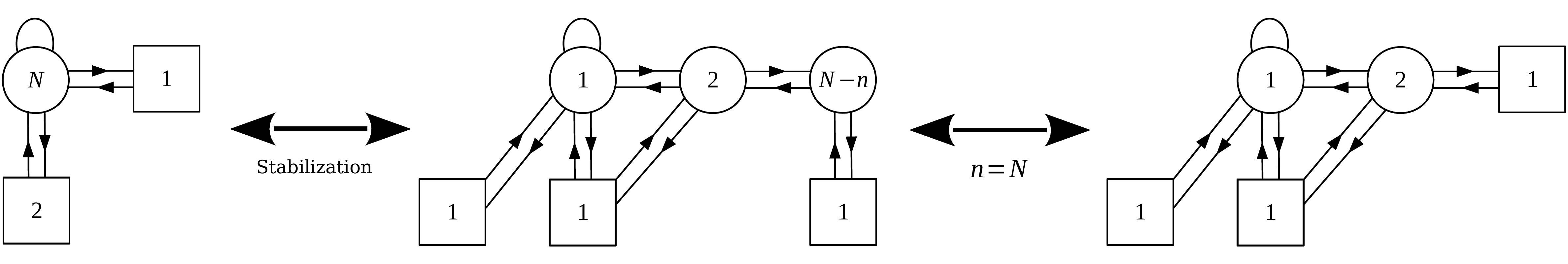}
	\caption{Sketch of the derivation of the rank stabilization duality in the $k=2$ case. In the middle we have the stable configuration.}
	\label{stable2}
\end{figure}

For example, for the case $k=2$ depicted in Figure \ref{stable2}, we will see that combining various fundamental dualities we can reach a configuration which is stable under the sequential application of the one-monopole and the Aharony duality. After $n$ iterations of these two dualities, we find the original theory with rank $N-n$ and 2 flavors glued via gauging to the $FM[U(2)]$ theory, without the adjoint chiral at the $U(2)$ node and plus various singlets. Setting $n=N$ we obtain the $G[U(2)]$ theory with the extra singlets.

We prove the equality of the $S^3_b$ partition functions for the cases $k=1,2$ following precisely this logic and using iteratively  the integral identities for each duality. We conjecture that this pattern carries on for generic $k$ and that we can reach a dual frame which is the $G[U(k)]$ theory with $3N-2k$ singlets depicted in Figure \ref{generalk}. We motivate this by mapping the operators of the chiral rings for arbitrary $k$ and by matching various orders of the perturbative expansion of the superconformal index for $k=3$.

Also for generic $k$, as we did in \cite{US1} for $k=0$, we can give a meaning to the analytic continuation in the number of screening charges in the gauge theory context. Indeed, we can re-express the partition function of the theory dual to the $U(N)$ theory with an adjoint and $k+1$ flavors in such a way that the rank $N$ enters only parametrically. For the contribution of the singlet fields, we can use the periodicity properties of some special functions to rewrite it in terms of the $5d$ $T_2$ theory describing free hypers. Instead, $N$ already enters parametrically in the $G[U(k)]$ sector, which describes a $3d$ defect theory interacting with the $5d$ part.

\vspace{5pt}


The philosophy of the present paper and of its companion \cite{US1} can be profitably used to find several new $3d$ IR dualities by uplifting the many duality identities for free field correlators that appear in the $2d$ CFT literature. 
We will  continue this program in \cite{chiung} where we focus on the Toda free field correlators obtaining
 new $3d$  dualities involving the $FM[SU(N)]$ theory  with one or both of its $SU(N)$ symmetries gauged.

Analogous constructions are known for $T[SU(N)]$, which is for several aspects similar to $FM[SU(N)]$. Indeed, $T[SU(N)]$ can be used as a building block for constructing several interesting theories, such as the star-shaped quivers  \cite{Benini:2010uu} which are mirror dual to the dimensional reduction of class $\mathcal{S}$ theories \cite{Gaiotto:2009we} and the $S$-fold CFTs \cite{Terashima:2011qi,Gang:2015bwa,Gang:2015wya,Assel:2018vtq,Gang:2018huc,Garozzo:2018kra,Garozzo:2019hbf,noppy}.
It is also known that $T[SU(N)]$ is the $S$-duality wall for the $4d$ $\mathcal{N}=4$ SYM \cite{Gaiotto:2008ak}. 
The similarity of our $FM[SU(N)]$ to $T[SU(N)]$, specifically the fact that it possesses two $SU(N)$ symmetries that are exchanged under its self-duality, suggests that it might as well be the theory living on some $3d$ duality wall between two $4d$ theories. For example, in \cite{Teschner:2012em,Floch:2015hwo,Benini:2017dud,Garozzo:2019xzi} the duality wall in $4d$ $\mathcal{N}=2$ $SU(N)$ gauge theory with $2N$ flavors has been considered, which consists of the $3d$ $\mathcal{N}=2$ $U(N)$ gauge theory with $2N$ flavors and $\mathcal{W}=\mathfrak{M}^++\mathfrak{M}^-$. In the case $N=1$ this corresponds to a deformation of our $FM[SU(2)]$ theory. 

Another interesting aspect of the $FM[SU(N)]$ theory is its relation to the kernel function appearing in the integral representations of  $q$-deformed hypergeometric functions, which can be expressed in terms of Macdonald polynomials.
We plan to explore further this connection in \cite{fra}.

\section{The \boldmath $FM[SU(N)]$ theory}


\subsection{Superpotential, operators and sphere partition function}
\label{fmsuntheory}

$FM[SU(N)]$ is the $3d$ $\mathcal{N}=2$ linear quiver gauge theory represented in Figure \ref{mtsun}. More precisely, the chiral fields of this theory are:
\begin{enumerate}[$\bullet$]
\item $V^{(k)}$, $\tilde{V}^{(k)}$: fundamental flavors connecting the $U(k)$ gauge node with a $U(1)$ flavor node vertically;
\item $D^{(k)}$, $\tilde{D}^{(k)}$: fundamental flavors connecting the $U(k)$ node with a $U(1)$ flavor node diagonally;
\item $Q^{(k,k+1)}$, $\tilde{Q}^{(k,k+1)}$: bifundamental flavors connecting the $k$-th node with the $(k+1)$-th one.\footnote{In our conventions, $Q^{(k,k+1)}$ transforms in the representation $\Box\otimes\bar{\Box}$ of $U(k)\times U(k+1)$, while $\tilde{Q}^{(k,k+1)}$ transforms in $\bar{\Box}\otimes\Box$ of $U(k+1)\times U(k)$, so some color indices are understood. For example, for $k=2$ we have $Q^{(k,k+1)}_{ia}$ and $\tilde{Q}^{(k,k+1)}_{ai}$, with $i=1,2$ and $a=1,2,3$.} For $k=N-1$ it connects the last $U(N-1)$ gauge node with the $SU(N)$ flavor symmetry on the very right;
\item $\Gp^{(k)}$: adjoint chiral corresponding to the $k$-th gauge node.
 For $k=N$ the adjoint chiral is  on the $SU(N)$ flavor node.
 \end{enumerate}

In order to write the superpotential of the theory in a compact form, we introduce the following notation. From the bifundamentals $Q^{(k,k+1)}_{ia}$ and $\tilde{Q}^{(k,k+1)}_{bj}$ we construct a tensor that represents a chiral field in the representation $(\Box\otimes\bar{\Box})\otimes(\Box\otimes\bar{\Box})$ of $U(k)\times U(k+1)$:
\be
\mathbb{Q}^{(k,k+1)}_{ijab}\equiv Q^{(k,k+1)}_{ia}\tilde{Q}^{(k,k+1)}_{bj},\qquad i,j=1,\cdots,k,\quad a,b=1,\cdots,k+1\, .
\ee
Moreover, we denote with $\mathrm{Tr}_k$ the trace over the color indices of the $U(k)$ gauge group.
The superpotential of $FM[SU(N)]$ contains the standard $\mathcal{N}=4$ cubic superpotential coupling bifundamental and adjoints, a linear  monopole superpotential turned on at each node and a cubic interaction term coupling the fields in the {\it saw} to the bifundamentals:
\be
\mathcal{W}_{FM[SU(N)]}=\mathcal{W}_{mono}+\mathcal{W}_{FT[SU(N)]}+\mathcal{W}_{cub}\, .
\label{wfmun}
\ee
The first term is a linear monopole superpotential containing monopoles with magnetic flux $\pm1$ with respect to only one of the factors in the gauge group\footnote{In \cite{Pasquetti:2019hxf} it was shown that $FM[SU(N)]$ has a $4d$ ancestor called $E[USp(2N)]$ with $USp(2n)$ gauge groups. More precisely, $FM[SU(N)]$ can be obtained from $E[USp(2N)]$ upon dimensional reduction followed by a real mass deformation that higgses the gauge groups from $USp(2n)$ to $U(n)$. Similarly to what was discussed in \cite{Aharony:2013dha,Aharony:2013kma,Benini:2017dud}, the monopole superpotential \eqref{Wmono} is dynamically generated in the dimensional reduction and the requirement in $4d$ that $U(1)_R$ is non-anomalous translates in $3d$ in the constraint on the R-charges due to the marginality of the monopoles.}
\be
\mathcal{W}_{mono}&=&\mathfrak{M}^{(1,0,\cdots,0)}+\mathfrak{M}^{(-1,0,\cdots,0)}+\mathfrak{M}^{(0,1,0,\cdots,0)}+\nn\\
&+&\mathfrak{M}^{(0,-1,0,\cdots,0)}+\cdots\mathfrak{M}^{(0,\cdots,0,1)}+\mathfrak{M}^{(0,\cdots,0,-1)}\, .
\label{Wmono}
\ee
The second term is the superpotential of the $FT[SU(N)]$ theory \cite{Aprile:2018oau}, where the adjoint chiral $\Gp^{(N)}$  {\it flips}  the mesons matrix  $\mathrm{Tr}_{N-1}\mathbb{Q}^{(N-1,N)}$ made out of the last bifundamental
\be
\mathcal{W}_{FT[SU(N)]}=\sum_{k=1}^N\mathrm{Tr}_k\left[\Gp^{(k)}\left(\mathrm{Tr}_{k+1}\mathbb{Q}^{(k,k+1)}-\mathrm{Tr}_{k-1}\mathbb{Q}^{(k-1,k)}\right)\right]\, ,
\ee
where we define $\mathbb{Q}^{(0,1)}=\mathbb{Q}^{(N,N+1)}=0$. Finally the last term is given by
\be
\mathcal{W}_{cub}=\sum_{k=1}^{N-1}\sum_{i=1}^k\sum_{a=1}^{k+1}\left(D^{(k+1)}_a\tilde{Q}^{(k,k+1)}_{ai}V^{(k)}_i+\tilde{V}^{(k)}_iQ^{(k,k+1)}_{ia}\tilde{D}^{(k+1)}_a\right)\, ,
\ee

\begin{table}[t]
\centering
\scalebox{0.9}{
\begin{tabular}{c|ccccc|c}
{} & $U(1)_{T_k}$ & $U(1)_{T_N}$ & $SU(N)_M$ & $U(1)_{m_A}$ & $U(1)_\Gd$ & $U(1)_R$ \\ \hline
$Q^{(k-1,k)}$ & 0 & 0 & 0 & -1 & 0 & $1-R_A$ \\
$\tilde{Q}^{(k-1,k)}$ & 0 & 0 & 0 & -1 & 0 & $1-R_A$ \\ 
$Q^{(N-1,N)}$ & 0 & 0 & $\bar{\Box}$ & -1 & 0 & $1-R_A$ \\
$\tilde{Q}^{(N-1,N)}$ & 0 & 0 & $\Box$ & -1 & 0 & $1-R_A$ \\ \hline
$V^{(k-1)}$ & 1 & 0   & 0 & $k-N+1$ & -1 & $2+(N-k-1)(1-R_A)-R_\Gd$ \\
$\tilde{V}^{(k-1)}$ & -1 & 0 & 0 & $k-N+1$ & -1 & $2+(N-k-1)(1-R_A)-R_\Gd$ \\ 
$V^{(N-1)}$ & 0 & 1 & 0 & $1$ & -1 & $1+R_A-R_\Gd$ \\
$\tilde{V}^{(N-1)}$ & 0 & -1 & 0 & $1$ & -1 & $1+R_A-R_\Gd$ \\ \hline
$D^{(k)}$ & -1 & 0 & 0 & $N-k$ & 1 & $(k-N)(1-R_A)+R_\Gd$ \\
$\tilde{D}^{(k)}$ & 1 & 0 & 0 & $N-k$ & 1 & $(k-N)(1-R_A)+R_\Gd$ \\
$D^{(N)}$ & 0 & -1 & $\Box$ & 0 & 1 & $R_\Gd$  \\
$\tilde{D}^{(N)}$ & 0 & 1 & $\bar{\Box}$ & 0 & 1 & $R_\Gd$  \\ \hline
$\Gp^{(k)}$ & 0 & 0 & 0 & 2 & 0 & $2R_A$ \\
$\Gp^{(N)}_{ab}$ & 0 & 0 & adj & 2 & 0 & $2R_A$ \\
\end{tabular}}
\caption{Representations and charges under the global symmetries of all the chiral fields of the $FM[SU(N)]$ theory. In the table, $k$ runs from 1 to $N-1$. By definition, $Q^{(0,1)}=\tilde{Q}^{(0,1)}=0$ and $V^{(0)}=\tilde{V}^{(0)}=0$.}
\label{mtsuncharges}
\end{table}
The manifest global symmetry of this theory is:
\be
SU(N)_M\times\frac{\prod_{k=1}^{N}U(1)_{T_k}}{U(1)}\times U(1)_{m_A}\times U(1)_{\Gd}\, ,
\label{mtsunglobal}
\ee
which enhances in the IR to
\be
SU(N)_M\times SU(N)_T\times U(1)_{m_A}\times U(1)_{\Gd}\, .
\label{mtsunglobalIR}
\ee
Our main argument to support this claim is the self-duality which we discuss in the following section,  that swaps the $SU(N)_M$ and the $SU(N)_T$ symmetries. Another evidence of the symmetry enhancement comes from the fact that the operators in the chiral ring with the same charges under the other global symmetries, included the R-symmetry, re-organize into representations of the full $SU(N)_T$ symmetry, as we will show below.

Hence, at low energies we only have two abelian global symmetries $U(1)_{m_A}$ and $U(1)_\Gd$ that can mix with the R-symmetry. We denote with $R_A$ and $R_\Gd$ respectively the parameters that quantify this mixing. 
The R-charges of the fields will then be parameterized by these two coefficients as follows.
We assign R-charge $R_\Gd$ to the last diagonal flavor $D^{(N)}$, $\tilde{D}^{(N)}$ and $1-R_A$ to the last bifundamental $Q^{(N-1,N)}$, $\tilde{Q}^{(N-1,N)}$. Because of the superpotential terms $\mathcal{W}_{FT[SU(N)]}$ also all the other bifundamentals will have R-charge $1-R_A$, while the adjoint chirals $\Gp^{(k)}$ will have R-charge $R_A$. 
The cubic superpotential $\mathcal{W}_{cub}$ then fixes the R-charge of the last vertical flavor to be $R[V^{(N)},\tilde{V}^{(N)}]=2-(1-R_A)-R_\Gd=1+R_A-R_\Gd$. Then, we have to take into account the monopole superpotential. Requiring that the fundamental monopole operators of the $U(N-1)$ node are exactly marginal, we find that the next diagonal flavor must have R-charge $R[D^{(N-1)},\tilde{D}^{(N-1)}]=-1+R_A+R_\Gd$. Following this procedure along the whole tail, we can fix the R-charges of all the chiral fields in terms of the parameters $R_A$ and $R_\Gd$ only. In Table \ref{mtsuncharges} we summarize the charges of the chiral fields under all the global symmetries and we specify their R-charges.
\begin{figure}[t]
	\centering
	\includegraphics[scale=0.5]{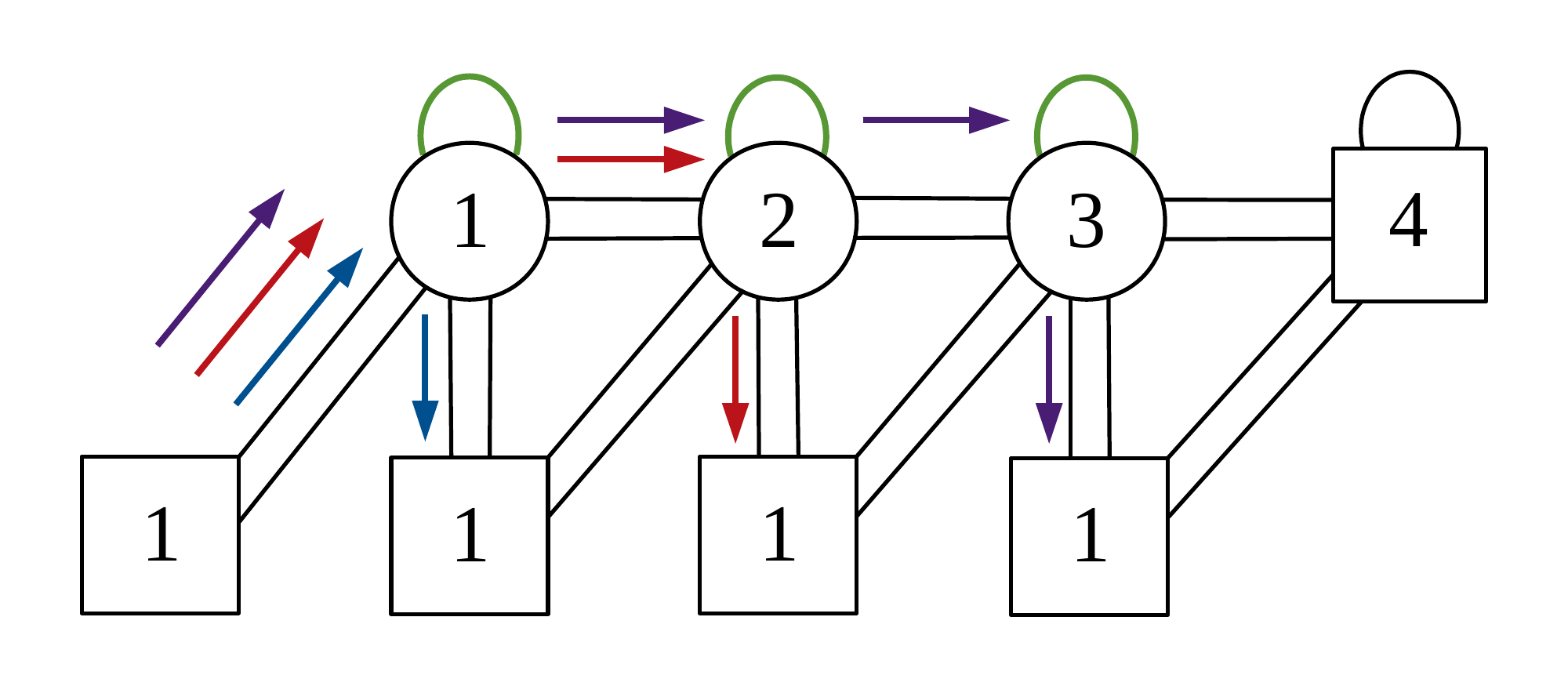}
	\caption{Diagrammatic representation of the operators in the first row of the matrix $\mathcal{M}$. Arrows of the same color represent chiral fields that we assemble to construct an element of the matrix. In order to have a gauge invariant operator, we have to consider sequences of arrows that start and end on a squared node. In this case, this is achieved starting with one diagonal flavor, going along the tail with an arbitrary number of bifundamentals and ending on a vertical flavor.}
	\label{Mmatrix}
\end{figure}

The chiral ring of the theory is generated by the following operators.
First of all, we have the chiral $\Gp^{(N)}$ in the adjoint representation of $SU(N)_M$. The charges of this operator under the global symmetries can be read from the last line of Table \ref{mtsuncharges}.

Then, we can construct an operator which transform in the adjoint representation of $SU(N)_T$ combining the traces of the adjoints at each gauge node on the diagonal and some mixed mesons on the off-diagonal elements. These mesons are built starting from one of the diagonal chirals, moving along the tail with the bifundamentals and ending on a vertical chiral (see Figure \ref{Mmatrix}). Explicitly, for $N=3$ it takes the form
\be 
\mathcal{M}=\begin{pmatrix}
0 & V^{(1)}D^{(1)} & V^{(2)}_i\tilde{Q}^{(1,2)}_iD^{(1)} \\
\tilde{D}^{(1)}\tilde{V}^{(1)} & 0 & V^{(2)}_iD^{(2)}_i \\
\tilde{D}^{(1)}Q^{(1,2)}_i\tilde{V}^{(2)}_i & \tilde{D}^{(2)}_i\tilde{V}^{(2)}_i & 0
\end{pmatrix}+\sum_{i=1}^2\Tr_i\Gp^{(i)}\mathcal{D}_i\, ,
\ee
where $\mathcal{D}_i$ are traceless diagonal generators of $SU(N)_T$. This operator has exactly the same charges under the two axial symmetries and the same R-charge as $\Gp^{(N)}$.

\begin{figure}[t]
	\centering
	\includegraphics[scale=0.5]{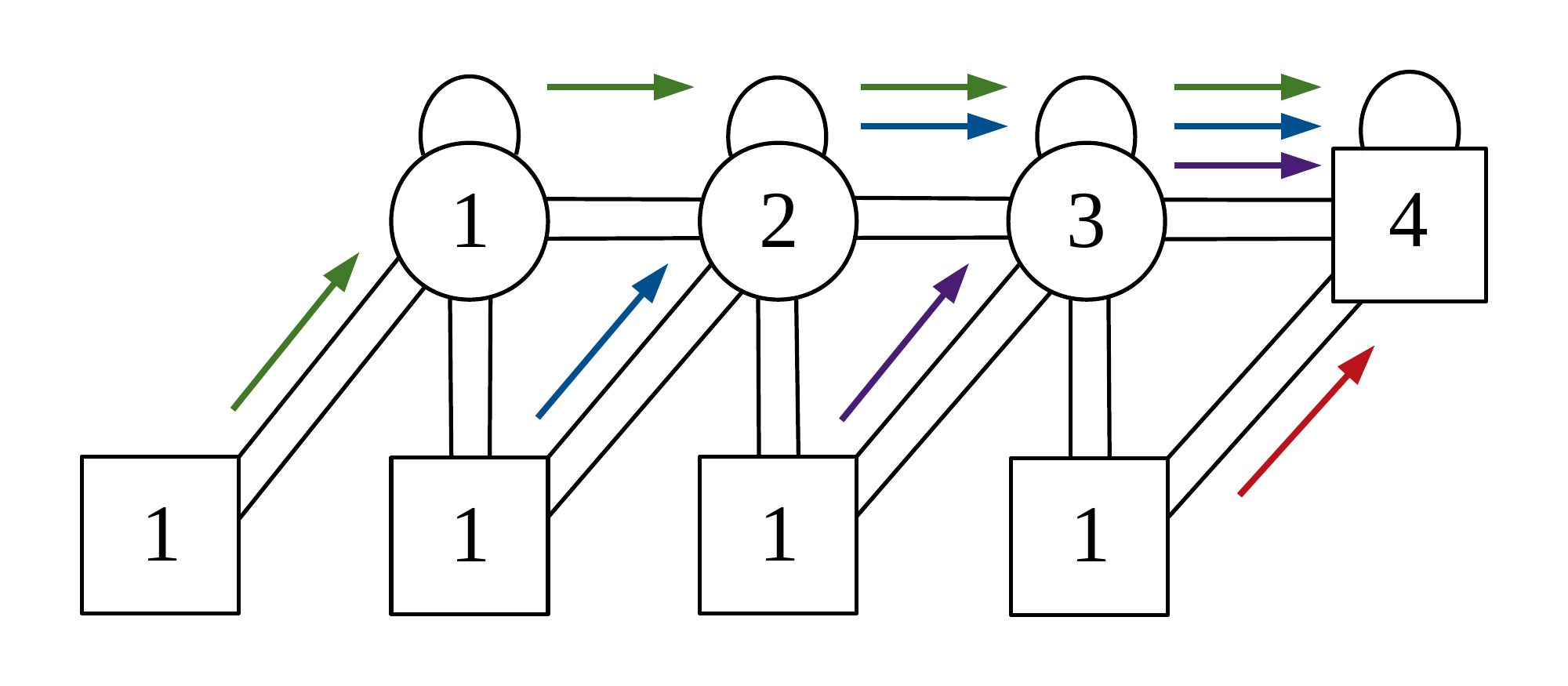}
	\caption{Diagrammatic representation of the operator $\Pi$. In this case, gauge invariant operators are obtained starting with one diagonal flavor, going along the tail with all the remaining bifundamentals and ending on the bifundamental connected to the last flavor node.}
	\label{Pimatrix}
\end{figure}

There are two other gauge invariant mixed mesons that one can construct from the chiral fields of the theory. In this case, we still start with a diagonal flavor and move along the tail, but we have to include all the bifundamentals and end with $Q^{(N-1,N)}$ (see Figure \ref{Pimatrix}). Such operators can be collected in two vectors that we denote with $\Pi$ and $\tilde{\Pi}$. Explicitly, for $N=3$ these operators take the form
\be
\Pi=\begin{pmatrix}
\tilde{Q}^{(2,3)}_{i,a}\tilde{Q}^{(1,2)}_iD^{(1)}\\ \tilde{Q}^{(2,3)}_{i,a}D^{(2)}_i \\D^{(3)}_a
\end{pmatrix},
\qquad
\tilde{\Pi}=\begin{pmatrix}
\tilde{D}^{(1)}Q^{(1,2)}_iQ^{(2,3)}_{i,a}\\ \tilde{D}^{(2)}_iQ^{(2,3)}_{i,a}\\ \tilde{D}^{(3)}_a
\end{pmatrix}\, .
\ee
They are uncharged under the axial symmetry $U(1)_{m_A}$, have charge 1 under the other axial symmetry $U(1)_\Gd$, have R-charge $R_\Gd$ and transform respectively in the bifundamental $\Box\otimes\bar{\Box}$ and anti-bifundamental $\bar{\Box}\otimes\Box$ representation of the flavor symmetries $SU(N)_M\times SU(N)_T$. 

Finally, we have some mesons obtained combining the flavors of the saw. These operators are all uncharged under the flavor symmetries $SU(N)_M$ and $SU(N)_T$. For example, we can consider the mesons constructed with the diagonal chirals with opposite charge under the same gauge node, which can be dressed with the corresponding adjoint chiral
\be
\tilde{D}^{(k)}\left(\Gp^{(k)}\right)^sD^{(k)}\equiv\Tr_k\left[\tilde{D}^{(k)}\left(\Gp^{(k)}\right)^sD^{(k)}\right],\qquad k=1,\cdots,N-1,\quad s=0,\cdots,k-1\, .\nn\\
\ee
We can also consider the (dressed) mesons made of the vertical chirals
\be
\tilde{V}^{(k)}\left(\Gp^{(k)}\right)^sV^{(k)}\equiv\Tr_k\left[\tilde{V}^{(k)}\left(\Gp^{(k)}\right)^sV^{(k)}\right],\qquad k=1,\cdots,N-1,\quad s=0,\cdots,k-1\, .\nn\\
\ee
The list of the chiral ring generators with the corresponding charges under the global symmetries is
\begin{table}[h]
\centering
\makebox[\linewidth][c]{
\scalebox{0.8}{
\begin{tabular}{c|cccc|c}
{} & $SU(N)_M$ & $SU(N)_T$ & $U(1)_{m_A}$ & $U(1)_\Gd$ & $U(1)_R$ \\ \hline
$\Gp^{(N)}$ & adj & 0 & 2 & 0 & $2R_A$ \\
$\mathcal{M}$ & 0 & adj & 2 & 0 & $2R_A$\\
$\Pi$ & $\Box$ & $\bar{\Box}$ & 0 & 1 & $R_\Gd$ \\
$\tilde{\Pi}$ & $\bar{\Box}$ & $\Box$ & 0 & 1 & $R_\Gd$ \\
$\tilde{D}^{(k)}\left(\Gp^{(k)}\right)^sD^{(k)}$ & 0 & 0 & $2(N-k+s)$ & 2 & $2(k-N)(1-R_A)+2sR_A+2R_\Gd$ \\
$\tilde{V}^{(k)}\left(\Gp^{(k)}\right)^sV^{(k)}$ & 0 & 0 & $2(k-N+s+2)$ & $-2$ & $2+2(N-k-2)(1-R_A)+2sR_A-2R_\Gd$
\end{tabular}}}
\label{mtsunop}
\end{table}

Finally, we can write down the partition function of the theory on the squashed three-sphere $S^3_b$ \cite{Jafferis:2010un,Hama:2010av,Hama:2011ea}, which is, together with the map of the chiral ring generators and the superconformal index, our main tool to test dualities. We turn on real masses in the Cartan of all the factors in the global symmetry group \eqref{mtsunglobalIR}, that we denote respectively with $M_a$, $T_a$, $\mathrm{Re}(m_A)$ and $\mathrm{Re}(\Gd)$. The parameters for the two $U(1)$ axial symmetries are defined as holomorphic combinations of the corresponding real masses with the R-symmetry mixing parameters $R_A$ and $R_\Gd$ \cite{Jafferis:2010un}
\be
m_A=\mathrm{Re}(m_A)+i\frac{Q}{2}R_A,\qquad \Gd=\mathrm{Re}(\Gd)+i\frac{Q}{2}R_\Gd\, ,
\ee
Then, the partition function can be written iteratively as (we follow the same conventions used in \cite{US1})
\be
&&\mathcal{Z}_{FM[U(N)]}(M_a,T_a,m_A,\Gd)=\underbrace{\prod_{a,b=1}^Ns_b\left(i\frac{Q}{2}+(M_a-M_b)-2m_A\right)}_{\Gp^{(N)}}\times\nn\\
&&\qquad\quad\times\underbrace{\prod_{a=1}^Ns_b\left(i\frac{Q}{2}\pm(M_a-T_N)-\Gd\right)}_{D^{(N)},\tilde{D}^{(N)}}\int\frac{\udl{x_{N-1}}}{\prod_{i<j}^{N-1}s_b\left(i\frac{Q}{2}\pm(x^{(N-1)}_i-x^{(N-1)}_j)\right)}\times\nn\\
&&\qquad\quad\times \underbrace{\prod_{i=1}^{N-1}s_b\left(\pm(x^{(N-1)}_i-T_N)+\Gd-m_A\right)}_{V^{(N-1)},\tilde{V}^{(N-1)}}\underbrace{\prod_{a=1}^Ns_b\left(\pm(x^{(N-1)}_i-M_a)+m_A\right)}_{Q^{(N-1,N)},\tilde{Q}^{(N-1,N)}}\times\nn\\
&&\qquad\quad\times \mathcal{Z}_{FM[U(N-1)]}\left(x^{(N-1)}_1,\cdots,x^{(N-1)}_{N-1},T_1,\cdots,T_{N-1},m_A,\Gd+m_A-i\frac{Q}{2}\right)\, ,\nn\\
\label{pffmsun}
\ee
where the integration measure is defined including the Weyl symmetry factor of the gauge group
\be
\udl{x_{k}}=\frac{1}{k!}\prod_{i=1}^k\udl{x^{(k)}_i}\, .
\ee
In order to make sense of the recursive definition we also specify
\be
&&\mathcal{Z}_{FM[U(1)]}(M,T,m_A,\Gd)=s_b\left(i\frac{Q}{2}-2m_A\right)s_b\left(i\frac{Q}{2}\pm(M-T)-\Gd\right)\, .
\label{pffmsu1}
\ee
The $\mathcal{Z}_{FM[SU(N)]}(M_a,T_a,m_A,\Gd)$ partition function is simply $\mathcal{Z}_{FM[U(N)]}(M_a,T_a,m_A,\Gd)$ with the tracelessness condition enforced for the fugacities of the $SU(N)_M$ and $SU(N)_T$ symmetries:
\be
\sum_{a=1}^NM_a=\sum_{a=1}^NT_a=1\, .
\label{tracecond}
\ee
For later convenience we also define the partition function $\mathcal{Z}_{FM[U(N)]}'(M_a,T_a,m_A,\Gd)$
where the adjoint  $\Gp^{(N)}$ associated to the flavor node is not present:
\be
&&\mathcal{Z}_{FM[U(N)]}'(M_a,T_a,m_A,\Gd)=\prod_{a=1}^Ns_b\left(i\frac{Q}{2}\pm(M_a-T_N)-\Gd\right)\times\nn\\
&&\qquad\quad\times \int\udl{x_{N-1}}\frac{\prod_{i,j=1}^{N-1}s_b\left(i\frac{Q}{2}\pm(x^{(N-1)}_i-x^{(N-1)}_j)-2m_A\right)}{\prod_{i<j}^{N-1}s_b\left(i\frac{Q}{2}\pm(x^{(N-1)}_i-x^{(N-1)}_j)\right)}\times\nn\\
&&\qquad\quad\times \prod_{i=1}^{N-1}s_b\left(\pm(x^{(N-1)}_i-T_N)+\Gd-m_A\right)\prod_{a=1}^Ns_b\left(\pm(x^{(N-1)}_i-M_a)+m_A\right)\times\nn\\
&&\qquad\quad\times \mathcal{Z}_{FM[U(N-1)]}'\left(x^{(N-1)}_1,\cdots,x^{(N-1)}_{N-1},T_1,\cdots,T_{N-1},m_A,\Gd+m_A-i\frac{Q}{2}\right)\, ,\nn\\
\label{pffmsunprime}
\ee
where the case $N=1$ is defined as
\be
&&\mathcal{Z}_{FM[U(1)]}'(M,T,\Gd)=s_b\left(i\frac{Q}{2}\pm(M-T)-\Gd\right)\, .
\label{pffmsu1prime}
\ee

\subsection{Self-duality}
\label{selfduality}

In this section we  provide evidences of the self-duality of the $FM[SU(N)]$ theory
which acts trivially on $U(1)_{m_A}$ and $U(1)_{\Gd}$ and exchanges
\be
SU(N)_M\leftrightarrow SU(N)_T\, ,
\ee
hence implying that the flavor symmetry $U(1)_T^{N-1}$ on the teeth of the saw enhances in the IR to the full $SU(N)_T$.
The map of the generators of the chiral ring is then an immediate guess. By looking at their charges under the global symmetries, we see that the adjoint chiral $\Gp^{(N)}$ gets exchanged with the matrix $\mathcal{M}$ and that also the two bifundamental mesons $\Pi$, $\tilde{\Pi}$ are exchanged
\be
\Gp^{(N)}\leftrightarrow\mathcal{M},\qquad\Pi\leftrightarrow\tilde{\Pi}\, .
\ee
All other operators constructed with the flavors of the saw are simply mapped into themselves, since they are uncharged under $SU(N)_M$ and $SU(N)_T$.

At the level of the $S^3_b$ partition function, the statement of the self-duality translates into the following integral identity:\footnote{
As explained in \cite{Pasquetti:2019hxf} this identity can also be obtained from the  $S^3\times S^1$ identity  encoding the self-duality of the $4d$ $E[USp(2N)]$ theory
(in the $S^1\to 0$ limit). Actually this and many other non-trivial identities for the dualities involving $4d$ $E[USp(2N)]$ theory have been proven in \cite{2014arXiv1408.0305R}.}

\be
\mathcal{Z}_N(M_a,T_a,m_A,\Gd)=\mathcal{Z}_N(T_a,M_a,m_A,\Gd)
\label{fmsunself}
\ee
For $N=2$, when the gauge group is abelian, this identity can be proved analytically with a piecewise procedure that we explicitly show in Appendix \ref{piecwisefm2}. When $N>2$, one can compute the superconformal index as a power series in the R-symmetry fugacity and verify that the coefficients of the expansion are invariant under $m_a\leftrightarrow t_a$ order by order. In Sec.~\ref{scifmsun} we present the results of this test for $N=2,3$.


\subsection{Real mass deformation to $FT[SU(N)]$}

In this section we show that, by taking a real mass deformation  associated to the $U(1)_\Gd$ symmetry, the $FM[SU(N)]$ theory reduces to the $FT[SU(N)]$ theory, which is the $T[SU(N)]$ theory with an additional set of singlets flipping the Higgs branch moment map \cite{Aprile:2018oau}. When this deformation is turned on, the chirals  $D^{(k)}$, $\tilde{D}^{(k)}$ and $V^{(k)}$, $\tilde{V}^{(k)}$ that form the saw of the quiver and are charged under $U(1)_\Gd$ become massive. Integrating out these fields, mixed CS-like couplings between the gauge symmetry and the $U(1)_T^{N-1}$ symmetry are generated, so that this is now identified with the restored topological symmetry. This in turns implies that the monopole operators are no longer turned on in the superpotential and that they are part of the chiral ring. We can then organize them, together with the traces of the adjoints at each gauge node, in a matrix transforming in the adjoint representation of the $SU(N)_T$ symmetry that enhances at low energies. This is the monopole matrix parameterizing the Coulomb branch of the $FT[SU(N)]$ theory. For example, for $N=3$ this matrix takes the form:
\be
\mathcal{M}=\begin{pmatrix}
0 & \mathfrak{M}^{(1,0)} & \mathfrak{M}^{(1,1)} \\
\mathfrak{M}^{(-1,0)} & 0 & \mathfrak{M}^{(0,1)} \\
\mathfrak{M}^{(-1,-1)} & \mathfrak{M}^{(0,-1)} & 0
\end{pmatrix}+\sum_{i=1}^2\Tr_i\Gp^{(i)}\mathcal{D}_i\, .
\ee
After the real mass deformation  the operators $\Pi$, $\tilde{\Pi}$ as well as all the mesons of the saw are integrated out, since they are charged under the $U(1)_\Gd$ symmetry. 
Finally the chiral $\Gp^{(N)}$ in the adjoint representation of the flavor symmetry $SU(N)_M$ maps to the same operator which parameterizes the Higgs branch of $FT[SU(N)]$. Hence, the chiral ring of $FM[SU(N)]$ reduces to that of $FT[SU(N)]$.

We can also look at the effect of the real mass deformation at the level of the sphere partition function, where it is implemented by taking  the limit $\Gd\rightarrow\infty$. This limit gives:
\be
\underset{\Gd\rightarrow\infty}{\lim}\mathcal{Z}_{FM[U(N)]}(M_a,T_a,m_A,\Gd)=C_N(\Gd,m_A)\e^{-i\pi\sum_{a=1}^N\left(M_a^2+T_a^2\right)}\mathcal{Z}_{FT[U(N)]}(M_a,T_a,m_A)\, ,\nn\\
\label{realmasstoftsun}
\ee
where the prefactor $C_N(m_A,\Gd)$,  independent from the flavor fugacities $M_a,T_a$, diverges for $\Gd\rightarrow\infty$ and the partition function of $FT[U(N)]$ can be defined iteratively as
\be
&&\mathcal{Z}_{FT[U(N)]}(M_a,T_a,m_A)=\e^{2\pi iT_N\sum_{a=1}^NM_a}\prod_{a,b=1}^N\sbfunc{i\frac{Q}{2}+(M_a-M_b)-2m_A}\times\nn\\
&&\qquad\qquad\times\int\udl{x_{N-1}}\e^{2\pi i(T_{N-1}-T_N)\sum_{i=1}^{N-1}x^{(N-1)}_i}\frac{\prod_{i=1}^{N-1}\prod_{a=1}^N\sbfunc{\pm(x^{(N-1)}_i-M_a)+m_A}}{\prod_{i<j}^N\sbfunc{i\frac{Q}{2}\pm(x^{(N-1)}_i-x^{(N-1)}_j)}}\times\nn\\
&&\qquad\qquad\times\mathcal{Z}_{FT[U(N-1)]}(x^{(N-1)}_1,\cdots,x^{(N-1)}_{N-1},T_1,\cdots,T_{N-1},m_A)\,,
\label{ftun}
\ee
with the case $N=1$ defined as
\be
&&\mathcal{Z}_{FT[U(1)]}(M,T,m_A)=\e^{2\pi iTM}\sbfunc{i\frac{Q}{2}-2m_A}\,.
\label{ftu1}
\ee

The proof of \eqref{realmasstoftsun} proceeds by induction. We prove it first for $FM[U(2)]$, whose partition function we recall being
\be
&&\mathcal{Z}_2\equiv\mathcal{Z}_{FM[U(2)]}(M_1,M_2,T_1,T_2,m_A,\Gd)= \prod_{a,b=1}^2s_b\left(i\frac{Q}{2}+(M_a-M_b)-2m_A\right)\times\nn\\
&&\times \prod_{a=1}^2s_b\left(i\frac{Q}{2}\pm(M_a-T_2)-\Gd\right)s_b\left(i\frac{Q}{2}-2m_A\right)\int\udl{x}s_b\left(iQ\pm(x-T_1)-\Gd-m_A\right)\times\nn\\
&&\times s_b\left(\pm(x-T_2)+\Gd-m_A\right)\prod_{a=1}^2s_b\left(\pm(x-M_a)+m_A\right)\, .\nn\\
\ee
We focus on the limit of the following block of double-sine functions depending on $\Delta$:
\be
&&\mathcal{B}_2=\prod_{a=1}^2s_b\left(i\frac{Q}{2}\mp(M_a-T_2)-\Gd\right)s_b\left(iQ\pm(x-T_1)-\Gd-m_A\right)\times\nn\\
&&\qquad\qquad\qquad\qquad\qquad\qquad\qquad\qquad\qquad\times s_b\left(\pm(x-T_2)+\Gd-m_A\right)\, .\nn\\
\ee
Using the asymptotic behaviour of the double-sine function
\be
\underset{x\rightarrow\pm\infty}{\lim}\sbfunc{x}=\e^{\pm i\frac{\pi}{2}x^2}\, ,
\label{sbasympotic}
\ee
we find
\be
\underset{\Gd\rightarrow\infty}{\lim}\mathcal{B}_2&=&\exp\left[i\pi\left(\frac{3}{2}Q^2+2im_A(Q+2i\Gd)+4iQ\Gd-2\Gd^2+\right.\right.\nn\\
&-&\left.\left.\sum_{a=1}^2(M_a^2+T_a^2)+2x(T_1-T_2)+2T_2\sum_{a=1}^2M_a\right)\right]\, .\nn\\
\label{realmassB2}
\ee
The rest of the partition function is independent from $\Gd$, so we find
\be
&&\underset{\Gd\rightarrow\infty}{\lim}\mathcal{Z}_2=C_2(m_A,\Gd)\e^{-i\pi\sum_{a=1}^2\left(M_a^2+T_a^2\right)}\e^{2\pi iT_2\sum_{a=1}^2M_a}\prod_{a,b=1}^2\sbfunc{i\frac{Q}{2}+(M_a-M_b)-2m_A}\times\nn\\
&&\qquad\qquad\times s_b\left(i\frac{Q}{2}-2m_A\right)\int\udl{x}\e^{2\pi i(T_1-T_2)x}\prod_{a=1}^2s_b\left(\pm(x-M_a)+m_A\right)=\nn\\
&&\qquad\qquad=C_2(m_A,\Gd)\e^{-i\pi\sum_{a=1}^2\left(M_a^2+T_a^2\right)}\mathcal{Z}_{FT[U(2)]}(M_1,M_2,T_1,T_2,m_A)\, ,\nn\\
\ee

Now we consider the recursive definition of the partition function of $FM[U(N+1)]$
\be
&&\mathcal{Z}_{N+1}\equiv\mathcal{Z}_{FM[U(N+1)]}(M_a,T_a,m_A,\Gd)=\prod_{a,b=1}^{N+1}s_b\left(i\frac{Q}{2}+(M_a-M_b)-2m_A\right)\times\nn\\
&&\qquad\qquad\times\prod_{a=1}^{N+1}s_b\left(i\frac{Q}{2}\pm(M_a-T_{N+1})-\Gd\right)\int\frac{\udl{x_{N}}}{\prod_{i<j}^{N}s_b\left(i\frac{Q}{2}\pm(x_i-x_j)\right)}\times\nn\\
&&\qquad\qquad\times \prod_{i=1}^{N}s_b\left(\pm(x_i-T_{N+1})+\Gd-m_A\right)\prod_{a=1}^{N+1}s_b\left(\pm(x_i-M_a)+m_A\right)\times\nn\\
&&\qquad\qquad\times \mathcal{Z}_{FM[U(N)]}(x_i,T_i,m_A,\Gd+m_A-i\frac{Q}{2})\, ,
\ee
Only two pieces of this partition function are affected by the $\Gd\rightarrow\infty$ limit. The first one is the partition function of the $FM[U(N)]$ subquiver, whose limit is given by the inductive hypothesis  \eqref{realmasstoftsun}.
The second one is the block of double-sine functions representing the last flavors of the saw $D^{(N+1)}$, $\tilde{D}^{(N+1)}$ and $V^{(N)}$, $\tilde{V}^{(N)}$
\be
&&\mathcal{B}_{N+1}=\prod_{a=1}^{N+1}s_b\left(i\frac{Q}{2}\mp(M_a-T_{N+1})-\Gd\right)\prod_{i=1}^{N}s_b\left(\pm(x_i-T_{N+1})+\Gd-m_A\right)\nn\\
&&\qquad\qquad\rightarrow\exp\left[i\pi\left( Nm_A^2+\frac{N+1}{4}Q^2-2Nm_A\Gd+(N+1)iQ\Gd-\Gd^2+\right.\right.\nn\\
&&\qquad\qquad\left.\left.-\sum_{a=1}^{N+1}M_a^2-T_{N+1}^2+2T_{N+1}\sum_{a=1}^{N+1}M_a-2T_{N+1}\sum_{i=1}^Nx_i+\sum_{i=1}^Nx_i^2\right)\right]\, .\nn\\
\label{realmassBN+1}
\ee
Notice that we have a quadratic term in the integration variable, which represents a CS coupling for the gauge field of the last node of the quiver. This precisely cancels with the corresponding term in \eqref{realmasstoftsun}. Hence, combining \eqref{realmasstoftsun} and \eqref{realmassBN+1} we get
\be
\underset{\Gd\rightarrow\infty}{\lim}\mathcal{Z}_{N+1}&=&C_{N+1}(\Gd,m_A)\e^{-i\pi\sum_{a=1}^{N+1}(M_a^2+T_a^2)}\e^{2\pi iT_{N+1}\sum_{a=1}^{N+1}M_a}\times\nn\\
&\times&\prod_{a,b=1}^{N+1}\sbfunc{i\frac{Q}{2}+(M_a-M_b)-2m_A}\int\udl{x_N}\e^{-2\pi iT_{N+1}\sum_{i=1}^Nx_i}\times\nn\\
&\times&\prod_{i=1}^N\prod_{a=1}^{N+1}\sbfunc{\pm(x_i-M_a)+m_A}\mathcal{Z}_{FT[U(N)]}(x_i,T_i,m_A)=\nn\\
&=&C_{N+1}(\Gd,m_A)\e^{-i\pi\sum_{a=1}^{N+1}(M_a^2+T_a^2)}\mathcal{Z}_{FT[U(N+1)]}(M_a,T_a,T_{N+1},m_A)\, .\nn\\
\ee
where in the last step we used the recursive definition \eqref{ftun} of the $FT[U(N)]$. This concludes the proof of \eqref{realmasstoftsun} for arbitrary $N$.

If we take the real mass deformation on the two sides of the self-duality identity \eqref{fmsunself}, the divergent prefactor and the mixed CS terms  $C_N(\Gd,m_A)\e^{-i\pi\sum_{a=1}^N\left(M_a^2+T_a^2\right)}$ cancel out since they are symmetric under $M_a\leftrightarrow T_a$ and we obtain the  identity for the spectral duality of $FT[SU(N)]$ \cite{Aprile:2018oau}
\be
\mathcal{Z}_{FT[SU(N)]}(M_a,T_a,m_A)=\mathcal{Z}_{FT[SU(N)]}(T_a,M_a,m_A)\, ,
\ee
Notice that, unlike mirror-symmetry,  spectral duality  swaps the fugacities $M_a$ and $T_a$ without changing the sign to $m_A$.

\section{\boldmath $G[U(N)]$ and its recombination dual frames }
\label{recombinationduality}

In this section we  introduce $G[U(N)]$, a quiver theory closely related to $FM[SU(N)]$, which enjoys various amusing dualities
that we are going to discuss.

\subsection{The $\gun$ theory}
\label{gun}

The $\gun$ theory, depicted in Figure \ref{gunfig}, is obtained from  $FM[U(N)]$ by gauging the last flavor node
(with no adjoint) and adding one fundamental flavor $P$, $\tilde{P}$.
 The superpotential is
\be
\mathcal{W}_{G[U(N)]}=\mathcal{W}_{mono}+\mathcal{W}_{T[U(N)]}+\mathcal{W}_{cub}\, ,
\label{Wgun}
\ee
where\footnote{The reason why in this case we have $\mathcal{W}_{T[U(N)]}$ rather than $\mathcal{W}_{FT[U(N)]}$ as in \eqref{wfmun} is precisely because we don't have the adjoint $\Gp^{(N)}$.}
\be
\mathcal{W}_{T[U(N)]}=\sum_{k=1}^{N-1}\mathrm{Tr}_k\left[\Gp^{(k)}\left(\mathrm{Tr}_{k+1}\mathbb{Q}^{(k,k+1)}-\mathrm{Tr}_{k-1}\mathbb{Q}^{(k-1,k)}\right)\right]\,.
\ee
Since the extra flavor doesn't interact with any other field, we have an additional $U(1)_\mu$ flavor symmetry. Moreover, we have no monopole superpotential associated to the $U(N)$ node, which means that its topological symmetry $U(1)_{\zeta}$ is not broken. Hence, the complete global symmetry group of Theory A is\footnote{Since we used the freedom due to the gauge symmetry to fix the baryonic symmetry of the flavor $P$, $\tilde{P}$, the flavor symmetry associated to the saw is now the full $U(N)_z$ group.}
\be
U(N)_z\times U(1)_{m_A}\times U(1)_\Gd\times U(1)_\mu\times U(1)_\zeta\,,
\label{recombglobalsymm}
\ee
where the $U(N)_z$ symmetry is not manifest in the UV, but it enhances in the IR. This can be understood from the fact that the chiral ring generators of $\gun$ re-organize into representations of $U(k)_z$, as we will show below, but it will become evident also in Sec.~\ref{rankstabilizationduality} where we will discuss a dual frame for $\gun$ in which the full $U(k)_z$ symmetry is manifest. 

Since $U(1)_{m_A}$, $U(1)_\Gd$ and  $U(1)_\mu$  are abelian symmetries that can mix with the R-symmetry, the corresponding parameters are actually defined as the holomorphic combinations
\be
m_A=\mathrm{Re}(m_A)+i\frac{Q}{2}R_A\,,\qquad\Gd=\mathrm{Re}(\Gd)+i\frac{Q}{2}R_\Gd\, ,\qquad\mu=\mathrm{Re}(\mu)+i\frac{Q}{2}r
\label{axialmassesgun}
\ee
where $R_A$, $R_\Gd$, $r$ are the mixing coefficients. In Table \ref{chargesgaugedmun} we summarize the charges under these symmetries of all the chiral fields of the theory.

\begin{table}[t]
\centering
\scalebox{0.96}{
\begin{tabular}{c|cccc|c}
{} & $U(1)_{z_a}$ & $U(1)_{m_A}$ & $U(1)_\Gd$ & $U(1)_\mu$ & $U(1)_R$ \\ \hline
$Q^{(a-1,a)}$ & 0 & -1 & 0 & 0 & $1-R_A$ \\
$\tilde{Q}^{(a-1,a)}$ & 0 & -1 & 0 & 0 & $1-R_A$ \\ 
$P$ & 0 & 0 & 0 & 1 & $r$ \\
$\tilde{P}$ & 0 & 0 & 0 & 1 & $r$ \\
$V^{(a-1)}$ & 1 & $a-N+1$ & -1 & 0 & $2+(N-a-1)(1-R_A)-R_\Gd$ \\
$\tilde{V}^{(a-1)}$ & -1 & $a-N+1$ & -1 & 0 & $2+(N-a-1)(1-R_A)-R_\Gd$ \\ 
$D^{(a)}$ & -1 & $N-a$ & 1 & 0 & $(a-N)(1-R_A)+R_\Gd$ \\
$\tilde{D}^{(a)}$ & 1 & $N-a$ & 1 & 0 & $(a-N)(1-R_A)+R_\Gd$ \\
$\Gp^{(a)}$ & 0 & 2 & 0 & 0 & $2R_A$ \\
\end{tabular}}
\caption{In the table, $a$ runs from 1 to $N$. By definition, $Q^{(0,1)}=\tilde{Q}^{(0,1)}=0$, $V^{(0)}=\tilde{V}^{(0)}=0$ and $\Gp^{(N)}=0$.}
\label{chargesgaugedmun}
\end{table}

Some of the chiral ring generators of $\gun$ are similar to those of the $FM[U(N)]$ theory. Firstly, we have the operator $\mathcal{M}$, which is constructed exactly as for $FM[U(N)]$. We then have the operators $\Omega$, $\tilde{\Omega}$ which are constructed by attaching the new chiral fields $P$, $\tilde{P}$ to the  $\Pi$, $\tilde{\Pi}$ operators of $FM[U(N)]$ so to have gauge invariant objects. For example, for $N=3$ we have
\be
\Omega=\begin{pmatrix}
P_a\tilde{Q}^{(2,3)}_{i,a}\tilde{Q}^{(1,2)}_iD^{(1)}\\P_a\tilde{Q}^{(2,3)}_{i,a}D^{(2)}_i\\P_aD^{(3)}_a
\end{pmatrix},
\qquad
\tilde{\Omega}=\begin{pmatrix}
\tilde{D}^{(1)}Q^{(1,2)}_iQ^{(2,3)}_{i,a}\tilde{P}_a\\ \tilde{D}^{(2)}_iQ^{(2,3)}_{i,a}\tilde{P}_a\\ \tilde{D}^{(3)}_a\tilde{P}_a
\end{pmatrix}\, .
\ee

Then, we have the dressed mesons and the dressed monopoles \cite{Cremonesi:2013lqa}
\be
\mathfrak{M}^{\pm}_{\mathbb{M}^s},\qquad\Tr_N\left(\tilde{P}\mathbb{M}^sP\right),\qquad s=0,\cdots,N-1\,,
\ee
where $\mathfrak{M}^\pm$ are the fundamental monopoles associated to the $U(N)$ gauge node, which are not turned on in the superpotential. The dressing is performed with the  meson matrix constructed with the last bifundamental of the tail
\be
\mathbb{M}=\Tr_{N-1}Q^{(N-1,N)}\tilde{Q}^{(N-1,N)}\,,
\ee
which transforms in the adjoint representation of $U(N)$.

In $FM[U(N)]$ we also have a bunch of mesonic operators made with the flavors of the saw, that are singlets with respect to the flavor symmetry. We claim that, among those discussed in Sec.~\ref{fmsuntheory}, only the ones constructed from the diagonal flavors
\be
\tilde{D}^{(k)}D^{(k)},\qquad k=1,\dots,N\, ,
\ee
are chiral ring generators of $\gun$, while we expect the others to be composite operators because of non-trivial quantum effects. This statement is supported by the several dualities involving the $G[U(N)]$ theory that we will present.

The charges under the global symmetries of the chiral ring generators are
\begin{table}[h]
\centering
\scalebox{0.88}{
\begin{tabular}{c|ccccc|c}
{} & $U(N)_z$ & $U(1)_{m_A}$ & $U(1)_\Gd$ & $U(1)_\mu$ & $U(1)_\zeta$ & $U(1)_R$ \\ \hline
$\mathcal{M}$ & adj & 2 & 0 & 0 & 0 & $2R_A$\\
$\Omega$ & $\bar{\Box}$ & 0 & 1 & 1 & 0 & $R_\Gd+r$ \\
$\tilde{\Omega}$ & $\Box$ & 0 & 1 & 1 & 0 & $R_\Gd+r$ \\
$\tilde{D}^{(a)}D^{(a)}$ & 0 & $2(N-a)$ & 2 & 0 & 0 & $2(a-N)(1-R_A)+2R_\Gd$ \\
$\mathfrak{M}^{\pm}_{\mathbb{M}^s}$ & 0 & $N-2s-1$ & $-1$ & $-1$ & $\pm1$ & $2-(N-2s-1)(1-R_A)-R_\Gd-r$ \\
$\Tr_N\left(\tilde{P}\mathbb{M}^sP\right)$ & 0 & $-2s$ & 0 & 2 & 0 & $2s(1-R_A)+2r$
\end{tabular}}
\end{table}

\subsection{Recombination dual}

We propose a recombination property of $\gun$, which actually provides a set of several duality frames for the theory. These dual theories are obtained from a $G[U(N-k)]$ and a $G[U(k)]$ tail, where $k\le N$, glued together with a bifundamental flavor $q_{LR}$. The fundamental flavors $p_L$, $\tilde{p}_L$ and $p_R$, $\tilde{p}_R$ attached to the ends of the two tails transform under the same symmetry $U(1)_\mu$. Moreover, all the $U(1)_{z_{N-n+1}}$ nodes, for $n=1,\cdots,k$, are connected to the $U(1)_\mu$ node by some gauge singlets $\chi_n$, $\tilde{\chi}_n$\footnote{Notice that the $U(1)_{z_i}$ symmetries of the $G[U(N-k)]$ tail are ordered in the usual way, that is $U(1)_{z_1}$ corresponds to the leftmost square node of the $G[U(N-k)]$ subquiver of Figure \ref{gluedmun} and $U(1)_{z_{N-k}}$ to the rightmost one, while the $U(1)_{z_{N-n+1}}$ symmetries of the $G[U(k)]$ tail are ordered in the opposite way, that is $U(1)_{z_N}$ corresponds rightmost square node of the $G[U(k)]$ subquiver (which appears reversed in Figure \ref{gluedmun}) and $U(1)_{z_{N-k+1}}$ to the leftmost one.}. The complete structure of the theory is represented in the quiver of Figure \ref{gluedmun}. On top of this, we also have $4k$ gauge singlets that we denote by $S^\pm_n$, $\ga_n$ and $\gb_n$.

The superpotential of the dual theory is
\be
\mathcal{W}_{recomb}&=&\mathcal{W}_{G[U(N-k)]}+\mathcal{W}_{G[U(k)]}+\mathcal{W}_{mid}+\mathcal{W}_{flips}\, .
\label{recombinationsuperpotB}
\ee
The first two terms are the usual superpotential \eqref{Wgun} for the two tails $G[U(N-k)]$ and $G[U(k)]$. The third term contains some cubic and quartic couplings and a monopole superpotential that relate the tails
\be
\mathcal{W}_{mid}&=&\Tr_{N-k}\left(\Tr_kq_{LR}\tilde{q}_{LR}\right)\left(\Tr_{N-k-1}q_R^{(N-k-1,N-k)}\tilde{q}_R^{(N-k-1,N-k)}\right)+\nn\\
&-&\Tr_{k}\left(\Tr_{N-k}q_{LR}\tilde{q}_{LR}\right)\left(\Tr_{k-1}q_L^{(k-1,k)}\tilde{q}_L^{(k-1,k)}\right)+\nn\\
&+&\Tr_{k}\left(p_R\Tr_{N-k}\left(q_{LR}\tilde{p}_L\right)\right)+\Tr_{N-k}\left(p_L\Tr_{k}\left(\tilde{q}_{LR}\tilde{p}_R\right)\right)+\nn\\
&+&\mathfrak{M}^{(0,\cdots,0,1,1,0,\cdots,0)}+\mathfrak{M}^{(0,\cdots,0,-1,-1,0,\cdots,0)}\, .
\ee
The last term involves the monopoles with non-vanishing magnetic fluxes corresponding to the $U(N-k)$ and $U(k)$ gauge nodes only. This has the effect of breaking the two topological symmetries of these nodes to their anti-diagonal combination, which is mapped to the $U(1)_{\zeta}$ symmetry of the dual $G[U(N)]$ theory.
\begin{figure}[t]
	\centering
	\makebox[\linewidth][c]{
	\includegraphics[scale=0.4]{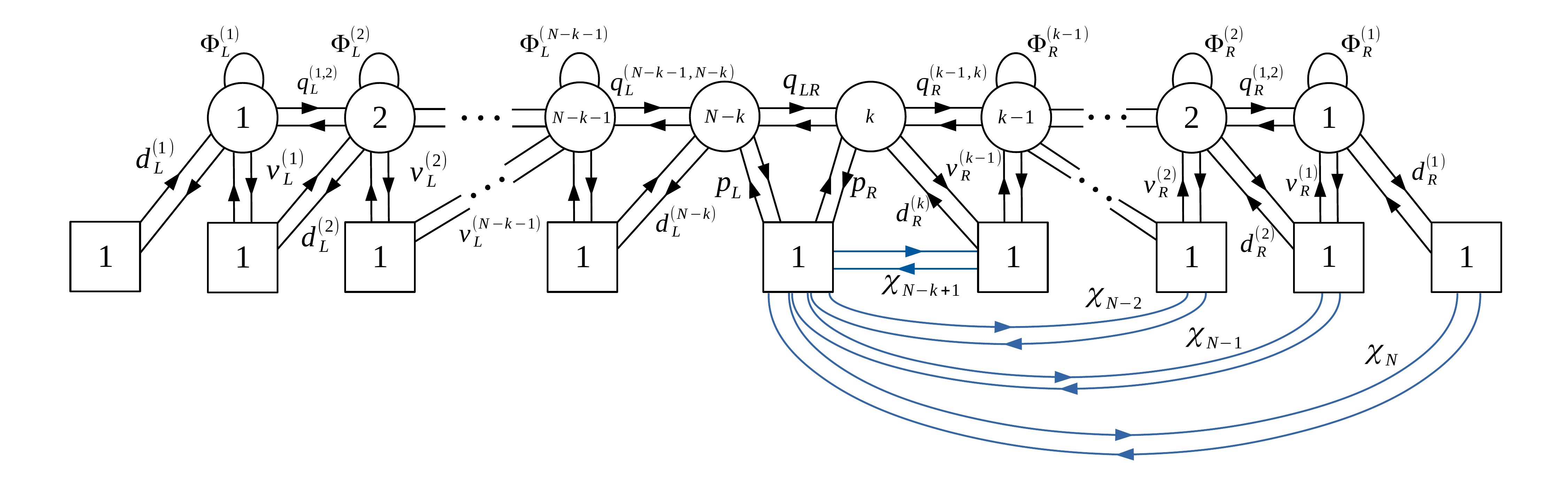}
	}
	\caption{Quiver diagram of the magnetic theory. The blue lines represent gauge singlets that transform under the flavor symmetries of the nodes they connect.}
	\label{gluedmun}
\end{figure}
Finally, we have some flip terms
\be
&&\mathcal{W}_{flips}=\sum_{n=1}^k\left(S_n^\pm \mathfrak{M}^\pm_{(k) \mathbb{M}_{R}^{k-n}}+\ga_n\Tr_k\left(\tilde{p}_R\mathbb{M}_R^{n-1}p_R\right)+\gb_n\tilde{d}_R^{(n)}d_R^{(n)}+\right.\nn\\
&&\qquad\qquad\qquad\qquad\qquad\qquad\qquad\qquad+\left.\chi_{N-n+1}\Omega_{R,n}+\tilde{\chi}_{N-n+1}\tilde{\Omega}_{R,n}\right)\, ,
\ee
where $\mathfrak{M}^\pm_{(k)}$ denote the fundamental monopoles of the $U(k)$ gauge node, which can be dressed with the meson matrix
\be
\mathbb{M}_{R}=\Tr_{k-1}q_R^{(k-1,k)}\tilde{q}_R^{(k-1,k)}
\ee
transforming in the adjoint representation of $U(k)$, and $\Omega_{R,n}$ denotes the $n$-th component of the vector $\Omega$ associated to the right $G[U(k)]$ tail. In Table \ref{chargesgluedmun} we summarize the charges under the global symmetries of all the chiral fields of the theory.
\begin{table}[t]
\centering
\makebox[\linewidth][c]{
\scalebox{0.82}{
\begin{tabular}{c|cccccc|c}
{} & $U(1)_{z_i}$ & $U(1)_{z_{N-n+1}}$ & $U(1)_{m_A}$ & $U(1)_\Gd$ & $U(1)_\mu$ & $U(1)_\zeta$ & $U(1)_R$ \\ \hline
$S_n^\pm$ & 0 & 0 & $N-2n+1$ & $-1$ & $-1$ & $\pm1$ & $2-(N-2n+1)(1-R_A)-R_\Gd-r$ \\
$\ga_n$ & 0 & 0 & $-2(n-1)$ & 0 & 2 & 0 & $2(n-1)(1-R_A)+2r$ \\
$\gb_n$ & 0 & 0 & $2(n-1)$ & 2 & 0 & 0 & $-2(n-1)(1-R_A)+2R_\Gd$ \\
$\chi_{N-n+1}$ & 0 & 1 & 0 & 1 & 1 & 0 & $R_\Gd+r$ \\
$\tilde{\chi}_{N-n+1}$ & 0 & -1 & 0 & 1 & 1 & 0 & $R_\Gd+r$ \\
$q_L^{(i-1,i)}$ & 0 & 0 & $-1$ & 0 & 0 & 0 & $1-R_A$ \\
$\tilde{q}_L^{(i-1,i)}$ & 0 & 0 & $-1$ & 0 & 0 & 0 & $1-R_A$ \\
$q_R^{(n-1,n)}$ & 0 & 0 & $-1$ & 0 & 0 & 0 & $1-R_A$ \\
$\tilde{q}_R^{(n-1,n)}$ & 0 & 0 & $-1$ & 0 & 0 & 0 & $1-R_A$ \\
$q_{LR}$ & 0 & 0 & 1 & 0 & 0 & 0 & $R_A$ \\
$\tilde{q}_{LR}$ & 0 & 0 & 1 & 0 & 0 & 0 & $R_A$ \\
$p_L$ & 0 & 0 & $-k$ & 0 & 1 & 0 & $k(1-R_A)+r$ \\
$\tilde{p}_L$ & 0 & 0 & $-k$ & 0 & 1 & 0 & $k(1-R_A)+r$ \\
$p_R$ & 0 & 0 & $k-1$ & 0 & $-1$ & 0 & $1-(k-1)(1-R_A)-r$ \\
$\tilde{p}_R$ & 0 & 0 & $k-1$ & 0 & $-1$ & 0 & $1-(k-1)(1-R_A)-r$ \\
$v_L^{(i-1)}$ & 1 & 0 & $i-N+1$ & $-1$ & 0 & 0 & $2+(N-i-1)(1-R_A)-R_\Gd$ \\
$\tilde{v}_L^{(i-1)}$ & $-1$ & 0 & $i-N+1$ & $-1$ & 0 & 0 & $2+(N-i-1)(1-R_A)-R_\Gd$ \\
$v_R^{(n-1)}$ & 0 & 1 & $n$ & 1 & 0 & 0 & $1-n(1-R_A)+R_\Gd$ \\
$\tilde{v}_R^{(n-1)}$ & 0 & $-1$ & $n$ & 1 & 0 & 0 & $1-n(1-R_A)+R_\Gd$ \\
$d_L^{(i)}$ & $-1$ & 0 & $N-i$ & 1 & 0 & 0 & $(i-N)(1-R_A)+R_\Gd$ \\
$\tilde{d}_L^{(i)}$ & $1$ & 0 & $N-i$ & 1 & 0 & 0 & $(i-N)(1-R_A)+R_\Gd$ \\
$d_R^{(n)}$ & 0 & $-1$ & $1-n$ & $-1$ & 0 & 0 & $1+(n-1)(1-R_A)-R_\Gd$ \\
$\tilde{d}_R^{(n)}$ & 0 & $1$ & $1-n$ & $-1$ & 0 & 0 & $1+(n-1)(1-R_A)-R_\Gd$ \\
$\Gp^{(j)}_L$ & 0 & 0 & 2 & 0 & 0 & 0 & $2R_A$ \\
$\Gp^{(m)}_R$ & 0 & 0 & 2 & 0 & 0 & 0 & $2R_A$ \\
\end{tabular}}}
\caption{In the table, $i$ runs from 1 to $N-k$, $j$ from 1 to $N-k-1$, $n$ from 1 to $k$ and $m$ from 1 to $k-1$.}
\label{chargesgluedmun}
\end{table}

\begin{figure}[t]
	\centering
	\includegraphics[scale=0.6]{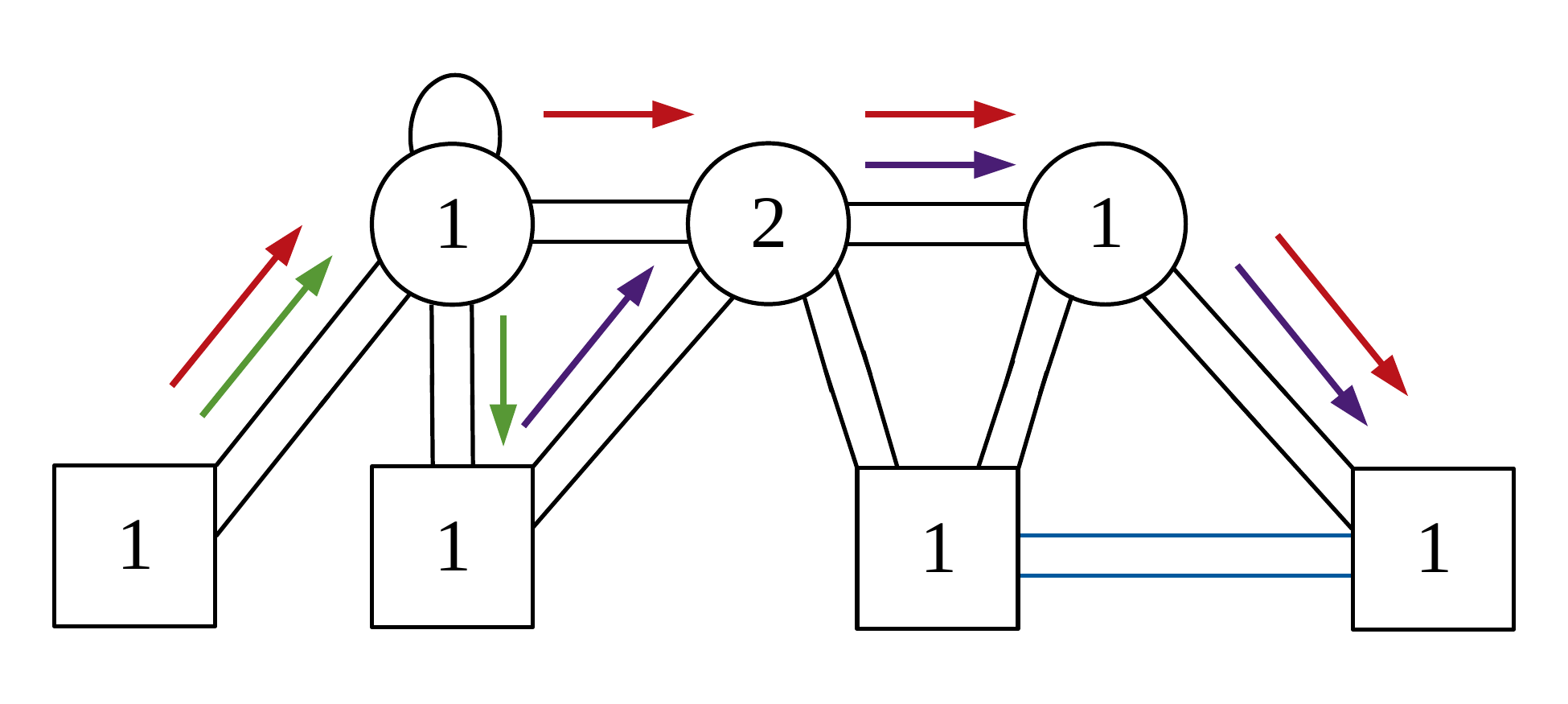}
	\caption{Diagrammatic representation of the operator $\hat{\mathcal{M}}$ in the case $N=3$ and $k=1$. Arrows of the same color represent chiral fields that we assemble to construct an element of the matrix. In order to have a gauge invariant operators, we have to consider sequences of arrows that start and end on a squared node.}
	\label{Mmatrixgluedmun}
\end{figure}

The chiral ring generators are basically obtained by gluing those of the two $G[U(N)]$ tails. First, we have an operator that we denote $\hat{\mathcal{M}}$ which transforms in the adjoint representation of $U(N)_z$. This consists of four blocks. The two on the diagonal are respectively $(N-k)\times (N-k)$ and $k\times k$ matrices that correspond to the usual $\mathcal{M}$ operator of $G[U(N-k)]$ and $G[U(k)]$. Recall that these are constructed starting with one of the diagonal flavor, moving along the tail following the bifundamentals and then ending on one of the vertical flavors. On the diagonal we still have the traces of the adjoint chirals, but since we have only $N-2$ of them one element has to be
\be
\Tr \mathbb{M}_{LR}=\Tr_{N-k}\Tr_k q_{LR}\tilde{q}_{LR}\, .
\ee
The off-diagonal blocks are built in a similar way, but going from one tail to the other using the bifundamental $q_{LR}$ as a link and ending on one of the diagonal flavors of the opposite tail rather than a vertical one (see Figure \ref{Mmatrixgluedmun}). For example, for $N=3$ and $k=1$ this matrix takes the form
\be 
\hat{\mathcal{M}}=\begin{pmatrix}
0 & v_L^{(1)}d_L^{(1)} &  d_R^{(1)}\tilde{q}_{LR,i}\tilde{q}^{(1,2)}_id_L^{(1)} \\
\tilde{d}_L^{(1)}\tilde{v}_L^{(1)} & 0 & d_R^{(1)}\tilde{q}_{LR,i}d_{L,i}^{(2)} \\
\tilde{d}_L^{(1)}q^{(1,2)}_i q_{LR,i}\tilde{d}_R^{(1)} & \tilde{d}_{L,i}^{(2)} q_{LR,i}\tilde{d}_R^{(1)} & 0
\end{pmatrix}+\Gp^{(1)}\mathcal{D}_1+\Tr \mathbb{M}_{LR}\mathcal{D}_2\, ,\nn\\
\ee

Then, we have the operators $\hat{\Omega}$, $\tilde{\hat{\Omega}}$. One may think that they are obtained by simply juxtaposing the vectors $\Omega_L$ and $\Omega_R$ of the two tails, but this is not possible since they have not the same charges under the global symmetries. Moreover, the operators of the right tail are set to zero in the chiral ring by the equations of motion of the flipping fields $\chi_n$. The correct operators are then
\be
\hat{\Omega}=\tilde{\Omega}_R\oplus \begin{pmatrix}\chi_{N-k+1} \\ \vdots \\ \chi_N \end{pmatrix},\qquad\tilde{\hat{\Omega}}=\Omega_R\oplus \begin{pmatrix}\tilde{\chi}_{N-k+1} \\ \vdots \\ \chi_N \end{pmatrix}\, .
\ee
These transform in the fundamental and anti-fundamental representation of the flavor symmetry $U(N)_z$ respectively. 

Something similar happens for the mesonic operators of the saw. Those in the left tail are truly generators of the chiral ring, but those of the right tail are flipped by the singlets $\gb_n$. Hence, the complete tower of $N$ generators of this type is
\be
\begin{cases}
d_L^{(i)}\tilde{d}_L^{(i)} & i=1,\cdots,N-k \\
\gb_n & n=1,\cdots,k
\end{cases}\, ,
\ee

Let's now consider the monopole operators and their dressings. Only those associated to the $U(N-k)$ node are generators, since those at the $U(k)$ node are flipped by the singlets $S_n^\pm$ (recall that the monopoles of the other gauge nodes are turned on in the superpotential). Hence, we have the following $2N$ generators
\be
\begin{cases}
\mathfrak{M}^\pm_{\mathbb{M}_{L}^s} & s=0,\cdots,N-k-1 \\
S^\pm_n & n=1,\cdots,k
\end{cases}\, ,
\ee
where $\mathfrak{M}^\pm$ denotes the fundamental monopoles of the $U(N-k)$ node, which are dressed with the field
\be
\mathbb{M}_{L}=\Tr_{N-k-1}q_L^{(N-k-1,N-k)}\tilde{q}_L^{(N-k-1,N-k)}
\ee
that transforms in the adjoint representation of $U(N-k)$. 

Finally, we have the (dressed) mesons associated to the extra flavors of the two tails $p_L$, $\tilde{p}_L$, $p_R$, $\tilde{p}_R$, where the dressing is made using the matrices $\mathbb{M}_L$ and $\mathbb{M}_R$. Again, these operators are flipped in the right tail by the gauge singlets $\ga_n$. Thus, the last set of $N$ chiral ring generators is
\be
\begin{cases}
\Tr_{N-k}\left(\tilde{p}_L\mathbb{M}_L^sp_L\right) & s=0,\cdots,N-k-1 \\
\ga_n & n=1,\cdots,k
\end{cases}\, ,
\ee

Summing up, the chiral ring generators and their charges under the global symmetries are
\begin{table}[h]
\centering
\scalebox{0.82}{
\begin{tabular}{c|ccccc|c}
{} & $U(N)_{z}$ & $U(1)_{m_A}$ & $U(1)_\Gd$ & $U(1)_\mu$ & $U(1)_\zeta$ & $U(1)_R$ \\ \hline
$\hat{\mathcal{M}}$ & adj & 2 & 0 & 0 & 0 & $2R_A$\\
$\hat{\Omega}$ & $\bar{\Box}$ & 0 & 1 & 1 & 0 & $R_\Gd+r$ \\
$\tilde{\hat{\Omega}}$ & $\Box$ & 0 & 1 & 1 & 0 & $R_\Gd+r$ \\
$\gb_n$ & 0 & $2(n-1)$ & 2 & 0 & 0 & $-2(n-1)(1-R_A)+2R_\Gd$ \\
$d_L^{(i)}\tilde{d}_L^{(i)}$ & 0 & $2(N-i)$ & 2 & 0 & 0 & $-2(N-i)(1-R_A)+2R_\Gd$ \\
$S_n^\pm$ & 0 & $N-2n+1$ & $-1$ & $-1$ & $\pm1$ & $2-(N-2n+1)(1-R_A)-R_\Gd-r$ \\
$\mathfrak{M}^\pm_{\mathbb{M}_{L}^s}$ & 0 & $N-2k-2s-1$ & $-1$ & $-1$ & $\pm1$ & $2-(N-2k-2s-1)(1-R_A)-R_\Gd-r$ \\
$\ga_n$ & 0 & $-2(n-1)$ & 0 & 2 & 0 & $2(n-1)(1-R_A)+2r$ \\
$\Tr_{N-k}\left(p_L\mathbb{M}_L^s\tilde{p}_L\right)$ & 0 & $-2(k+s)$ & 0 & 2 & 0 & $2(k+s-1)(1-R_A)+2r$
\end{tabular}}
\end{table}

\newpage

As a first check of the duality we can map the generators of the chiral ring of the two theories
\be
\mathcal{M}\quad&\leftrightarrow&\quad\hat{\mathcal{M}}\nn\\
\Omega\quad&\leftrightarrow&\quad\hat{\Omega}\nn\\
\tilde{\Omega}\quad&\leftrightarrow&\quad\hat{\tilde{\Omega}}\nn\\
D^{(a)}\tilde{D}^{(a)}\quad&\leftrightarrow&\quad\begin{cases}
d_L^{(a)}\tilde{d}_L^{(a)} & a=1,\cdots,N-k \\
\gb_{N-a+1} & a=N-k+1,\cdots,N
\end{cases}\nn\\
\mathfrak{M}^{\pm}_{\mathbb{M}^{a-1}}\quad&\leftrightarrow&\quad\begin{cases}
S^\pm_a & a=1,\cdots,k \\
\mathfrak{M}^\pm_{\mathbb{M}_{L}^{a-k-1}} & a=k+1,\cdots,N
\end{cases}\nn\\
\Tr_N\left(P\mathbb{M}^{a-1}\tilde{P}\right)\quad&\leftrightarrow&\quad\begin{cases}
\ga_a & a=1,\cdots,k \\
\Tr_{N-k}\left(p_L\mathbb{M}_L^{a-k-1}\tilde{p}_L\right) & a=k+1,\cdots,N
\end{cases}
\ee

At the level of $S^3_b$ partition functions, the recombination duality is represented by the following integral identity
\be
&&\mathcal{Z}_{G[U(N)]}(z_a,\zeta,\mu,m_A,\Gd)=\int\udl{x_N}\e^{2\pi i\zeta\sum_{a=1}^Nx^{(N)}_a}\frac{\prod_{a=1}^N\sbfunc{i\frac{Q}{2}\pm x^{(N)}_a-\mu}}{\prod_{a<b}^N\sbfunc{i\frac{Q}{2}\pm(x^{(N)}_a-x^{(N)}_b)}}\times\nn\\
&&\qquad\qquad\qquad\qquad\qquad\quad\times\mathcal{Z}'_{FM[U(N)]}(x^{(N)}_a,z_a,m_A,\Gd)=\nn\\
&&\qquad=\Gl^N_k(m_A,\Gd,\zeta,\mu)\prod_{n=N-k+1}^N\e^{2\pi i\zeta z_n}\sbfunc{i\frac{Q}{2}\pm z_n-\mu-\Gd}\times\nn\\
&&\qquad\times\int\udl{x_{N-k}}\e^{2\pi i\zeta\sum_{i=1}^{N-k}x^{(N-k)}_i}\frac{\prod_{i=1}^{N-k}\sbfunc{i\frac{Q}{2}\pm x^{(N-k)}_i-\mu-k(i\frac{Q}{2}-m_A)}}{\prod_{i<j}^{N-k}\sbfunc{i\frac{Q}{2}\pm(x^{(N-k)}_i-x^{(N-k)}_j)}}\times\nn
\ee
\be
&&\qquad\times\mathcal{Z}'_{FM[U(N-k)]}\left(x^{(N-k)}_1,\cdots,x^{(N-k)}_{N-k},z_1,\cdots,z_{N-k},m_A,\Gd-k\left(i\frac{Q}{2}-m_A\right)\right)\times\nn\\
&&\qquad\times\int\udl{y_k}\e^{-2\pi i\zeta\sum_{n=1}^ky^{(k)}_n}\frac{\prod_{n=1}^k\sbfunc{\pm y^{(k)}+\mu+(k-1)\left(i\frac{Q}{2}-m_A\right)}}{\prod_{n<m}^k\sbfunc{i\frac{Q}{2}\pm(y^{(k)}_n-y^{(k)}_m)}}\times\nn\\
&&\qquad\times\prod_{n=1}^k\prod_{i=1}^{N-k}\sbfunc{i\frac{Q}{2}\pm(y^{(k)}_n-x^{(N-k)}_n)-m_A}\nn\\
&&\qquad\times\mathcal{Z}'_{FM[U(k)]}\left(y^{(k)}_1,\cdots,y^{(k)}_k,z_N,\cdots,z_{N-k+1},m_A,m_A-\Gd+k\left(i\frac{Q}{2}-m_A\right)\right)\nn\\
\label{recombination}
\ee
where $\Gl^N_k$ is the contribution of the $4k$ flipping singlets $S_n^\pm$, $\ga_n$ and $\gb_n$
\be
\Gl^N_k(m_A,\Gd,\zeta,\mu)&=&\prod_{n=1}^k\sbfunc{\pm\zeta+\mu+\Gd-m_A+(N-2n)\left(i\frac{Q}{2}-m_A\right)}\times\nn\\
&\times&\sbfunc{i\frac{Q}{2}-2\mu-2(n-1)\left(i\frac{Q}{2}-m_A\right)}\times\nn\\
&\times&\sbfunc{i\frac{Q}{2}-2\Gd+2(n-1)\left(i\frac{Q}{2}-m_A\right)}\, .
\ee
The parameters on which the partition function depends are the real masses $z_a$ for the flavor symmetry $U(N)_z$, the axial masses $m_A$, $\Gd$, $\mu$ for the axial symmetries $U(1)_{m_A}\times U(1)_\Gd\times U(1)_\mu$ and the FI parameter $\zeta$ corresponding to the topological symmetry $U(1)_\zeta$. This identity will be proven in Appendix \ref{partfuncrecomb} following the procedure we are going to describe in the next section from the field theory point of view.

\subsection{Derivation}
\label{piecewisefieldtheory}

In this section, we show how the recombination duality can be derived by sequentially applying Aharony duality \cite{Aharony:1997gp} (see Appendix \ref{funddualities}), starting from the last $U(N)$ node whose monopoles are not turned on in the superpotential. As we discussed in \cite{US1}, the effect of the contact terms of Aharony duality is to modify the quantum numbers of the monopole operators of the adjacent nodes and, in this particular case, to remove those of the $U(N-1)$ node from the superpotential. This allows us to apply again Aharony duality on the next node. Repeating this procedure for an arbitrary number $k$ of iterations, we obtain exactly the claimed duality. 


\begin{figure}[t]
	\centering
	\includegraphics[scale=0.65]{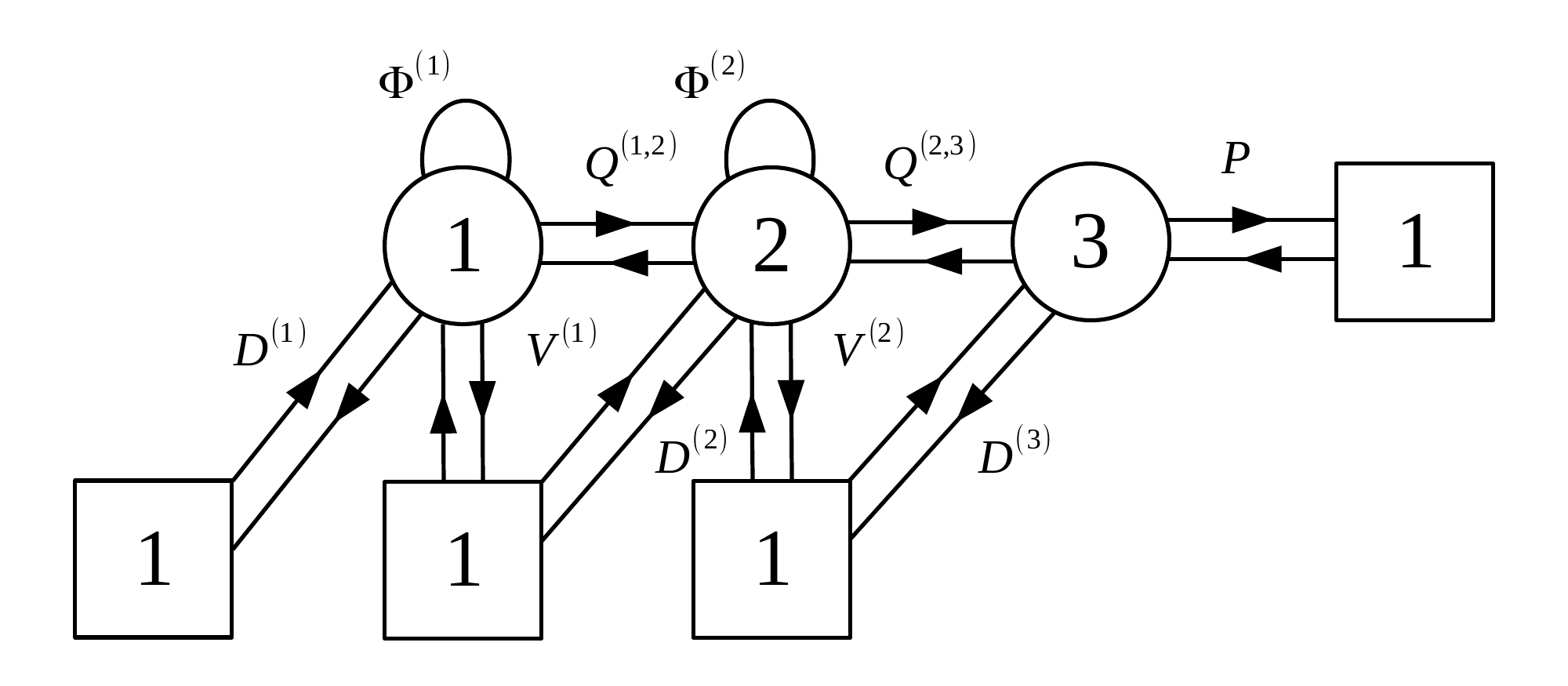}
	\caption{Quiver diagram of the $G[U(3)]$ theory, which is the starting point of the piecewise derivation of the recombination duality.}
	\label{gaugedmu3}
\end{figure}

We will explicitly present the derivation in the $N=3$ case. The starting point is the quiver theory of Figure \ref{gaugedmu3} with superpotential
\be
\mathcal{W}&=&\mathfrak{M}^{(1,0,0)}+\mathfrak{M}^{(-1,0,0)}+\mathfrak{M}^{(0,1,0)}+\mathfrak{M}^{(0,-1,0)}+\nn\\
&+&\Gp^{(1)}Q^{(1,2)}_i\tilde{Q}^{(1,2)}_i-\Gp^{(2)}_{ij}Q^{(1,2)}_i\tilde{Q}^{(1,2)}_j+\Gp^{(2)}Q^{(2,3)}_{ia}\tilde{Q}^{(2,3)}_{aj}+\nn\\
&+&D^{(2)}_i\tilde{Q}^{(1,2)}_iV^{(1)}+\tilde{V}^{(1)}Q^{(1,2)}_i\tilde{D}^{(2)}_i+D^{(3)}_a\tilde{Q}^{(2,3)}_{ai}V^{(2)}_i+\tilde{V}^{(2)}_iQ^{(1,2)}_{ia}\tilde{D}^{(3)}_a\, .\nn\\
\ee
We can apply Aharony duality on the $U(3)$ node since we have no monopole superpotential associated to it. This node is attached to $N_f=4$ flavors and will thus be replaced by a $U(1)$ node. The mesons that will be mapped to the matrix of gauge singlets take the explicit form
\be
\begin{pmatrix}
Q^{(2,3)}_{1a}\tilde{Q}^{(2,3)}_{a1} & Q^{(2,3)}_{1a}\tilde{Q}^{(2,3)}_{a2} & Q^{(2,3)}_{1a}\tilde{D}^{(3)}_a & Q^{(2,3)}_{1a}\tilde{P}_a \\
Q^{(2,3)}_{2a}\tilde{Q}^{(2,3)}_{a1} & Q^{(2,3)}_{2a}\tilde{Q}^{(2,3)}_{a2} & Q^{(2,3)}_{2a}\tilde{D}^{(3)}_a & Q^{(2,3)}_{2a}\tilde{P}_a \\
D^{(3)}_a\tilde{Q}^{(2,3)}_{a1} & D^{(3)}_a\tilde{Q}^{(2,3)}_{a2} & D^{(3)}_a\tilde{D}^{(3)}_a & D^{(3)}_a\tilde{P}_a \\
P_a\tilde{Q}^{(2,3)}_{a1}  & P_a\tilde{Q}^{(2,3)}_{a2} & P_a\tilde{D}^{(3)}_a & P_a\tilde{P}_a
\end{pmatrix}
\leftrightarrow
\begin{pmatrix}
M_{11} & M_{12} & v_1 & p_{L,1} \\
M_{21} & M_{22} & v_2 & p_{L,2} \\
\tilde{v}_1 & \tilde{v}_2 & \gb_1 & \chi_3 \\
\tilde{p}_{L,1} & \tilde{p}_{L,2} & \tilde{\chi}_3 & \ga_1
\end{pmatrix}\, .
\ee
With this piece of information we can find how the old superpotential is mapped and adding the superpotential dictated by Aharony duality we have
\be
\mathcal{W}&=&\mathfrak{M}^{(1,0,0)}+\mathfrak{M}^{(-1,0,0)}+\mathfrak{M}^{(0,1,1)}+\mathfrak{M}^{(0,-1,-1)}+\Gp^{(1)}Q^{(1,2)}_i\tilde{Q}^{(1,2)}_i-\Gp^{(2)}_{ij}Q^{(1,2)}_i\tilde{Q}^{(1,2)}_j+\nn\\
&+&D^{(2)}_i\tilde{Q}^{(1,2)}_iV^{(1)}+\tilde{V}^{(1)}Q^{(1,2)}_i\tilde{D}^{(2)}_i+\Gp^{(2)}_{ij}M_{ij}+v_iV^{(2)}_i+\tilde{v}_i\tilde{V}^{(2)}+\nn\\
&+&S^-_1\mathfrak{M}^{(0,0,1)}+S^+_1\mathfrak{M}^{(0,0,-1)}+M_{ij}\tilde{q}_{LR,i}q_{LR,j}+v_i\tilde{q}_{LR,i}d_R^{(1)}+\tilde{d}^{(1)}_Rq_{LR,i}\tilde{v}_i+\nn\\
&+&p_{L,i}\tilde{q}_{LR,i}\tilde{p}_R+p_Rq_{LR,i}\tilde{p}_{L,i}+\gb_1\tilde{d}^{(1)}_Rd^{(1)}_R+\chi_3p_Rd^{(1)}_R+\tilde{\chi}_3\tilde{d}^{(1)}_R\tilde{p}_R+\ga_1p_R\tilde{p}_R\, ,\nn\\
\label{recombstep1}
\ee
where the bifundamental $q_{LR,i}$, $\tilde{q}_{LR,i}$ carries only one index since in this case it connects the $U(2)$ node with the new $U(1)$ node. Notice that the monopoles of the $U(2)$ node are not turned on in the superpotential, while the monopoles $\mathfrak{M}^{(0,\pm1,\pm1)}$ are. This is due to the contact terms predicted by Aharony duality. As explained in \cite{US1}, these are actually BF couplings for the $U(2)$ node since the symmetry is gauged and they have the effect of charging the corresponding monopoles under the $U(1)_\zeta$ topological symmetry, preventing them from appearing in the superpotential. On the other hand, the monopoles $\mathfrak{M}^{(0,\pm1,\pm1)}$ are uncharged under the topological symmetry as well as under all the other global symmetries and are exactly marginal (see Appendix \ref{partfuncrecomb} for a partition function perspective on this point). Moreover, many of the fields appearing in \eqref{recombstep1} are massive and can be integrated out. 
\begin{enumerate}[$\bullet$]
\item If we focus on the part of the superpotential involving $\Gp^{(2)}$ and $M$
\be
\gd\mathcal{W}=-\Gp^{(2)}_{ij}Q^{(1,2)}_i\tilde{Q}^{(1,2)}_j+\Gp^{(2)}_{ij}M_{ij}+M_{ij}\tilde{q}_{LR,i}q_{LR,j}
\ee
we see that they are massive. Using their equations of motion we find that integrating them out this piece of the superpotential becomes
\be
\gd\mathcal{W}=q_{LR,j}\tilde{q}_{LR,i}Q^{(1,2)}_i\tilde{Q}^{(1,2)}_j=\Tr_2\left(q_{LR}\tilde{q}_{LR}Q^{(1,2)}\tilde{Q}^{(1,2)}\right)\, ,
\ee
where we have only the trace over the $U(2)$ color indices since those on the $U(1)$ nodes are trivial.
\item From the terms
\be
\gd\mathcal{W}=v_iV^{(2)}_i+\tilde{v}_i\tilde{V}^{(2)}+v_i\tilde{q}_{LR,i}d_R^{(1)}+\tilde{d}^{(1)}_Rq_{LR,i}\tilde{v}_i
\ee
we see that the fields $v_i$, $\tilde{v}_i$ and $V^{(2)}_i$, $\tilde{V}^{(2)}_i$ are massive and integrating them out we have no contribution to the superpotential left.
\end{enumerate}
\begin{figure}[t]
	\centering
	\includegraphics[scale=0.65]{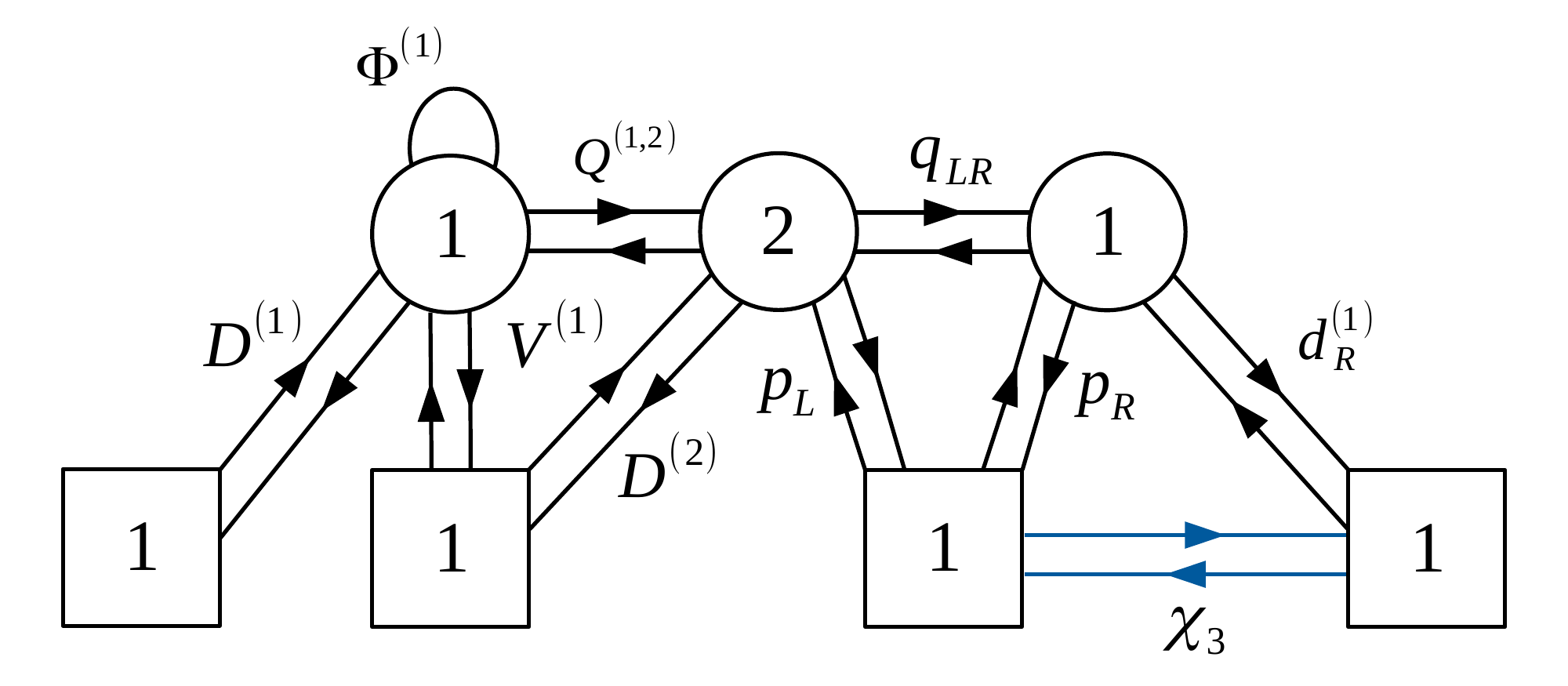}
	\caption{Quiver diagram after the application of Aharony duality on the right node. This coincides with the one of Figure \ref{gluedmun} predicted by the recombination duality in the case $N=3$ and $k=1$.}
	\label{recombstep1fig}
\end{figure}
\noindent Finally, we can recognize
\be
p_Rd^{(1)}_R=\Omega_R,\qquad \tilde{d}^{(1)}_R\tilde{p}_R=\tilde{\Omega}\, .
\ee
If we collect all these results, we find that the dual theory is the quiver of Figure \ref{recombstep1fig} with superpotential
\be
\label{pippo}
\mathcal{W}&=&\mathfrak{M}^{(1,0,0)}+\mathfrak{M}^{(-1,0,0)}+\mathfrak{M}^{(0,1,1)}+\mathfrak{M}^{(0,-1,-1)}+\Gp^{(1)}Q^{(1,2)}_i\tilde{Q}^{(1,2)}_i+\nn\\
&+&D^{(2)}_i\tilde{Q}^{(1,2)}_iV^{(1)}+\tilde{V}^{(1)}Q^{(1,2)}_i\tilde{D}^{(2)}_iq_{LR,i}\tilde{q}_{LR,j}Q^{(1,2)}_j\tilde{Q}^{(1,2)}_i+p_{L,i}\tilde{q}_{LR,i}\tilde{p}_R+p_Rq_{LR,i}\tilde{p}_{L,i}+\nn\\
&+&S^-_1\mathfrak{M}^{(0,0,1)}+S^+_1\mathfrak{M}^{(0,0,-1)}+\ga_1p_R\tilde{p}_R+\gb_1\tilde{d}^{(1)}_Rd^{(1)}_R+\chi_3\Omega_R+\tilde{\chi}_3\tilde{\Omega}_R\, .\nn\\
\ee
This is exactly our claimed dual theory in the case $N=3$ and $k=1$. It's worth analyzing the role of monopole operators in more details. Recall that in the original theory we had six possible dressed monopoles
\be
\mathfrak{M}^{\pm}_{\mathbb{M}^s}\qquad s=0,1,2\, ,
\ee
where
\be
\mathbb{M}_{ab}=Q^{(2,3)}_{ia}\tilde{Q}^{(2,3)}_{bi}\, .
\ee
All the other monopole operators corresponding to the other gauge nodes are not in the chiral ring because of the monopole superpotential. In the dual theory, only the monopoles of the left $U(1)$ node are in the superpotential, while those of the right $U(1)$ node are set to zero in the chiral ring because of the equations of motion of $S^\pm_1$. Hence, we are left with the monopoles of the $U(2)$ node
\be
\mathfrak{M}^{(0,\pm1,0)},\qquad\mathfrak{M}^{(0,\pm1,0)}_{\mathbb{M}_L},
\ee
where
\be
\mathbb{M}_{L,ij}=Q^{(1,2)}_{i}\tilde{Q}^{(1,2)}_{j}\, .
\ee
To complete the map we have to combine these four operators with the gauge singlets $S^\pm_1$, as we pointed out at the end of the previous section.

At this point, since the monopoles of the $U(2)$ gauge node are not in the superpotential anymore and its adjoint chiral has been flipped away, we can apply Aharony duality again. The number of flavors attached to this node in four, so after Aharony duality it will remain a $U(2)$ node. As before, we first need to understand how the old superpotential is mapped. Let's discuss separately the monopole part and the polynomial part. For the latter, we need to use the fact that the meson matrix is mapped under Aharony duality into a matrix of gauge singlets\footnote{We will denote with a prime all the new fields that play the same role of some of the old ones. At the end of the day, the old fields will be integrated out and we will then drop the prime index without any confusion.}
\be
\begin{pmatrix}
Q^{(1,2)}_i\tilde{Q}^{(1,2)}_i & Q^{(1,2)}_i\tilde{q}_{LR,i} & Q^{(1,2)}_i\tilde{D}^{(2)}_i & Q^{(1,2)}_i\tilde{p}_{L,i} \\
q_{LR,i}\tilde{Q}^{(1,2)}_i & q_{LR,i}\tilde{q}_{LR,i} & q_{LR,i}\tilde{D}^{(2)}_i & q_{LR,i}\tilde{p}_{L,i} \\
D^{(2)}_i\tilde{Q}^{(1,2)}_i & D^{(2)}_i\tilde{q}_{LR,i} & D^{(2)}_i\tilde{D}^{(2)}_i & D^{(2)}_i\tilde{p}_{L,i} \\
p_{L,i}\tilde{Q}^{(1,2)}_i & p_{L,i}\tilde{q}_{LR,i} & p_{L,i}\tilde{D}^{(2)}_i & p_{L,i}\tilde{p}_{L,i}
\end{pmatrix}
\leftrightarrow
\begin{pmatrix}
M_{11} & M_{12} & v & p_L' \\
M_{21} & \Gp^{(1)'} & v^{(1)}_R & u \\
\tilde{v} & \tilde{v}^{(1)}_R & \gb_2 & \chi_2 \\
\tilde{p}_L' & \tilde{u} & \tilde{\chi}_2 & \ga_2
\end{pmatrix}\, .
\ee

The fate of the monopoles is a little bit more subtle. Once again, the contact terms predicted by Aharony duality modify the quantum numbers of the monopole operators of the two $U(1)$ nodes, with the effect of removing from the superpotential $\mathfrak{M}^{(\pm1,0,0)}$ and $\mathfrak{M}^{(0,\pm1,\pm 1)}$ while turning on  $\mathfrak{M}^{(0,0,\pm1)}$ and  $\mathfrak{M}^{(\pm1,\pm1,0)}$. Moreover, in order to map the terms $S_1^\pm\mathfrak{M}^{(0,0,\mp1)}$ in \eqref{pippo} we need to understand how  the monopoles $\mathfrak{M}^{(0,0,\mp1)}$ are mapped in the Aharony dual. We claim that they are mapped into the dressed monopoles of the middle $U(2)$ node
\be
\mathfrak{M}^{(0,0,\pm1)}\quad\leftrightarrow\quad\mathfrak{M}^{(0,\pm1,0)}_{\mathbb{M}_R}\, ,
\ee
where this time the dressing is performed using the right meson matrix
\be
\mathbb{M}_{R,ij}=q_{R,i}^{(1,2)}\tilde{q}^{(1,2)}_{R,j}\, .
\ee
One can indeed check that their charges under all the global symmetries match. Moreover, this is consistent with the operator map for $k=2$, since the six dressed monopole operators of the original theory are mapped into the two monopoles of the left $U(1)$ node and the four gauge singlets $S_1^\pm$, $S_2^\pm$, which flip the fundamental and the dressed monopoles of the $U(2)$ node.

We now have all we need to write the superpotential of the dual theory
\be
\mathcal{W}&=&\mathfrak{M}^{(0,0,1)}+\mathfrak{M}^{(0,0,-1)}+\mathfrak{M}^{(1,1,0)}+\mathfrak{M}^{(-1,-1,0)}+\Gp^{(1)}M_{11}+vV^{(1)}+\tilde{v}\tilde{V}^{(1)}+M_{12}M_{21}+\nn\\
&+&\tilde{u}\tilde{p}_R+up_R+S_1^-\mathfrak{M}^{(0,1,0)}_{\mathbb{M}_R}+S_1^+\mathfrak{M}^{(0,-1,0)}_{\mathbb{M}_R}+S_2^-\mathfrak{M}^{(0,1,0)}+S_2^+\mathfrak{M}^{(0,-1,0)}+\nn\\
&+&\ga_1p_R\tilde{p}_R+\gb_1\tilde{d}^{(1)}_Rd^{(1)}_R+\chi_3p_Rd^{(1)}_R+\tilde{\chi}_3\tilde{d}^{(1)}_R\tilde{p}_R+M_{11}q'_{LR,i}\tilde{q}'_{LR,i}+M_{12}q'_{LR,i}\tilde{q}^{(1,2)}_{R,i}+\nn\\
&+&M_{21}q^{(1,2)}_{R,i}\tilde{q}'_{LR,i}+\Gp^{(1)'}q^{(1,2)}_{R,i}\tilde{q}^{(1,2)}_{R,i}v\,d^{(2)}_R\tilde{q}'_{LR,i}+\tilde{v}\,q'_{LR,i}\tilde{d}^{(2)}_R+d^{(2)}_{R,i}\tilde{q}^{(1,2)}_{R,i}v^{(1)}_R+\nn\\
&+&\tilde{v}^{(1)}_Rq^{(1,2)}_{R,i}\tilde{d}^{(2)}_{R,i}+p'_{L}\tilde{q}'_{LR,i}\tilde{p}'_{R,i}+p'_{R,i}q'_{LR,i}\tilde{p}'_{L}up'_{R,i}\tilde{q}^{(1,2)}_{R,i}+\tilde{u}q^{(1,2)}_{R,i}\tilde{p}'_{R,i}+\gb_2\tilde{d}^{(2)}_{R,i}d^{(2)}_{R,i}+\nn\\
&+&\ga_2p'_{R,i}\tilde{p}'_{R,i}+\chi_2p'_{R,i}d^{(2)}_{R,i}+\tilde{\chi}_2\tilde{d}^{(2)}_{R,i}\tilde{p}'_{R,i}\, .
\ee
Many of the fields appearing in this superpotential are massive and can be integrated out.
\begin{figure}[t]
	\centering
	\includegraphics[scale=0.65]{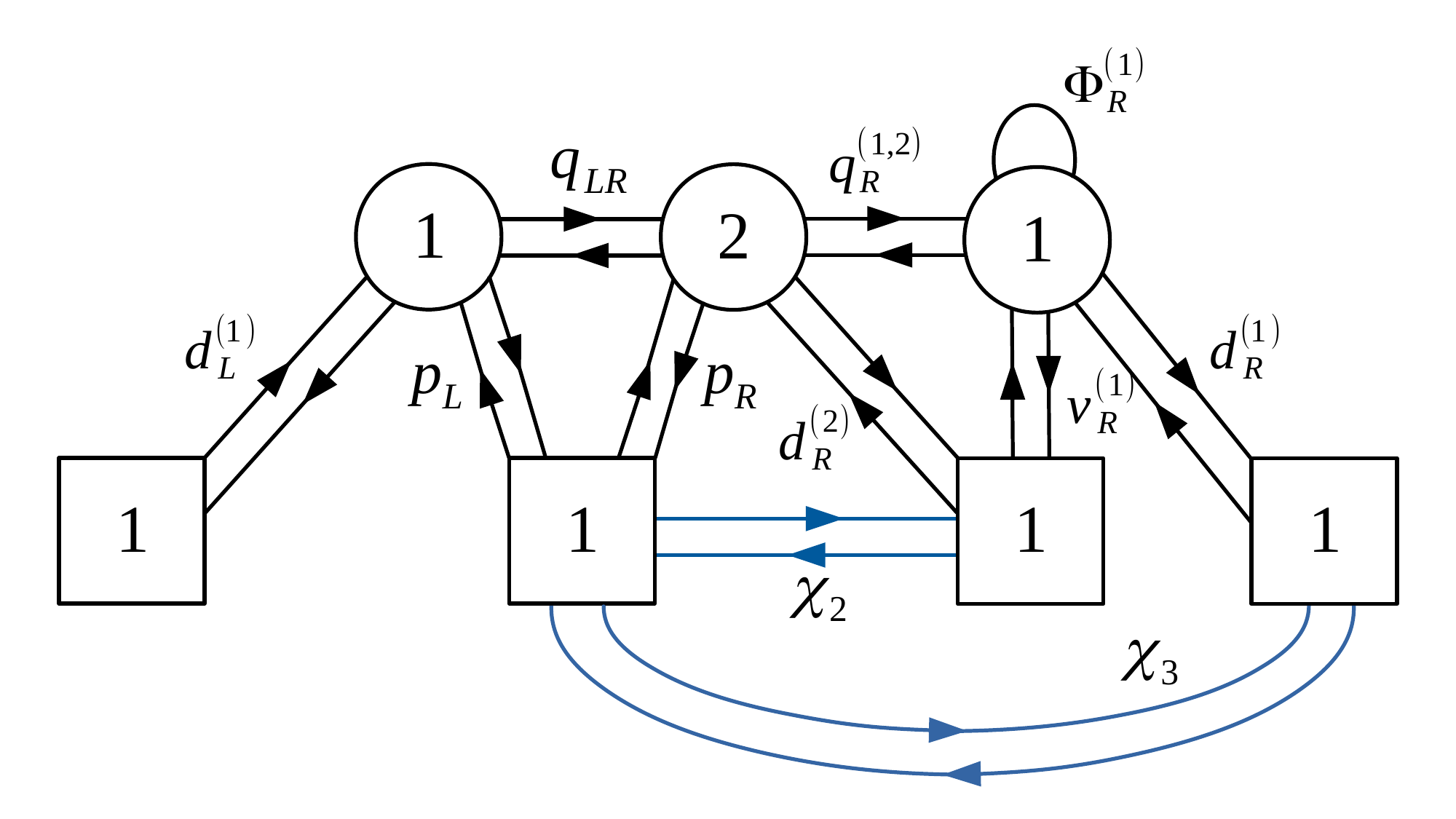}
	\caption{Quiver diagram after the application of Aharony duality on the middle node. This coincides with the one of Figure \ref{gluedmun} predicted by the recombination duality in the case $N=3$ and $k=2$.}
	\label{recombstep2fig}
\end{figure}
\noindent 
\begin{enumerate}[$\bullet$]
\item If we look at the terms
\be
\gd\mathcal{W}=\Gp^{(1)}M_{11}+M_{11}q'_{LR,i}\tilde{q}'_{LR,i}
\ee
we see that the field $\Gp^{(1)}$, $M_{11}$ are massive and that the equations of motion of $\Gp^{(1)}$ simply set this part of the superpotential to zero.
\item From the piece
\be
\gd\mathcal{W}=vV^{(1)}+\tilde{v}\tilde{V}^{(1)}+v\,d^{(2)}_R\tilde{q}'_{LR,i}+\tilde{v}\,q'_{LR,i}\tilde{d}^{(2)}_R
\ee
we see that $v$, $\tilde{v}$ and $V^{(1)}$, $\tilde{V}^{(1)}$ are massive and integrating them out we have no contribution to the superpotential left.
\item If we focus on
\be
\gd\mathcal{W}=M_{12}M_{21}+M_{12}q'_{LR,i}\tilde{q}^{(1,2)}_{R,i}+M_{21}q^{(1,2)}_{R,i}\tilde{q}'_{LR,i}
\ee
we see that $M_{12}$, $M_{21}$ are massive and using the equation of motion of any of the two we find
\be
\gd\mathcal{W}=-q'_{LR,i}\tilde{q}'_{LR,j}q^{(1,2)}_{R,j}\tilde{q}^{(1,2)}_{R,i}=-\Tr_2\left(q'_{LR}\tilde{q}'_{LR}q^{(1,2)}_R\tilde{q}^{(1,2)}_R\right)\, .
\ee
\item Finally, we consider
\be
\gd\mathcal{W}=\tilde{u}\tilde{p}_R+up_R+up'_{R,i}\tilde{q}^{(1,2)}_{R,i}+\tilde{u}q^{(1,2)}_{R,i}\tilde{p}'_{R,i}+\chi_3p_Rd^{(1)}_R+\tilde{\chi}_3\tilde{d}^{(1)}_R\tilde{p}_R\, .
\ee
We see that $u$, $\tilde{u}$ and $p_R$, $\tilde{p}_R$ are massive and using the equations of motion of the former 
\be
p_R=-p'_{R,i}\tilde{q}^{(1,2)}_{R,i},\qquad \tilde{p}_R=-q^{(1,2)}_{R,i}\tilde{p}'_{R,i}
\ee
we get
\be
\gd\mathcal{W}=\chi_3p'_{R,i}\tilde{q}^{(1,2)}_{R,i}d^{(1)}_R+\tilde{\chi}_3\tilde{d}^{(1)}_Rq^{(1,2)}_{R,i}\tilde{p}'_{R,i}\, .
\ee
\end{enumerate}
Using all these results and recalling that
\be
&&p'_{R,i}\tilde{q}^{(1,2)}_{R,i}d^{(1)}_R=\Omega'_{R,1}, \qquad p'_{R,i}d^{(2)}_{R,i}=\Omega'_{R,2}\nn\\
&&\tilde{d}^{(1)}_Rq^{(1,2)}_{R,i}\tilde{p}'_{R,i}=\tilde{\Omega}'_{R,1}\qquad \tilde{d}^{(2)}_{R,i}\tilde{p}'_{R,i}=\tilde{\Omega}'_{R,2}\, ,
\ee
we find that the dual theory is the quiver of Figure \ref{recombstep2fig} with superpotential (at this point we can safely drop the prime indices)
\be
\mathcal{W}&=&\mathfrak{M}^{(0,0,1)}+\mathfrak{M}^{(0,0,-1)}+\mathfrak{M}^{(1,1,0)}+\mathfrak{M}^{(-1,-1,0)}+\Gp^{(1)}q^{(1,2)}_{R,i}\tilde{q}^{(1,2)}_{R,i}+d^{(2)}_{R,i}\tilde{q}^{(1,2)}_{R,i}v^{(1)}_R+\nn\\
&+&\tilde{v}^{(1)}_Rq^{(1,2)}_{R,i}\tilde{d}^{(2)}_{R,i}-q_{LR,i}\tilde{q}_{LR,j}q^{(1,2)}_{R,j}\tilde{q}^{(1,2)}_{R,i}+p_{L}\tilde{q}_{LR,i}\tilde{p}_{R,i}+p_{R,i}q_{LR,i}\tilde{p}_{L}+S_1^-\mathfrak{M}^{(0,1,0)}_{\mathbb{M}_R}+\nn\\
&+&S_1^+\mathfrak{M}^{(0,-1,0)}_{\mathbb{M}_R}+S_2^-\mathfrak{M}^{(0,1,0)}+S_2^+\mathfrak{M}^{(0,-1,0)}+\ga_1p_{R,i}\tilde{p}_{R,j}q^{(1,2)}_{R,j}\tilde{q}^{(1,2)}_{R,i}+\ga_2p_{R,i}\tilde{p}_{R,i}+\nn\\
&+&\gb_1\tilde{d}^{(1)}_Rd^{(1)}_R+\gb_2\tilde{d}^{(2)}_{R,i}d^{(2)}_{R,i}+\chi_3\Omega_{R,1}+\tilde{\chi}_3\tilde{\Omega}_{R,1}+\chi_2\Omega_{R,2}+\tilde{\chi}_2\tilde{\Omega}_{R,2}\, ,
\ee
which agrees with our claimed result \eqref{recombinationsuperpotB} in the case $N=3$ and $k=2$. 

We complete our derivation of all the recombination duality frames in the case $N=3$ applying one last time Aharony duality, which we are allowed to do since the monopoles associated to the left $U(1)$ node are not in the superpotential anymore. The number of flavors attached to this node is again four, so applying the duality we increase the rank of the gauge group to $U(3)$. In order to map the old superpotential we first have to make use of the map of the meson matrix
\be
\begin{pmatrix}
q_{LR,1}\tilde{q}_{LR,1} & q_{LR,1}\tilde{q}_{LR,2} & q_{LR,1}\tilde{D}^{(1)} & q_{LR,1}\tilde{p}_L \\
q_{LR,2}\tilde{q}_{LR,1} & q_{LR,2}\tilde{q}_{LR,2} & q_{LR,2}\tilde{D}^{(1)} & q_{LR,2}\tilde{p}_L \\
D^{(1)}\tilde{q}_{LR,1} & D^{(1)}\tilde{q}_{LR,2} & D^{(1)}\tilde{D}^{(1)} & D^{(1)}\tilde{p}_L \\
p_L\tilde{q}_{LR,1} & p_L\tilde{q}_{LR,2} & p_L\tilde{D}^{(1)} & p_L\tilde{p}_L
\end{pmatrix}
\leftrightarrow
\begin{pmatrix}
\Gp^{(2)}_{11} & \Gp^{(2)}_{12} & v^{(2)}_{R,1} & u_1 \\
\Gp^{(2)}_{21} & \Gp^{(2)}_{22} & v^{(2)}_{R,2} & u_2 \\
\tilde{v}^{(2)}_{R,1} & \tilde{v}^{(2)}_{R,2} & \gb_3 & \chi_1 \\
\tilde{u}_1 & \tilde{u}_2 & \tilde{\chi}_1 & \ga_3
\end{pmatrix}\, .
\ee
Then, we need to understand how the monopole operators of the $U(2)$ node are mapped. Indeed, the application of Aharony duality has modified their charges so that now the two fundamental monopoles of the middle $U(2)$ node are turned on in the superpotential. This means that the old $U(2)$ monopoles are not trivially mapped into themselves, since their new version is not in the chiral ring. We claim that
\be
\mathfrak{M}^{(0,\pm1,0)}\quad&\leftrightarrow&\quad\mathfrak{M}^{(\pm1,0,0)}_{\mathbb{M}_R}\nn\\
\mathfrak{M}^{(0,\pm1,0)}_{\mathbb{M}_R}\quad&\leftrightarrow&\quad\mathfrak{M}^{(\pm1,0,0)}_{\mathbb{M}_R^2}\, ,
\ee
where the new meson matrix used for the dressing is
\be
\mathbb{M}_{R,ab}=q^{(2,3)}_{ia}\tilde{q}^{(2,3)}_{bj}\, .
\ee
Indeed, since the left node is now $U(3)$ we can dress it with the matrix $\mathbb{M}_R$ up to the power of two. Hence, the new superpotential is
\be
\mathcal{W}&=&\mathfrak{M}^{(0,1,0)}+\mathfrak{M}^{(0,-1,0)}+\mathfrak{M}^{(0,0,1)}+\mathfrak{M}^{(0,0,-1)}+\Gp^{(1)}q^{(1,2)}_{R,i}\tilde{q}^{(1,2)}_{R,i}+d^{(2)}_{R,i}\tilde{q}^{(1,2)}_{R,i}v^{(1)}_R+\nn\\
&+&\tilde{v}^{(1)}_Rq^{(1,2)}_{R,i}\tilde{d}^{(2)}_{R,i}-\Gp^{(2)}_{ij}q^{(1,2)}_{R,i}\tilde{q}^{(1,2)}_{R,j}+u_i\tilde{p}_{R,i}+\tilde{u}_ip_{R,i}+S_1^-\mathfrak{M}^{(1,0,0)}_{\mathbb{M}_R^2}+S_1^+\mathfrak{M}^{(-1,0,0)}_{\mathbb{M}_R^2}+\nn\\
&+&S_2^-\mathfrak{M}^{(1,0,0)}_{\mathbb{M}_R}+S_2^+\mathfrak{M}^{(-1,0,0)}_{\mathbb{M}_R}+\ga_1p_{R,i}\tilde{p}_{R,j}q^{(1,2)}_{R,j}q^{(1,2)}_{R,i}+\ga_2p_{R,i}\tilde{p}_{R,i}+\gb_1\tilde{d}^{(1)}_Rd^{(1)}_R+\nn\\
&+&\gb_2\tilde{d}^{(2)}_Rd^{(2)}_R+\chi_3p_{R,i}\tilde{q}^{(1,2)}_{R,i}d^{(1)}_R+\tilde{\chi}_3\tilde{d}^{(1)}_Rq^{(1,2)}_{R,i}\tilde{p}_{R,i}+\chi_2p_{R,i}d^{(2)}_{R,i}+\tilde{\chi}_2\tilde{d}^{(2)}_{R,i}\tilde{p}_{R,i}+\nn\\
&+&S_3^-\mathfrak{M}^{(1,0,0)}+S_3^+\mathfrak{M}^{(-1,0,0)}+\Gp^{(2)}_{ij}q^{(2,3)}_{R,ia}\tilde{q}^{(2,3)}_{R,aj}+d^{(3)}_{R,a}\tilde{q}^{(2,3)}_{R,ai}v^{(2)}_{R,i}+\tilde{v}^{(2)}_{R,i}q^{(2,3)}_{R,ia}\tilde{d}^{(3)}_R+\nn\\
&+&p'_{R,a}\tilde{q}^{(2,3)}_{R,ai}u_i+\tilde{u}_iq^{(2,3)}_{R,ia}p'_{R,a}+\gb_3\tilde{d}^{(3)}_{R,a}d^{(3)}_{R,a}+\ga_3p'_{R,a}\tilde{p}'_{R,a}+\chi_1p'_{R,a}d^{(3)}_{R,a}+\tilde{\chi}\tilde{d}^{(3)}_{R,a}\tilde{p}'_{R,a}\, .\nn\\
\ee
\begin{figure}[t]
	\centering
	\includegraphics[scale=0.65]{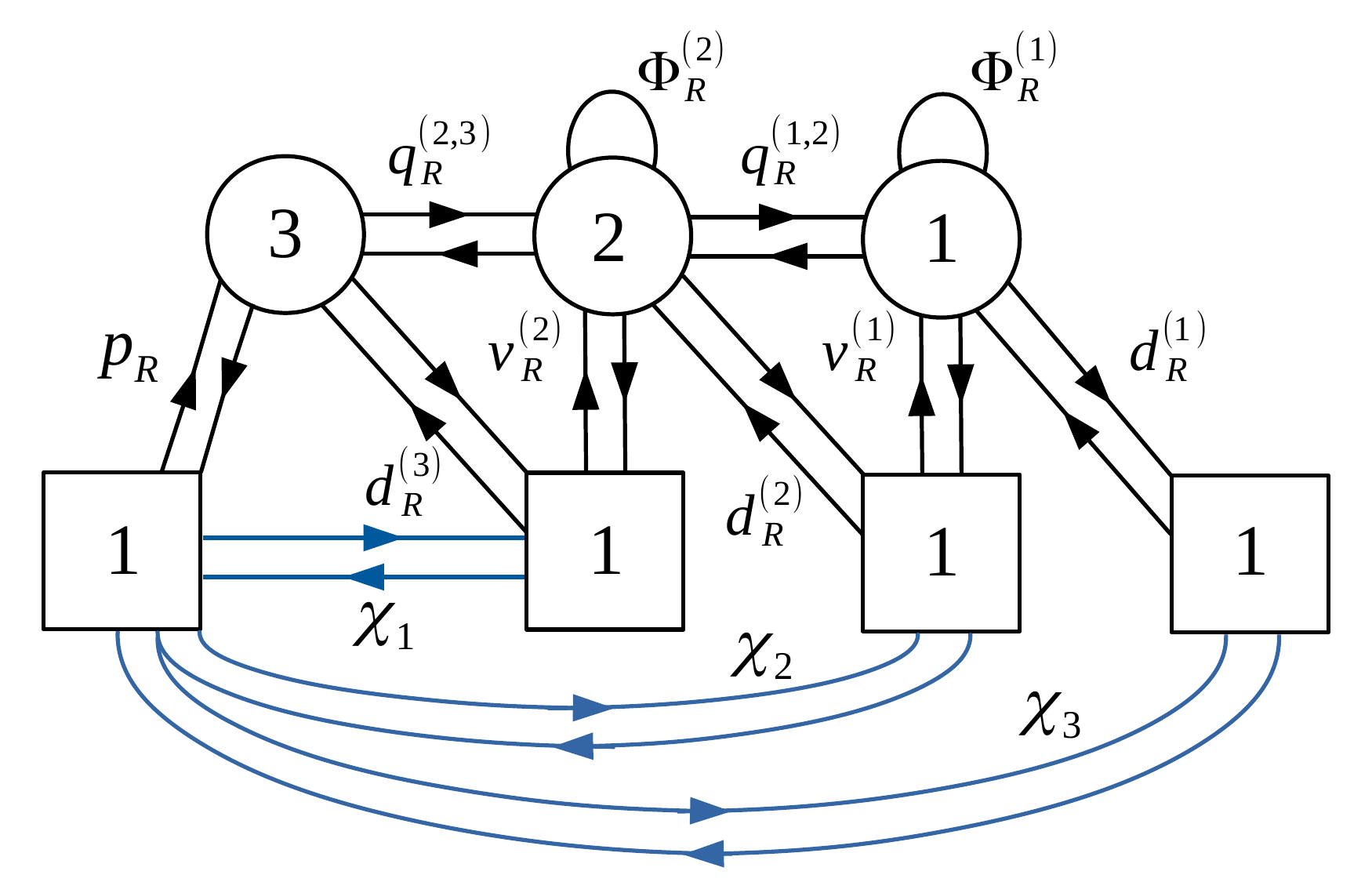}
	\caption{Quiver diagram after the application of Aharony duality on the left node. This coincides with the one of Figure \ref{gluedmun} predicted by the recombination duality in the case $N=3$ and $k=3$.}
	\label{recombstep3fig}
\end{figure}
\noindent In order to integrate out the massive fields, we need to focus on the following terms
\be
\gd\mathcal{W}&=&u_i\tilde{p}_{R,i}+\tilde{u}_ip_{R,i}+\ga_1p_{R,i}\tilde{p}_{R,j}q^{(1,2)}_{R,j}q^{(1,2)}_{R,i}+\ga_2p_{R,i}\tilde{p}_{R,i}+p'_{R,a}\tilde{q}^{(2,3)}_{R,ai}u_i+\tilde{u}_iq^{(2,3)}_{R,ia}p'_{R,a}+\nn\\
&+&\chi_3p_{R,i}\tilde{q}^{(1,2)}_{R,i}d^{(1)}_R+\tilde{\chi}_3\tilde{d}^{(1)}_Rq^{(1,2)}_{R,i}\tilde{p}_{R,i}+\chi_2p_{R,i}d^{(2)}_{R,i}+\tilde{\chi}_2\tilde{d}^{(2)}_{R,i}\tilde{p}_{R,i}\, .
\ee
We see that $u_i$, $\tilde{u}_i$ and $p_{R,i}$, $\tilde{p}_{R,i}$ are massive and the equations of motion of the first two give
\be
p_{R,i}=-p'_{R,a}\tilde{q}^{(2,3)}_{R,ai}, \qquad \tilde{p}_{R,i}=-q^{(2,3)}_{R,ia}\tilde{p}'_{R,a}\, .
\ee
Plugging this back in the superpotential we get
\be
\gd\mathcal{W}&=&\ga_1p'_{R,a}\tilde{p}'_{R,b}\tilde{q}^{(2,3)}_{R,ai}q^{(2,3)}_{R,jb}\tilde{q}^{(1,2)}_{R,i}q^{(1,2)}_{R,j}+\ga_2p'_{R,a}\tilde{p}'_{R,b}q^{(2,3)}_{R,ib}\tilde{q}^{(2,3)}_{R,ai}+\chi_3p'_{R,a}\tilde{q}^{(2,3)}_{R,ai}\tilde{q}^{(1,2)}_{R,i}d^{(1)}_R+\nn\\
&+&\tilde{\chi}_3\tilde{d}^{(1)}_Rq^{(1,2)}q^{(2,3)}_{R,ia}\tilde{p}'_{R,a}\chi_2p'_{R,a}\tilde{q}^{(2,3)}_{R,ai}d^{(2)}_{R,i}+\tilde{\chi}_2\tilde{d}^{(2)}_{R,i}q^{(2,3)}_{R,ia}\tilde{p}'_{R,a}\, .
\ee
The first term can also be rewritten using the equations of motion of $\Gp^{(2)}$
\be
q^{(1,2)}_{R,i}\tilde{q}^{(1,2)}_{R,j}=q^{(2,3)}_{R,ia}\tilde{q}^{(2,3)}_{R,aj}\, .
\ee
At the end of the day, we arrive at the reversed $G[U(3)]$ quiver of Figure \ref{recombstep3fig} plus a set of $3\times 3=9$ singlets and with superpotential (dropping the prime indices)
\be
\mathcal{W}&=&\mathfrak{M}^{(0,1,0)}+\mathfrak{M}^{(0,-1,0)}+\mathfrak{M}^{(0,0,1)}+\mathfrak{M}^{(0,0,-1)}+\Gp^{(1)}q^{(1,2)}_{R,i}\tilde{q}^{(1,2)}_{R,i}-\Gp^{(2)}_{ij}q^{(1,2)}_{R,i}\tilde{q}^{(1,2)}_{R,j}+\nn\\
&+&\Gp^{(2)}_{ij}q^{(2,3)}_{R,ia}\tilde{q}^{(2,3)}_{R,aj}+d^{(3)}_{R,a}\tilde{q}^{(2,3)}_{R,ai}v^{(2)}_{R,i}+\tilde{v}^{(2)}_{R,i}q^{(2,3)}_{R,ia}\tilde{d}^{(3)}_R+d^{(2)}_{R,i}\tilde{q}^{(1,2)}_{R,i}v^{(1)}_R+\tilde{v}^{(1)}_Rq^{(1,2)}_{R,i}\tilde{d}^{(2)}_{R,i}+\nn\\
&+&S_1^-\mathfrak{M}^{(1,0,0)}_{\mathbb{M}_R^2}+S_1^+\mathfrak{M}^{(-1,0,0)}_{\mathbb{M}_R^2}+S_2^-\mathfrak{M}^{(1,0,0)}_{\mathbb{M}_R}+S_2^+\mathfrak{M}^{(-1,0,0)}_{\mathbb{M}_R}+S_3^-\mathfrak{M}^{(1,0,0)}+S_3^+\mathfrak{M}^{(-1,0,0)}+\nn\\
&+&\ga_1\tilde{p}_{R,b}q^{(2,3)}_{R,jb}\tilde{q}^{(2,3)}_{R,cj}q^{(2,3)}_{R,ic}\tilde{q}^{(2,3)}_{R,ai}p_{R,a}+\ga_2\tilde{p}_{R,b}q^{(2,3)}_{R,ib}\tilde{q}^{(2,3)}_{R,ia}p_{R,a}+\ga_3p_{R,a}\tilde{p}_{R,b}+\gb_1\tilde{d}^{(1)}_Rd^{(1)}_R+\nn\\
&+&\gb_2\tilde{d}^{(2)}_Rd^{(2)}_R+\gb_3\tilde{d}^{(3)}_{R,a}d^{(3)}_{R,a}\chi_3\Pi_{R,1}+\tilde{\chi}_3\tilde{\Omega}_{R,1}+\chi_2\Omega_{R,2}+\tilde{\chi}_2\tilde{\Omega}_{R,2}+\chi_1\Omega_{R,3}+\tilde{\chi}_1\tilde{\Omega}_{R,3}\, ,\nn\\
\ee
which is exactly the recombination dual we claimed in the case $N=3$ and $k=3$. 

\subsection{Rank minimization}

This concludes the piecewise derivation of the recombination duality in the case $N=3$, where we can have three possible values of $k=1,2,3$. The same strategy can be applied to any tail of arbitrary length $N$. From this derivation it becomes clear an interesting property of the $G[U(N)]$ theory. As we go along the tail applying Aharony duality, we initially decrease the rank of the gauge node to which we apply it, until we reach the middle of the tail. From this point, the rank starts to increase back and when we finally arrive at the end of the tail we recover the same original $G[U(N)]$ theory, but reversed. Hence, for a particular number $k$ of iterations of Aharony duality we reach a configuration in which the dual theory has minimal rank. For even $N$ this happens exactly at $k=N/2$, while for odd $N$ we have two possibilities $k=(N\pm1)/2$.

The rank of the original theory was
\be
\mathrm{rank}(\mathcal{T}_{G[U(N)]})=\sum_{i=1}^N i=\frac{N(N+1)}{2}\, .
\ee
Instead, when we use the recombination duality to get to the configuration with minimal rank, we have
\be
\mathrm{rank}(\mathcal{T}_{min})=\begin{cases}\frac{N}{2}\left(\frac{N}{2}+1\right) & N \text{ even} \\ \frac{N-1}{2}\left(\frac{N-1}{2}+1\right)+\frac{N+1}{2} & N \text{ odd} \end{cases}\, .
\ee

\section{Rank Stabilization Duality}
\label{rankstabilizationduality}

In this section, we discuss the duality mentioned in the Introduction between the $U(N)$ gauge theory with one adjoint and $k+1$ flavors, $k$ of which interact with the adjoint chiral, and the $G[U(k)]$ theory plus $3N-2k$ gauge singlets. We call it \emph{rank stabilization duality} since it significantly relies on a stabilization property of the theory. We say that a theory is \emph{stable} if, after the sequential application of some basic dualities (see Appendix \ref{funddualities}), we recover the same theory but with the rank decreased by one unit and some additional gauge singlets. In \cite{US1} we considered the case $k=0$, where the original $U(N)$ theory was already in a stabilized form. Instead, for higher $k$ we need to manipulate the theory acting on it with some of the basic dualities in order to find a dual frame which is actually stable, as we will show in Sec.~\ref{rankstabderivation}.

\subsection{Theory A}

The first theory involved in the duality is the $U(N)$ gauge theory with $k+1$ fundamental flavors $Q$, $\tilde{Q}$, $P$, $\tilde{P}$ and one adjoint chiral $\Gp$ with superpotential
\be
\mathcal{W}_A=\Tr_N\left(\Gp\Tr_kQ\tilde{Q}\right)+\sum_{j=1}^{N-k}\gb_j\Tr_N\Gp^j=\sum_{i=1}^k\sum_{a,b=1}^NQ_{ia}\Gp_{ab}\tilde{Q}_{bi}+\sum_{j=1}^{N-k}\gb_j\Tr_N\Gp^j\, ,
\ee
with $k<N$. Recall that in the case $k=0$ all the Casimir operators are flipped by the $\gb$-fields since they are expected to violate the unitarity bound and decouple in the IR \cite{Benvenuti:2017lle}. Moreover, the $\gb$-fields can't acquire a VEV because of quantum effects \cite{Benvenuti:2017kud}. As we increase the number of flavors, the superconformal R-charge of the adjoint chiral $\Gp$ is expected to increase and the highest Casimir operators start to go above the unitarity bound. Hence, for a fixed value of $k$ we only need to flip the first $N-k$ Casimir operators. 

The global symmetry group of the theory is\footnote{In our convention, we choose to gauge the baryonic symmetry associated to the flavor $P$, $\tilde{P}$ that doesn't enter in the superpotential. For this reason, the symmetry associated to the flavors $Q_i$, $\tilde{Q}_i$ is $U(k)$ rather than $SU(k)$.}
\be
U(k)_z\times U(1)_\tau\times U(1)_\mu\times U(1)_\zeta\, .
\ee
The indices of each factor of the global symmetry group denote the fugacities that we can turn on and on which the three-sphere partition function will depend on. Since $U(1)_{\tau}$ and $U(1)_\mu$ are abelian symmetries that can mix with the R-symmetry, the corresponding parameters are actually defined as the holomorphic combinations
\be
\tau=\mathrm{Re}(\tau)+i\frac{Q}{2}(1-R),\qquad\mu=\mathrm{Re}(\mu)+i\frac{Q}{2}r\, ,
\ee
where $r$ and $R$ are the mixing coefficients. 
The charges of all the chiral fields of the theory under the global symmetries and their R-charges are
\begin{table}[h]
\centering
\scalebox{0.95}{
\begin{tabular}{c|ccc|c}
{} & $U(k)_{z}$ & $U(1)_{\tau}$ & $U(1)_\mu$ & $U(1)_R$ \\ \hline
$Q$ & $\Box$ & -1 & 0 & $R$ \\
$\tilde{Q}$ & $\bar{\Box}$ & -1 & 0 & $R$ \\
$P$ & 0 & 0 & 1 & $r$ \\
$\tilde{P}$ & 0 & 0 & 1 & $r$ \\
$\Gp$ & 0 & 2 & 0 & $2(1-R)$ \\
$\gb_j$ & 0 &$-2j$ & 0 & $2-2j(1-R)$
\end{tabular}}
\label{chargesTA}
\end{table}

The chiral ring of this theory is generated by several gauge invariant operators. First of all, we have the Casimirs of the gauge group built from the adjoint chiral $\Gp$. The first $N-k$ of these are actually flipped by the $\gb$-field, so that we only have $k$ operators of this kind
\be
\Tr_N\Gp^j,\qquad j=N-k+1,\cdots,N\, .
\ee

Then, we have the fundamental monopole operators $\mathfrak{M}^\pm$ which can also be dressed with $\Gp$ in the adjoint representation of the residual gauge group that survives in the monopole background \cite{Cremonesi:2013lqa}. In total, we have $2N$ independent operators of this form, which we denote by
\be
\mathfrak{M}^\pm_{\Gp^s},\qquad s=0,\cdots,N-1\,.
\ee

The mesonic operators can be of different types, depending on which flavor we use to construct them. We can have mesons built from the $P$, $\tilde{P}$ flavor, which can also be dressed with the adjoint chiral $\Gp$
\be
\Tr_N\left(\tilde{P}\Gp^sP\right),\qquad s=0,\cdots,N-1\, .
\ee
Another possibility is to combine the flavor $P$, $\tilde{P}$ with one of the flavors $Q$, $\tilde{Q}$. In this case, we can't have dressed mesons because the equations of motion of $Q$, $\tilde{Q}$ set them to zero. Hence, we only have $2k$ of them
\be
Q_i\tilde{P},\qquad P\tilde{Q}_i,\qquad i=1,\cdots,k\, ,
\ee
which can be collected in two vectors transforming in the fundamental and anti-fundamental representation of $U(k)_z$ respectively.

Finally, we have the meson obtained combining $Q$ and $\tilde{Q}$. Also such a meson can't be dressed because of the equations of motion of $Q$, $\tilde{Q}$. Hence, we have $k^2$ of them
\be
Q_i\tilde{Q}_j\, ,\qquad i,j=1,\cdots,k\, ,
\ee
which can be collected into a matrix transforming in the adjoint representation of $U(k)_z$.
The charges of these operators under the global symmetries are

\begin{table}[h]
\centering
\scalebox{0.9}{
\begin{tabular}{c|cccc|c}
{} & $U(k)_{z}$ & $U(1)_{\tau}$ & $U(1)_\mu$ & $U(1)_\zeta$ & $U(1)_R$ \\ \hline
$\Tr_N\Gp^j$ & 0 & $2j$ & 0 & 0 & $2j(1-R)$ \\
$\mathfrak{M}^\pm_{\Gp^s}$ & 0 & $-2N+k+2s+2$ & -1 & $\pm1$ & $1-r-(2N-k-2s-2)(1-R)$ \\
$\Tr_N\left(\tilde{P}\Gp^sP\right)$ & 0 & $2s$ & 2 & 0 & $2r+2s(1-R)$ \\
$Q\tilde{P}$ & $\Box$ & $-1$ & 1 & 0 & $r+R$ \\
$P\tilde{Q}$ & $\bar{\Box}$ & $-1$ & 1 & 0 & $r+R$ \\
$Q\tilde{Q}$ & adj & $-2$ & 0 & 0 & $2R$
\end{tabular}}
\label{chargesTA}
\end{table}

\subsection{Theory B}

The dual theory is $G[U(k)]$\footnote{We denote the fields of the $G[U(k)]$ theory with lower case letter, in contrast to the convention we used in Sec.~\ref{gun}, to avoid confusion with the fields of Theory A.} with $3(N-k)+k$ additional gauge singlets $\ga_i$, $T^+_j$, $T^-_{N-l+1}$, $\tilde{\gb}_a$, with $i,j,l=1,\cdots,N-k$ and $a=1,\cdots,k$, and superpotential (recall that we are limiting ourselves to the regime $k<N$)
\be
\mathcal{W}_B=\mathcal{W}_{G[U(k)]}+\mathcal{W}_{int}+\mathcal{W}_{flips}\, ,
\ee
where $\mathcal{W}_{int}$ is a cubic superpotential that encodes interactions between the extra singlets $\ga_i$, $T^+_j$, $T^-_l$ and the operators of the $G[U(k)]$ tail
\be
\mathcal{W}_{int}&=&\sum_{i,j,l=1}^{N-k}\ga_iT^+_jT^-_{N-l+1}\gd_{i+j+l,2N-k+1}+\sum_{j,l=1}^{N-k}\sum_{r=0}^{k-1}\Tr_k\left(\tilde{p}\mathbb{M}^rp\right)T^+_jT^-_{N-l+1}\gd_{r+j+l,N}+\nn\\
&+&\sum_{i,j=1}^{N-k}\sum_{s=0}^{k-1}\ga_i\mathfrak{M}^+_{\mathbb{M}^s}T^-_{N-l+1}\gd_{i+s+l,N}+\sum_{i,j=1}^{N-k}\sum_{t=0}^{k-1}\ga_iT^+_j\mathfrak{M}^-_{\mathbb{M}^t}\gd_{i+j+t,N}+\nn\\
&+&\sum_{l=1}^{N-k}\sum_{r,s=0}^{k-1}\Tr_k\left(\tilde{p}\mathbb{M}^rp\right)\mathfrak{M}^+_{\mathbb{M}^s}T^-_{N-l+1}\gd_{r+s+l,k-1}+\sum_{j=1}^{N-k}\sum_{r,t=0}^{k-1}\Tr_k\left(\tilde{p}\mathbb{M}^rp\right)T^+_j\mathfrak{M}^-_{\mathbb{M}^t}\gd_{r+j+t,k-1}+\nn\\
&+&\sum_{i=1}^{N-k}\sum_{s,t=0}^{k-1}\ga_i\mathfrak{M}^+_{\mathbb{M}^s}\mathfrak{M}^-_{\mathbb{M}^t}\gd_{i+s+t,k-1}+\sum_{r,s,t=0}^{k-1}\Tr_k\left(\tilde{p}\mathbb{M}^rp\right)\mathfrak{M}^+_{\mathbb{M}^s}\mathfrak{M}^-_{\mathbb{M}^t}\gd_{r+s+t,2k-N-2}\, ,\nn\\
\label{Wint}
\ee
while $\mathcal{W}_{flips}$ is a superpotential that involves the remaining gauge singlets $\tilde{\gb}_a$ flipping a set of operators of $G[U(k)]$
\be
\mathcal{W}_{flips}=\sum_{a=1}^k\tilde{\gb}_a\tilde{d}^{(a)}d^{(a)}\, .
\ee
Both the meson and the monopole operators of $G[U(k)]$ are dressed with the matrix
\be
\mathbb{M}=\Tr_{k-1}q^{(k-1,k)}\tilde{q}^{(k-1,k)}\, ,
\ee
which transforms in the adjoint representation of the $U(k)$ factor of the gauge group.

\begin{table}[t]
\centering
\scalebox{0.96}{
\begin{tabular}{c|cccc|c}
{} & $U(1)_{z_a}$ & $U(1)_{\tau}$ & $U(1)_\mu$ & $U(1)_\zeta$ & $U(1)_R$ \\ \hline
$\ga_i$ & 0 & $2(i-1)$ & 2 & 0 & $2r+2(i-1)(1-R)$ \\
$T^+_j$ & 0 & $-2N+k+2j$ & $-1$ & 1 & $1-r-(2N-k-2j)(1-R)$ \\
$T^-_{N-l+1}$ & 0 & $-2N+k+2l$ & $-1$ & $-1$ & $1-r-(2N-k-2l)(1-R)$ \\
$\tilde{\gb}_a$ & 0 & $2(N-k+a)$ & 0 & 0 & $2(N-k+a)(1-R)$ \\
$q^{(a-1,a)}$ & 0 & 1 & 0 & 0 & $1-R$ \\
$\tilde{q}^{(a-1,a)}$ & 0 & 1 & 0 & 0 & $1-R$ \\
$p$ & 0 & $N-k$ & 1 & 0 & $r+(N-k)(1-R)$ \\
$\tilde{p}$ & 0 & $N-k$ & 1 & 0 & $r+(N-k)(1-R)$ \\
$v^{(a-1)}$ & 1 & $N-a$ & 0 & 0 & $1+(N-a)(1-R)$ \\
$\tilde{v}^{(a-1)}$ & $-1$ & $N-a$ & 0 & 0 & $1+(N-a)(1-R)$ \\
$d^{(a)}$ & $-1$ & $-N+a-1$ & 0 & 0 & $1-(N-a+1)(1-R)$ \\
$\tilde{d}^{(a)}$ & 1 & $-N+a-1$ & 0 & 0 & $1-(N-a+1)(1-R)$ \\
$\Gp^{(a)}$ & 0 & $-2$ & 0 & 0 & $2R$
\end{tabular}}
\caption{Representations and charges under the global symmetries of all the chiral fields of Theory B. In the table the indices $i$, $j$, $l$ run from 1 to $N-k$, while $a$ from 1 to $k-1$. By convention, $q^{(0,1)}=\tilde{q}^{(0,1)}=0$, $v^{(0)}=\tilde{v}^{(0)}=0$ and $\Gp^{(k)}=0$.}
\label{chargesTB}
\end{table}

The last term in the superpotential \eqref{Wint} involves only the operators of the $G[U(k)]$ part of the theory and has the effect of breaking one of the $U(1)$ axial symmetries of $G[U(k)]$ \eqref{recombglobalsymm},  so now the  global symmetries of Theory A and Theory B  match (at least at the level of the Cartan subalgebra). Indeed, in order for such a term to be uncharged under all the global symmetries and have R-charge 2, the axial masses of $G[U(k)]$ \eqref{axialmassesgun} have to satisfy the constraint
\be
\Gd=(N-k+1)m_A-i\frac{Q}{2}(N-k)\, ,
\label{gaugeonshell}
\ee
which can be consistently solved in terms of a single parameter $\tau$
\be
m_A=i\frac{Q}{2}-\tau,\qquad\Gd=i\frac{Q}{2}-(N-k+1)\tau\,,
\ee
Hence, we see that the two axial symmetries are broken to this particular combination
\be
U(1)_{m_A}\times U(1)_\Gd\quad\rightarrow\quad U(1)_\tau
\ee

Taking into account this the global symmetry group of Theory B is
\be
\prod_{a=1}^kU(1)_{z_a}\times U(1)_\tau\times U(1)_\mu\times U(1)_\zeta\, .
\ee
On this side of the duality, the full flavor symmetry $U(k)_z$ is not visible in the UV, but it enhances at low energies, so that the global symmetry group coincides with that of Theory A
\be
U(k)_z\times U(1)_\tau\times U(1)_\mu\times U(1)_\zeta\, .
\ee
This feature is motivated by the validity of the duality, but also by the fact that the chiral ring generators of $G[U(k)]$ re-organize into representations of $U(k)_z$, as showed in Sec.~\ref{gun}. We list all the charges of the chiral fields under the global symmetries and their R-charges in Table \ref{chargesTB}.

\begin{table}[t]
\centering
\scalebox{0.96}{
\begin{tabular}{c|cccc|c}
{} & $U(k)_z$ & $U(1)_{\tau}$ & $U(1)_\mu$ & $U(1)_\zeta$ & $U(1)_R$ \\ \hline
$\ga_i$ & 0 & $2(i-1)$ & 2 & 0 & $2r+2(i-1)(1-R)$ \\
$T^+_j$ & 0 & $-2N+k+2j$ & $-1$ & 1 & $1-r-(2N-k-2j)(1-R)$ \\
$T^-_{N-l+1}$ & 0 & $-2N+k+2l$ & $-1$ & $-1$ & $1-r-(2N-k-2l)(1-R)$ \\
$\tilde{\gb}_a$ & 0 & $2(N-k+a)$ & 0 & 0 & $2(N-k+a)(1-R)$ \\
$\mathcal{M}$ & adj & $-2$ & 0 & 0 & $2R$ \\
$\Omega$ & $\bar{\Box}$ & $-1$ & 1 & 0 & $r+R$ \\
$\tilde{\Omega}$ & $\Box$ & $-1$ & 1 & 0 & $r+R$ \\
$\mathfrak{M}^\pm_{\mathbb{M}^s}$ & 0 & $-k+2s+2$ & $-1$ & $\pm1$ & $1-r-(k-2s-2)(1-R)$ \\
$\Tr\left(\tilde{p}\mathbb{M}^sp\right)$ & 0 & $2(N-k+s)$ & 2 & 0 & $2r+2(N-k+s)(1-R)$
\end{tabular}}
\caption{Chiral ring generators of Theory B.}
\label{operatorsTB}
\end{table}

The chiral ring generators are those of $G[U(k)]$, except for the operators $\tilde{d}^{(a)}d^{(a)}$ which are set to zero by the F-term equations of the fields $\tilde{\gb}_a$ and with the addition of the $3N-2k$ gauge singlets. They are summarized in Table \ref{operatorsTB}, where we also specify their charges under the global symmetries and their R-charges.  From this, we can find the map between the chiral ring generators of the dual theories, which provides a first non-trivial test of the duality
\be
\Tr_N\Gp^{N-k+a}\quad&\leftrightarrow&\quad\tilde{\gb}_a, \qquad a=1,\cdots,k \nn\\
\mathfrak{M}^+_{\Gp^s}\quad&\leftrightarrow&\quad \begin{cases}T^+_{s+1} & s=0,\cdots,N-k-1 \\ \mathfrak{M}^+_{\mathbb{M}^{k-N+s}} & s=N-k,\cdots,N \end{cases} \nn\\
\mathfrak{M}^-_{\Gp^s}\quad&\leftrightarrow&\quad \begin{cases}T^-_{N-s} & s=0,\cdots,N-k-1 \\ \mathfrak{M}^-_{\mathbb{M}^{k-N+s}} & s=N-k,\cdots,N \end{cases} \nn\\
\Tr_N\left(\tilde{P}\Gp^sP\right)\quad&\leftrightarrow&\quad \begin{cases}\ga_{s+1} & s=0,\cdots,N-k-1 \\ \Tr_k\left(\tilde{p}\mathbb{M}^{k-N+s}p\right) & s=N-k,\cdots,N-1 \end{cases} \nn\\
Q\tilde{P}\quad&\leftrightarrow&\quad\tilde{\Omega} \nn\\
P\tilde{Q}\quad&\leftrightarrow&\quad\Omega \nn\\
Q\tilde{Q}\quad&\leftrightarrow&\quad\mathcal{M}\, .
\ee
At the level of the three-sphere partition functions the duality is expressed by the identity
\be
\mathcal{Z}_{\mathcal{T}_A}&=&\prod_{j=1}^{N-k}\underbrace{\sbfunc{-i\frac{Q}{2}+2j\tau}}_{\gb_j}\int\udl{x_N}\e^{2\pi i\zeta\sum_\ga x_\ga}\frac{\prod_{\ga,\gb=1}^Ns_b\left(i\frac{Q}{2}+(x_\ga-x_\gb)-2\tau\right)}{\prod_{\ga<\gb}^Ns_b\left(i\frac{Q}{2}\pm(x_\ga-x_\gb)\right)}\times\nn\\
&&\qquad\times\prod_{\ga=1}^Ns_b\left(i\frac{Q}{2}\pm x_\ga-\mu\right)\prod_{a=1}^k s_b\left(\pm(x_\ga-z_a)+\tau\right)=\nn
\ee
\be
&=&\prod_{a=1}^k\underbrace{\sbfunc{i\frac{Q}{2}-2(N-k+a)\tau}}_{\tilde{\gb}_a}\prod_{j=1}^{N-k}\underbrace{\sbfunc{i\frac{Q}{2}-2\mu-2(j-1)\tau}}_{\ga_j}\times\nn\\
&&\qquad\times\underbrace{\sbfunc{-\zeta+\mu+(2N-k-2j)\tau}}_{T^+_j}\prod_{j=k+1}^N\underbrace{\sbfunc{\zeta+\mu+(-k+2j-2)\tau}}_{T^-_j}\times\nn\\
&&\qquad\times\mathcal{Z}_{G[U(k)]}\left(z_a,\zeta,\mu+(N-k)\tau,i\frac{Q}{2}-\tau,i\frac{Q}{2}-(N-k+1)\tau\right)=\mathcal{Z}_{\mathcal{T}_B}\, .\nn\\
\label{rankstabid}
\ee
which we prove in  Sec.~\ref{rankstabderivation} for $k=1,2$.

An additional test of duality is  provided in the  Appendix \ref{scirankstab}, where we match the superconformal indices perturbatively in the R-symmetry fugacity for various order for $k=1,2,3$. 

Finally it is a tedious but straightforward exercise to show that, starting from the duality identity for the superconformal indices $\mathcal{I}_{\mathcal{T}_A}=\mathcal{I}_{\mathcal{T}_B}$,
and taking the Coulomb limit as explained in \cite{US1}, we recover the duality identity for the free-field correlator 
\eqref{m3point} of \cite{Fateev:2007qn}.

\subsection{Rank analytic continuation}
\label{rankanalcont}

As we discussed in the Introduction, the rank stabilization duality relating the $U(N)$ theory with an adjoint and $k+1$ flavors to the $G[U(k)]$ quiver theory (with various flipping fields) can be considered the $3d$ uplift of the duality relation \eqref{m3point} for the free field representation with $N$
screening charges of the correlator with $3$ primaries and $k$ degenerate operators in the Liouville theory.
The duality relation \eqref{m3point} provides a form suitable for analytic continuation in $N$ which allows us to reconstruct the correlator for generic values of the momenta lifting the screening condition  \eqref{onshell}.

The $3d$ partition functions enjoys a similar property. Indeed the partition function of Theory B \eqref{rankstabid} consists of two  blocks, the partition function of $G[U(k)]$ and the contribution of the gauge singlets. In the former $N$ enters as  a parameter  inside the charges of the various fields, while in the latter it counts the number of singlets:
\be 
&&\mathcal{Z}_{\mathcal{T}_B}=\prod_{j=1}^N\Sfunc{Q+2ij\tau}\prod_{j=1}^{N-k}\Sfunc{Q+2i\mu+2i(j-1)\tau}\times\nn\\
&&\qquad\times\Sfunc{\frac{Q}{2}+i\zeta-i\mu-i(2N-k-2j)\tau}\prod_{j=k+1}^{N}\Sfunc{\frac{Q}{2}-i\zeta-i\mu-i(2j-k-2)\tau}\times\nn\\
&&\qquad\times\mathcal{Z}_{G[U(k)]}\left(z_a,\zeta,\mu+(N-k)\tau,i\frac{Q}{2}-\tau,i\frac{Q}{2}-(N-k+1)\tau\right)\, ,
\ee
where we moved to this side of the duality the contribution of the $\beta$-fields and used that  $\sbfunc{x}=\Sfunc{\frac{Q}{2}-ix|b,b^{-1}}\equiv\Sfunc{\frac{Q}{2}-ix}$.
Now we can use the periodicity property of the triple-sine function:
\be
S_3(z+\omega_3|\omega_1,\omega_2,\omega_3)=\frac{S_3(z|\omega_1,\omega_2,\omega_3)}{S_{2}(z|\omega_1,\omega_2)}
\ee
 to move the dependence on $N$ inside the argument of the triple-sine function (this allows for analytic continuation) so that the $3d$ partition function can be expressed as:
\be
&&\mathcal{Z}_{\mathcal{T}_B}=\underset{N\in\mathbb{N}}{\mathrm{Res}}\,\Big\{\frac{S_3'(0)\SFunc{-2i\mu+2i\tau}\SFunc{\frac{Q}{2}\pm i\zeta-i\mu-i(2N-k-2)\tau}}{\SFunc{-2iN\tau}\SFunc{-2i\mu-2i(N-k-1)\tau}\SFunc{\frac{Q}{2}\pm i\zeta-i\mu-i(k-2)\tau}}\Big\}\times\nn\\
&&\qquad\quad\times\mathcal{Z}_{G[U(k)]}\left(z_a,\zeta,\mu+(N-k)\tau,i\frac{Q}{2}-\tau,i\frac{Q}{2}-(N-k+1)\tau\right)\, ,
\ee
 where $S_3(x)\equiv S_3(x|b,b^{-1},2i\tau)$.
Inside the brackets we recognize the  five-sphere partition function of the $5d$ $T_2$ theory, which can be realized
on the toric CY geometry $\mathbb{C}^3/\mathbb{Z}_2\times \mathbb{Z}_2$ \cite{Benini:2009gi},
with  quantized  K\"ahler parameters. This is the result that we got in the  $k=0$ case  in \cite{US1} (to which we refer the reader for more details and for the definition of the multiple-sine functions). The analytic continuation in $N$ is then reinterpreted  as geometric transition with the $3d$ theory  appearing  as a codimension-two defect theory at the point  in the moduli space of the $5d$ $T_2$ theory specialized by the quantized values of the K\"ahler parameters as proposed  in
 \cite{Aganagic:2013tta,Aganagic:2014oia}.

The  $(k+3)$-point correlator
corresponds via the AGT map \cite{Alday:2009fs,Drukker:2009id} to the  $T_2$ theory (two M5 wrapping the 3-punctured sphere)
coupled to $k$ co-dimension-two defects ($k$ M2 branes which are points on the 3-punctured sphere).

In  our case the $5d$ theory emerging after the geometric transition can be realized as  the $5d$ $T_2$ 
geometry with the insertion of $k$ toric branes\footnote{In \cite{Kozcaz:2010af} the contribution of $k$ toric branes in the length-two strip geometry, which is closely related to the   $T_2$ geometry, was shown to reproduce the $(k+3)$-point conformal blocks. } and
the contribution of the $G[U(k)]$ theory captures how the defects interact among themselves.

\subsection{Derivation}
\label{rankstabderivation}

In this section we prove analytically the equality of the partition functions \eqref{rankstabid} for low number of flavors, namely $k=1,2$ (the case $k=0$ was discussed in \cite{US1}). This can be done through iterative applications of some basic dualities (see \cite{US1} or Appendix \ref{funddualities} for a quick review). 

The derivation highly relies on a \emph{stabilization} property of the theory, which  holds for $k<N$. We say that the theory is \emph{stable} if, after applying to it some of the fundamental dualities, we recover the same theory but with the rank decreased by one unit and possibly some modification in the parameters of the theory, such as the number of gauge singlets. In \cite{US1}, we showed that the $U(N)$ theory with one adjoint and one flavor, which corresponds to the case $k=0$, is stable and this allowed us to reduce it to a WZ model. We will see that for a higher number of flavors Theory A is not itself stable, but with some initial manipulations we can find a dual frame which actually is. From this point, one can significantly simplify the integrals using the stabilization property and get the partition function of the claimed dual.

\subsubsection{Two flavors}

We start considering the partition function of the $k=1$ case without the contribution of the $\gb$-fields, which we will add at the end for simplicity:
\be
\mathcal{Z}_N^1(z,\tau,\zeta,\mu)&\equiv&\frac{1}{N!}\int\prod_{\ga=1}^N\udl{x_\ga}\e^{2\pi i\zeta\sum_\ga x_\ga}\frac{\prod_{\ga,\gb=1}^Ns_b\left(i\frac{Q}{2}+(x_\ga-x_\gb)-2\tau\right)}{\prod_{\ga<\gb}^Ns_b\left(i\frac{Q}{2}\pm(x_\ga-x_\gb)\right)}\times\nn\\
&\times&\prod_{\ga=1}^Ns_b\left(i\frac{Q}{2}\pm x_\ga-\mu\right)s_b\left(\pm(x_\ga-z)+\tau\right)\, .
\ee
The approach is the same we used in \cite{US1} for the case $k=0$, that is we start by replacing the contribution of the adjoint chiral with an auxiliary $U(N-1)$ integral using the one-monopole duality \eqref{onemonopole}
\be
&&\mathcal{Z}_N^1(z,\tau,\zeta,\mu)=\sbfunc{i\frac{Q}{2}-2N\tau}\frac{1}{(N-1)!}\int\prod_{\ga'=1}^{N-1}\udl{y_{\ga'}}\frac{\e^{-2\pi iN\tau\sum_{\ga'}y_{\ga'}}}{\prod_{\ga'<\gb'}^{N-1}\sbfunc{i\frac{Q}{2}\pm(y_{\ga'}-y_{\gb'})}}\times\nn\\
&&\qquad\qquad\times\frac{1}{N!}\int\prod_{\ga=1}^N\udl{x_\ga}\e^{2\pi i\left(\zeta-(N-1)\tau\right)\sum_\ga x_\ga}\frac{\prod_{\ga=1}^Ns_b\left(i\frac{Q}{2}\pm x_\ga-\mu\right) s_b\left(\pm(x_\ga-z)+\tau\right)}{\prod_{\ga<\gb}^Ns_b\left(i\frac{Q}{2}\pm(x_\ga-x_\gb)\right)}\times\nn\\
&&\qquad\qquad\times\prod_{\ga=1}^{N}\prod_{\ga'=1}^{N-1}\sbfunc{i\frac{Q}{2}\pm(x_\ga+y_\ga')-\tau}\, .
\ee
This corresponds to the partition function of an auxiliary $U(N-1)\times U(N)$ quiver gauge theory with a single fundamental monopole turned on at the $U(N)$ node. Then, we apply Aharony duality on the original integral. In contrast to the $k=0$ case, because of the extra flavor, the identity \eqref{aha} is not an evaluation formula, but it actually yields a $U(1)$ integral
\be
&&\mathcal{Z}_N^1(z,\tau,\zeta,\mu)=\e^{2\pi i\left(\zeta-(N-1)\tau\right)z}\sbfunc{\pm z-\mu+\tau}\sbfunc{i\frac{Q}{2}-2N\tau}\times\nn\\
&&\qquad\qquad\times\sbfunc{i\frac{Q}{2}-2\mu}\sbfunc{-\zeta+\mu+(2N-3)\tau}\sbfunc{\zeta+\mu-\tau}\times\nn\\
&&\qquad\qquad\times\sbfunc{-i\frac{Q}{2}+2\tau}\int\udl{x}\e^{2\pi i\left(\zeta-(N-1)\tau\right)x}\sbfunc{\pm x+\mu}\sbfunc{i\frac{Q}{2}\pm(x+z)-\tau}\times\nn\\
&&\qquad\qquad\times\frac{1}{(N-1)!}\int\prod_{\ga=1}^{N-1}\udl{y_\ga}\e^{-2\pi i(\zeta+\tau)\sum_\ga y_\ga}\frac{\prod_{\ga,\gb=1}^{N-1}\sbfunc{i\frac{Q}{2}+(y_\ga-y_\gb)-2\tau}}{\prod_{\ga<\gb}^{N-1}\sbfunc{i\frac{Q}{2}\pm(y_\ga-y_\gb)}}\times\nn\\
&&\qquad\qquad\times\prod_{\ga=1}^{N-1}\sbfunc{i\frac{Q}{2}\pm y_\ga-\mu-\tau}\sbfunc{\pm(y_\ga-x)+\tau}\, .
\label{2flavstabilized}
\ee
Notice that the contact terms predicted by Aharony duality had the effect of restoring the topological symmetry at the $U(N-1)$ node and thus of removing the monopole superpotential (see \cite{US1} for a more exhaustive discussion of this phenomenon). 

From \eqref{2flavstabilized} we can also see that the original integral was not in a stabilized form since its structure has changed after the application of these two fundamental dualities. Nevertheless, after performing the change of variables $y_i\leftrightarrow -y_i$, we see that in \eqref{2flavstabilized} the last integral has the form of the original integral, but with shifted parameters, so we can still write an iterative relation:
\be
&&\mathcal{Z}_N^1(z,\tau,\zeta,\mu)=\e^{2\pi i\left(\zeta-(N-1)\tau\right)z}\sbfunc{\pm z-\mu+\tau}\sbfunc{i\frac{Q}{2}-2N\tau}\times\nn\\
&&\qquad\times\sbfunc{i\frac{Q}{2}-2\mu}\sbfunc{-\zeta+\mu+(2N-3)\tau}\sbfunc{\zeta+\mu-\tau}\sbfunc{-i\frac{Q}{2}+2\tau}\times\nn\\
&&\qquad\times\int\udl{x}\e^{2\pi i\left(\zeta-(N-1)\tau\right)x}\sbfunc{\pm x+\mu}\sbfunc{i\frac{Q}{2}\pm(x+z)-\tau}\mathcal{Z}_{N-1}^1(x,\tau,\zeta+\tau,\mu+\tau)\, .\nn\\
\label{2flav1iteration}
\ee
With this identity, we can show that the integral that is stabilized is actually \eqref{2flavstabilized}. Indeed, if we repeat the two previous steps, that is we iterate \eqref{2flav1iteration}, we produce a second $U(1)$ integral
\be
&&\mathcal{Z}_N^1(z,\tau,\zeta,\mu)=\e^{2\pi i\left(\zeta-(N-1)\tau\right)z}\sbfunc{\pm z-\mu+\tau}\prod_{j=1}^2\sbfunc{i\frac{Q}{2}-2(N-j+1)\tau}\times\nn\\
&&\quad\times\sbfunc{i\frac{Q}{2}-2\mu-2(j-1)\tau}\sbfunc{-\zeta+\mu+(2N-2j-1)\tau}\sbfunc{\zeta+\mu+(2j-3)\tau}\times\nn\\
&&\quad\times\sbfunc{-i\frac{Q}{2}+2\tau}\int\udl{y}\e^{2\pi i\left(\zeta-(N-3)\tau\right)y}\sbfunc{\pm y+\mu+\tau}\mathcal{Z}_{N-2}^1(y,\tau,\zeta+2\tau,\mu+2\tau)\times\nn\\
&&\quad\times\sbfunc{-i\frac{Q}{2}+2\tau}\int\udl{x}\e^{-4\pi i\tau x}\sbfunc{i\frac{Q}{2}\pm(x-y)-\tau}\sbfunc{i\frac{Q}{2}\pm(x+z)-\tau}\, ,\nn\\
\ee
but the $x$-integral can now be evaluated applying the one-monopole duality \eqref{onemonopole} in the confining case:
\be
&&\mathcal{Z}_N^1(z,\tau,\zeta,\mu)=\e^{2\pi i\left(\zeta-(N-2)\tau\right)z}\sbfunc{\pm z-\mu+\tau}\prod_{j=1}^2\sbfunc{i\frac{Q}{2}-2(N-j+1)\tau}\times\nn\\
&&\quad\times\sbfunc{i\frac{Q}{2}-2\mu-2(j-1)\tau}\sbfunc{-\zeta+\mu+(2N-2j-1)\tau}\sbfunc{\zeta+\mu+(2j-3)\tau}\times\nn\\
&&\quad\times\sbfunc{-i\frac{Q}{2}+4\tau}\int\udl{y}\e^{2\pi i\left(\zeta-(N-2)\tau\right)y}\sbfunc{\pm y+\mu+\tau}\sbfunc{i\frac{Q}{2}\pm(y+z)-2\tau} \times\nn\\
&&\quad\times \mathcal{Z}_{N-2}^1(y,\tau,\zeta+2\tau,\mu+2\tau)\, .
\ee
Hence, we recover precisely the same structure of \eqref{2flavstabilized}, but with a lower rank, some extra gauge singlets and a shift of the parameters. In particular, the shift of the FI parameter indicates that the oppositely charged fundamental monopoles have different topological charge and that charge conjugation is broken in this frame. This explicitly shows that \eqref{2flavstabilized} was indeed stable under the sequential application of one-monopole and Aharony dualities.

We can use this stabilization property to significantly simplify the integral. If we iterate \eqref{2flav1iteration} and \eqref{onemonopole} $n$ times, we get indeed
\be
&&\mathcal{Z}_N^1(z,\tau,\zeta,\mu)=\e^{2\pi i\left(\zeta-(N-n)\tau\right)z}\sbfunc{\pm z-\mu+\tau}\prod_{j=1}^n\sbfunc{i\frac{Q}{2}-2(N-j+1)\tau}\times\nn\\
&&\times\sbfunc{-\zeta+\mu+(2N-2j-1)\tau}\sbfunc{\zeta+\mu+(2j-3)\tau}\sbfunc{i\frac{Q}{2}-2\mu-2(j-1)\tau}\times\nn\\
&&\times\sbfunc{-i\frac{Q}{2}+2n\tau}\int\udl{x}\e^{2\pi i\left(\zeta-(N-n)\tau\right)x}\sbfunc{\pm x+\mu+(n-1)\tau}\sbfunc{i\frac{Q}{2}\pm(x+z)-n\tau}\times\nn\\
&&\times \mathcal{Z}_{N-n}^1(x,\tau,\zeta+n\tau,\mu+n\tau)\, .
\ee
In particular, if we set $n=N$ in the above expression, the original gauge node is completely confined
\be
&&\mathcal{Z}_N^1(z,\tau,\zeta,\mu)=\e^{2\pi i\zeta z}\sbfunc{\pm z-\mu+\tau}\prod_{j=1}^N\sbfunc{i\frac{Q}{2}-2j\tau}\times\nn\\
&&\times\sbfunc{-\zeta+\mu+(2N-2j-1)\tau}\sbfunc{\zeta+\mu+(2j-3)\tau}\sbfunc{i\frac{Q}{2}-2\mu-2(j-1)\tau}\times\nn\\
&&\times\sbfunc{-i\frac{Q}{2}+2N\tau}\int\udl{x}\e^{2\pi i \zeta x}\sbfunc{\pm x+\mu+(N-1)\tau}\sbfunc{i\frac{Q}{2}\pm(x+z)-N\tau}\, . \nn\\
\ee
Notice that the FI parameter of the remaining $U(1)$ node is no longer shifted. This means that the oppositely charged monopole operators have the same quantum numbers under all the global symmetries and that charge conjugation, which was broken in all the previous auxiliary dual frames, has been restored.

The partition function that we obtained is that of $G[U(1)]$ with some extra gauge singlets. In order to write the result in the desired form, we apply Aharony duality to the $U(1)$ integral. This gives back another $U(1)$ integral, but with different parameters and some of the extra gauge singlets flipped away. Essentially, what we are doing is applying the recombination duality we discussed in Sec.~\ref{recombinationduality} in the particular case $N=1$ and $k=1$. If we also add the contribution of the $N-1$ $\gb$-fields, the final result coincides with \eqref{rankstabid} for $k=1$
\be
\mathcal{Z}_{\mathcal{T}_A}&=&\prod_{j=1}^{N-1}\sbfunc{-i\frac{Q}{2}+2j\tau}\mathcal{Z}_N^1(z,\tau,\zeta,\mu)=\nn\\
&=&\sbfunc{i\frac{Q}{2}-2N\tau}\prod_{j=1}^{N-1}\sbfunc{i\frac{Q}{2}-2\mu-2(j-1)\tau}\times\nn\\
&\times&\sbfunc{-\zeta+\mu+(2N-2j-1)\tau}\prod_{j=2}^N\sbfunc{\zeta+\mu+(2j-3)\tau}\times\nn\\
&\times&\int\udl{x}\e^{2\pi i\zeta x}\sbfunc{i\frac{Q}{2}\pm x-\mu-(N-1)\tau}\sbfunc{\pm(x-z)+N\tau}=\nn
\ee
\be
&=&\sbfunc{i\frac{Q}{2}-2N\tau}\prod_{j=1}^{N-1}\sbfunc{i\frac{Q}{2}-2\mu-2(j-1)\tau}\times\nn\\
&\times&\sbfunc{-\zeta+\mu+(2N-2j-1)\tau}\prod_{j=2}^N\sbfunc{\zeta+\mu+(2j-3)\tau}\times\nn\\
&\times&\mathcal{Z}_{G[U(1)]}(z,\zeta,\mu+(N-1)\tau,i\frac{Q}{2}-N\tau)=\mathcal{Z}_{\mathcal{T}_B}\, .\nn\\
\ee

\subsubsection{Three flavors}

Again, we start considering the partition function of Theory A in the $k=2$ case without the contribution of the $\gb$-fields
\be
\mathcal{Z}_N^2(z_a,\tau,\zeta,\mu)&\equiv&\frac{1}{N!}\int\prod_{\ga=1}^N\udl{x_\ga}\e^{2\pi i\zeta\sum_\ga x_\ga}\frac{\prod_{\ga,\gb=1}^Ns_b\left(i\frac{Q}{2}+(x_\ga-x_\gb)-2\tau\right)}{\prod_{\ga<\gb}^Ns_b\left(i\frac{Q}{2}\pm(x_\ga-x_\gb)\right)}\times\nn\\
&\times&\prod_{\ga=1}^Ns_b\left(i\frac{Q}{2}\pm x_\ga-\mu\right)\prod_{a=1}^2s_b\left(\pm(x_\ga-z_a)+\tau\right)\, .
\ee
The first manipulations are still the same, that is we use the one-monopole duality \eqref{onemonopole} to replace the contribution of the adjoint chiral with a $U(N-1)$ integral
\be
\mathcal{Z}_N^2(z_a,\tau,\zeta,\mu)&=&\sbfunc{i\frac{Q}{2}-2N\tau}\frac{1}{(N-1)!}\int\prod_{\ga'=1}^{N-1}\udl{y_{\ga'}}\frac{\e^{-2\pi iN\tau\sum_{\ga'}y_{\ga'}}}{\prod_{\ga'<\gb'}^{N-1}\sbfunc{i\frac{Q}{2}\pm(y_{\ga'}-y_{\gb'})}}\times\nn\\
&\times&\frac{1}{N!}\int\prod_{\ga=1}^N\udl{x_\ga}\e^{2\pi i(\zeta-(N-1)\tau)\sum_\ga x_\ga}\frac{\prod_{\ga=1}^N\sbfunc{i\frac{Q}{2}\pm x_\ga-\mu}}{\prod_{\ga<\gb}^N\sbfunc{i\frac{Q}{2}\pm(x_\ga-x_\gb)}}\times\nn\\
&\times&\prod_{a=1}^2\sbfunc{\pm(x_\ga-z_a)+\tau}\prod_{\ga'=1}^{N-1}\sbfunc{i\frac{Q}{2}\pm(x_\ga+y_{\ga'})-\tau}
\ee
and we reduce the rank of the original integral using Aharony duality \eqref{aha}
\be
&&\mathcal{Z}_N^2(z_a,\tau,\zeta,\mu)=\e^{2\pi i(\zeta-(N-1)\tau)\sum_a z_a}\prod_{a,b=1}^2\sbfunc{-i\frac{Q}{2}+(z_a-z_b)+2\tau}\times\nn\\
&&\qquad\times\prod_{a=1}^2\sbfunc{\pm z_a-\mu+\tau}\sbfunc{i\frac{Q}{2}-2N\tau}\sbfunc{i\frac{Q}{2}-2\mu}\times\nn
\ee
\be
&&\qquad\times\sbfunc{-\zeta+\mu+2(N-2)\tau}\sbfunc{\zeta+\mu-2\tau}\int\frac{\udl{x_1}\udl{x_2}}{2}\e^{2\pi i(\zeta-(N-1)\tau)\sum_a x_a}\times\nn\\
&&\qquad\times\frac{\prod_{a=1}^2\sbfunc{\pm x_a+\mu}\prod_{b=1}^2\sbfunc{i\frac{Q}{2}\pm(x_a+z_b)-\tau}}{\sbfunc{i\frac{Q}{2}\pm(x_1-x_2)}}\times\nn\\
&&\qquad\times\frac{1}{(N-1)!}\int\prod_{\ga=1}^{N-1}\udl{y_\ga}\e^{-2\pi i(\zeta+\tau)\sum_\ga y_\ga}\frac{\prod_{\ga,\gb=1}^{N-1}\sbfunc{i\frac{Q}{2}\pm(y_\ga-y_\gb)-2\tau}}{\prod_{\ga<\gb}^{N-1}\sbfunc{i\frac{Q}{2}\pm(y_\ga-y_\gb)}}\times\nn\\
&&\qquad\times\prod_{\ga=1}^{N-1}\sbfunc{i\frac{Q}{2}\pm y_\ga-\mu-\tau}\prod_{a=1}^2\sbfunc{\pm(y_\ga-x_a)+\tau}\, .
\label{eq}
\ee
In the case $k=1$ that we considered in the previous section, it was at this point that we reached the stable form of the integral. This is not true anymore and we actually need some extra work to get the stable integral. Indeed, we can still recognize in the last integral of \eqref{eq} the same original structure and this allows us to write the iterative relation
\be
&&\mathcal{Z}_N^2(z_a,\tau,\zeta,\mu)=\e^{2\pi i(\zeta-(N-1)\tau)\sum_a z_a}\prod_{a,b=1}^2\sbfunc{-i\frac{Q}{2}+(z_a-z_b)+2\tau}\times\nn\\
&&\qquad\times\prod_{a=1}^2\sbfunc{\pm z_a-\mu+\tau}\sbfunc{i\frac{Q}{2}-2N\tau}\sbfunc{i\frac{Q}{2}-2\mu}\times\nn\\
&&\qquad\times\sbfunc{-\zeta+\mu+2(N-2)\tau}\sbfunc{\zeta+\mu-2\tau}\int\frac{\udl{x_1}\udl{x_2}}{2}\e^{2\pi i(\zeta-(N-1)\tau)\sum_a x_a}\times\nn\\
&&\qquad\times\frac{\prod_{a=1}^2\sbfunc{\pm x_a+\mu}\prod_{b=1}^2\sbfunc{i\frac{Q}{2}\pm(x_a+z_b)-\tau}}{\sbfunc{i\frac{Q}{2}\pm(x_1-x_2)}}\mathcal{Z}_{N-1}^2(x_a,\tau,\zeta+\tau,\mu+\tau)\, .\nn\\
\label{3flav1iteration}
\ee
If we iterate this identity once, we get
\be
&&\mathcal{Z}_N^2(z_a,\tau,\zeta,\mu)=\e^{2\pi i(\zeta-(N-1)\tau)\sum_a z_a}\prod_{a,b=1}^2\sbfunc{-i\frac{Q}{2}+(z_a-z_b)+2\tau}\times\nn\\
&&\qquad\times\prod_{a=1}^2\sbfunc{\pm z_a-\mu+\tau}\prod_{j=1}^2\sbfunc{i\frac{Q}{2}-2(N-j+1)\tau}\times\nn
\ee
\be
&&\qquad\times\sbfunc{i\frac{Q}{2}-2\mu-2(j-1)\tau}\sbfunc{-\zeta+\mu+2(N-j-1)\tau}\sbfunc{\zeta+\mu+2(j-2)\tau}\times\nn\\
&&\qquad\times\int\frac{\udl{y_1}\udl{y_2}}{2}\e^{2\pi i(\zeta-(N-3)\tau)\sum_a y_a}\frac{\prod_{a=1}^2\sbfunc{\pm y_a+\mu+\tau}}{\sbfunc{i\frac{Q}{2}\pm(y_1-y_2)}}\mathcal{Z}_{N-2}^2(y_a,\tau,\zeta+2\tau,\mu+2\mu)\times\nn\\
&&\qquad\times\int\frac{\udl{x_1}\udl{x_2}}{2}\e^{-4\pi i\tau\sum_a x_a}\frac{\prod_{a,b=1}^2\sbfunc{-i\frac{Q}{2}\pm(x_a-x_b)+2\tau}}{\sbfunc{i\frac{Q}{2}\pm(x_1-x_2)}}\times\nn\\
&&\qquad\times\prod_{a,b=1}^2\sbfunc{i\frac{Q}{2}\pm(x_a-y_b)-\tau}\sbfunc{i\frac{Q}{2}\pm(x_a+z_b)-\tau}\, ,
\ee
but now there is no evaluation formula for any of the two $U(2)$ integrals which allows us to get back to an integral of the form of \eqref{3flav1iteration}. This shows that the integral is not stable yet. Instead, we can at this point apply the intermezzo duality \eqref{intermezzo}
\be
&&\mathcal{Z}_N^2(z_a,\tau,\zeta,\mu)=\e^{2\pi i(\zeta-(N-2)\tau)\sum_a z_a}\prod_{a,b=1}^2\sbfunc{-i\frac{Q}{2}+(z_a-z_b)+2\tau}\times\nn\\
&&\qquad\times\prod_{a=1}^2\sbfunc{\pm z_a-\mu+\tau}\prod_{j=1}^2\sbfunc{i\frac{Q}{2}-2(N-j+1)\tau}\times\nn\\
&&\qquad\times\sbfunc{i\frac{Q}{2}-2\mu-2(j-1)\tau}\sbfunc{-\zeta+\mu+2(N-j-1)\tau}\sbfunc{\zeta+\mu+2(j-2)\tau}\times\nn\\
&&\qquad\times\sbfunc{-i\frac{Q}{2}+4\tau}\sbfunc{-\frac{3}{2}iQ+6\tau}\sbfunc{i\frac{Q}{2}-2\tau}\times\nn\\
&&\qquad\times\int\frac{\udl{y_1}\udl{y_2}}{2}\e^{2\pi i(\zeta-(N-2)\tau)\sum_a y_a}\frac{\prod_{a,b=1}^2\sbfunc{i\frac{Q}{2}+(y_a-y_b)-2\tau}}{\sbfunc{i\frac{Q}{2}\pm(y_1-y_2)}}\times\nn\\
&&\qquad\times\prod_{a=1}^2\sbfunc{\pm y_a+\mu+\tau}\sbfunc{i\frac{Q}{2}\pm(y_a-z_1)-2\tau}\mathcal{Z}_{N-2}^2(y_a,\tau,\zeta+2\tau,\mu+2\tau)\times\nn\\
&&\qquad\times\int\udl{x}\sbfunc{\pm(x+z_1)+\tau}\sbfunc{iQ\pm(x+z_2)-3\tau}\prod_{a=1}^2\sbfunc{\pm(x+y_a)+\tau}\, .\nn\\
\label{3falvstab}
\ee
This is the integral that is actually stable. To see this, we apply again \eqref{3flav1iteration}
\be
&&\mathcal{Z}_N^2(z_a,\tau,\zeta,\mu)=\e^{2\pi i(\zeta-(N-2)\tau)\sum_a z_a}\prod_{a,b=1}^2\sbfunc{-i\frac{Q}{2}+(z_a-z_b)+2\tau}\prod_{a=1}^2\sbfunc{\pm z_a-\mu+\tau}\times\nn\\
&&\quad\times\prod_{j=1}^3\sbfunc{i\frac{Q}{2}-2(N-j+1)\tau}\sbfunc{i\frac{Q}{2}-2\mu-2(j-1)\tau}\sbfunc{-\zeta+\mu+2(N-j-1)\tau}\times\nn
\ee
\be
&&\quad\times\sbfunc{\zeta+\mu+2(j-2)\tau}\sbfunc{-i\frac{Q}{2}+4\tau}\sbfunc{-\frac{3}{2}iQ+6\tau}\sbfunc{i\frac{Q}{2}-2\tau}\times\nn\\
&&\quad\times\int\frac{\udl{x_1}\udl{x_2}}{2}\e^{2\pi i(\zeta-(N-5)\tau)\sum_a x_a}\frac{\prod_{a=1}^2\sbfunc{\pm x_a+\mu+2\tau}}{\sbfunc{i\frac{Q}{2}\pm(x_1-x_2)}}\mathcal{Z}_{N-3}^2(x_a,\tau,\zeta+3\tau,\mu+3\tau)\times\nn\\
&&\quad\times\int\udl{x}\sbfunc{\pm(x-z_1)+\tau}\sbfunc{iQ\pm(x-z_2)-3\tau}\times\nn\\
&&\quad\times\int\frac{\udl{y_1}\udl{y_2}}{2}\e^{-6\pi i\tau\sum_a y_a}\frac{\prod_{a=1}^2\sbfunc{i\frac{Q}{2}\pm(y_a+z_1)-2\tau}\sbfunc{\pm(y_a+x)+\tau}}{\sbfunc{i\frac{Q}{2}\pm(y_1-y_2)}}\times\nn\\
&&\quad\times \prod_{b=1}^2\sbfunc{i\frac{Q}{2}\pm(y_a-x_b)-\tau}\, .
\ee
Then, we use the one-monopole duality \eqref{onemonopole} to replace the last integral with a $U(1)$ one
\be
&&\mathcal{Z}_N^2(z_a,\tau,\zeta,\mu)=\e^{2\pi i(\zeta-(N-3)\tau)z_1}\e^{2\pi i(\zeta-(N-2)\tau)z_2}\prod_{a,b=1}^2\sbfunc{-i\frac{Q}{2}+(z_a-z_b)+2\tau}\times\nn\\
&&\quad\times\prod_{a=1}^2\sbfunc{\pm z_a-\mu+\tau}\prod_{j=1}^3\sbfunc{i\frac{Q}{2}-2(N-j+1)\tau}\times\nn\\
&&\quad\times\sbfunc{i\frac{Q}{2}-2\mu-2(j-1)\tau}\sbfunc{-\zeta+\mu+2(N-j-1)\tau}\sbfunc{\zeta+\mu+2(j-2)\tau}\times\nn\\
&&\quad\times\sbfunc{-i\frac{Q}{2}+6\tau}\sbfunc{-\frac{3}{2}iQ+6\tau}\sbfunc{i\frac{Q}{2}-2\tau}\times\nn\\
&&\quad\times\int\frac{\udl{x_1}\udl{x_2}}{2}\e^{2\pi i(\zeta-(N-3)\tau)\sum_a x_a}\frac{\prod_{a,b=1}^2\sbfunc{i\frac{Q}{2}+(x_a-x_b)-2\tau}}{\sbfunc{i\frac{Q}{2}\pm(x_1-x_2)}}\times\nn\\
&&\quad\times\prod_{a=1}^2\sbfunc{\pm x_a+\mu+2\tau}\sbfunc{i\frac{Q}{2}\pm(x_a+z_1)-3\tau}\mathcal{Z}_{N-3}^2(x_a,\tau,\zeta+3\tau,\mu+3\tau)\times\nn\\
&&\quad\times \int\udl{y}\e^{i\pi(iQ-6\tau)y}\sbfunc{\pm(y-z_1)+2\tau}\prod_{a=1}^2\sbfunc{\pm(y+x_a)+\tau}\times\nn\\
&&\quad\times\int\udl{x}\e^{i\pi(iQ-8\tau)x}\sbfunc{iQ\pm(x+z_2)-3\tau}\sbfunc{i\frac{Q}{2}\pm(x+y)-\tau}
\ee
and finally we can evaluate the last $U(1)$ integral using again the one-monopole duality \eqref{onemonopole}
\be
&&\mathcal{Z}_N^2(z_a,\tau,\zeta,\mu)=\e^{2\pi i(\zeta-(N-3)\tau)\sum_a z_a}\prod_{a,b=1}^2\sbfunc{-i\frac{Q}{2}+(z_a-z_b)+2\tau}\prod_{a=1}^2\sbfunc{\pm z_a-\mu+\tau}\times\nn\\
&&\quad\times\prod_{j=1}^3\sbfunc{i\frac{Q}{2}-2(N-j+1)\tau}\sbfunc{i\frac{Q}{2}-2\mu-2(j-1)\tau}\sbfunc{-\zeta+\mu+2(N-j-1)\tau}\times\nn\\
&&\quad\times\sbfunc{\zeta+\mu+2(j-2)\tau}\sbfunc{-i\frac{Q}{2}+6\tau}\sbfunc{-\frac{3}{2}iQ+8\tau}\sbfunc{i\frac{Q}{2}-2\tau}\times\nn\\
&&\quad\times\int\frac{\udl{x_1}\udl{x_2}}\e^{2\pi i(\zeta-(N-3)\tau)\sum_a x_a}\frac{\prod_{a,b=1}^2\sbfunc{i\frac{Q}{2}+(x_a-x_b)-2\tau}}{\sbfunc{i\frac{Q}{2}\pm(x_1-x_2)}}\times\nn\\
&&\quad\times\prod_{a=1}^2\sbfunc{\pm x_a+\mu+2\tau}\sbfunc{i\frac{Q}{2}\pm(x_a+z_1)-3\tau}\mathcal{Z}_{N-3}^2(x_a,\tau,\zeta+3\tau,\mu+3\tau)\times\nn
\ee
\be
&&\quad\times\int\udl{y}\sbfunc{\pm(y-z_1)+2\tau}\sbfunc{iQ\pm(y-z_2)-4\tau}\prod_{a=1}^2\sbfunc{\pm(y+x_a)+\tau}\, .\nn\\
\ee
The result has exactly the same structure of \eqref{3falvstab}, which means that the integral is now stable. Hence, we can iterate the last three steps $n$ times to get
\be
&&\mathcal{Z}_N^2(z_a,\tau,\zeta,\mu)=\e^{2\pi i(\zeta-(N-n)\tau)\sum_a z_a}\prod_{a,b=1}^2\sbfunc{-i\frac{Q}{2}+(z_a-z_b)+2\tau}\times\nn\\
&&\quad\times\prod_{a=1}^2\sbfunc{\pm z_a-\mu+\tau}\prod_{j=1}^n\sbfunc{i\frac{Q}{2}-2(N-j+1)\tau}\times\nn\\
&&\quad\times\sbfunc{i\frac{Q}{2}-2\mu-2(j-1)\tau}\sbfunc{-\zeta+\mu+2(N-j-1)\tau}\sbfunc{\zeta+\mu+2(j-2)\tau}\times\nn\\
&&\quad\times\sbfunc{-i\frac{Q}{2}+2n\tau}\sbfunc{-\frac{3}{2}iQ+2(n+1)\tau}\sbfunc{i\frac{Q}{2}-2\tau}\times\nn\\
&&\quad\times\int\frac{\udl{x_1}\udl{x_2}}\e^{2\pi i(\zeta-(N-n)\tau)\sum_a x_a}\frac{\prod_{a,b=1}^2\sbfunc{i\frac{Q}{2}+(x_a-x_b)-2\tau}}{\sbfunc{i\frac{Q}{2}\pm(x_1-x_2)}}\times\nn
\ee
\be
&&\quad\times\prod_{a=1}^2\sbfunc{\pm x_a+\mu+(n-1)\tau}\sbfunc{i\frac{Q}{2}\pm(x_a+z_1)-n\tau}\mathcal{Z}_{N-n}^2(x_a,\tau,\zeta+n\tau,\mu+n\tau)\times\nn\\
&&\quad\times\int\udl{y}\sbfunc{\pm(y-z_1)+(n-1)\tau}\sbfunc{iQ\pm(y-z_2)-(n+1)\tau}\prod_{a=1}^2\sbfunc{\pm(y+x_a)+\tau}\, .\nn\\
\ee
This corresponds to the partition function of the quiver gauge theory represented in the middle of Figure \ref{stable2} with the addition of several gauge singlets, which were produced by the sequential application of the fundamental dualities.

As in the previous cases, we can use the stabilization property of the integral to significantly simplify the result. Indeed, if we set $n=N$, the original $U(N)$ gauge node is completely confined
\be
&&\mathcal{Z}_N^2(z_a,\tau,\zeta,\mu)=\e^{2\pi i\zeta\sum_a z_a}\prod_{a,b=1}^2\sbfunc{-i\frac{Q}{2}+(z_a-z_b)+2\tau}\prod_{a=1}^2\sbfunc{\pm z_a-\mu+\tau}\times\nn\\
&&\quad\times\prod_{j=1}^N\sbfunc{i\frac{Q}{2}-2j\tau}\sbfunc{i\frac{Q}{2}-2\mu-2(j-1)\tau}\sbfunc{-\zeta+\mu+2(N-j-1)\tau}\times\nn\\
&&\quad\times\sbfunc{\zeta+\mu+2(j-2)\tau}\sbfunc{-i\frac{Q}{2}+2N\tau}\sbfunc{-\frac{3}{2}iQ+2(N+1)\tau}\sbfunc{i\frac{Q}{2}-2\tau}\times\nn\\
&&\quad\times\int\frac{\udl{x_1}\udl{x_2}}\e^{2\pi i\zeta\sum_a x_a}\frac{\prod_{a,b=1}^2\sbfunc{i\frac{Q}{2}+(x_a-x_b)-2\tau}}{\sbfunc{i\frac{Q}{2}\pm(x_1-x_2)}}\prod_{a=1}^2\sbfunc{\pm x_a+\mu+(N-1)\tau}\times\nn\\
&&\quad\times\sbfunc{i\frac{Q}{2}\pm(x_a+z_1)-N\tau}\int\udl{y}\sbfunc{\pm(y-z_1)+(N-1)\tau}\times\nn\\
&&\quad\times \sbfunc{iQ\pm(y-z_2)-(N+1)\tau}\prod_{a=1}^2\sbfunc{\pm(y+x_a)+\tau}\, .
\ee
This integral is not the partition function of $G[U(2)]$ yet because of the contribution of the adjoint chiral corresponding to the $U(2)$ node. This problem can be solved by simply applying the two-monopole duality \eqref{twomonopoles} to the $U(1)$ integral
\be
&&\mathcal{Z}_N^2(z_a,\tau,\zeta,\mu)=\e^{2\pi i\zeta\sum_a z_a}\prod_{a=1}^2\sbfunc{\pm z_a-\mu+\tau}\prod_{j=1}^N\sbfunc{i\frac{Q}{2}-2(N-j+1)\tau}\times\nn\\
&&\quad\times\sbfunc{i\frac{Q}{2}-2\mu-2(j-1)\tau}\sbfunc{-\zeta+\mu+2(N-j-1)\tau}\sbfunc{\zeta+\mu+2(j-2)\tau}\times\nn\\
&&\quad\times\sbfunc{-i\frac{Q}{2}+2N\tau}\sbfunc{-i\frac{Q}{2}+2(N-1)\tau}\sbfunc{i\frac{Q}{2}-2\tau}\times\nn\\
&&\quad\times\int\frac{\udl{x_1}\udl{x_2}}{2}\frac{ \e^{2\pi i\zeta\sum_a x_a}  }{\sbfunc{i\frac{Q}{2}\pm(x_1-x_2)}}
\prod_{a=1}^2\sbfunc{\pm x_a+\mu+(N-1)\tau}\times\nn\\
&&\quad\times\sbfunc{i\frac{Q}{2}\pm(x_a+z_2)-N\tau}\int\udl{x}\sbfunc{i\frac{Q}{2}\pm(x+z_1)-(N-1)\tau}\times\nn\\
&&\quad\times \sbfunc{-i\frac{Q}{2}\pm(x+z_2)+(N+1)\tau}\prod_{a=1}^2\sbfunc{i\frac{Q}{2}\pm(y+x_a)-\tau}\, .\nn\\
\ee
Now we can apply the recombination duality \eqref{recombination} in the case $N=k=2$ to flip away some of the gauge singlets and obtain the desired form of the $G[U(2)]$. If we also restore the contribution of the $N-2$ $\gb$-fields, we get indeed
\be
&&\mathcal{Z}_{\mathcal{T}_A}=\prod_{j=1}^{N-2}\sbfunc{-i\frac{Q}{2}+2j\tau}\mathcal{Z}_N^2(z_a,\tau,\zeta,\mu)=\nn\\
&&\qquad=\prod_{a=1}^2\sbfunc{i\frac{Q}{2}-2(N+a-2)\tau}\prod_{j=1}^{N-2}\sbfunc{i\frac{Q}{2}-2\mu-2(j-1)\tau}\times\nn\\
&&\qquad\times\sbfunc{-\zeta+\mu+2(N-j-1)\tau}\prod_{j=3}^N\sbfunc{\zeta+\mu+2(2j-2)\tau}\times\nn\\
&&\qquad\times\int\frac{\udl{x_1}\udl{x_2}}{2}\e^{2\pi i\zeta \sum_a x_a}\frac{\prod_{a=1}^2\sbfunc{i\frac{Q}{2}\pm x_a-\mu-(N-2)\tau}}{\sbfunc{i\frac{Q}{2}\pm(x_1-x_2)}}\times\nn\\
&&\qquad\times\sbfunc{\pm(x_a-z_1)+(N-1)\tau}\int\udl{x}\sbfunc{\pm(x-z_1)-(N-2)\tau}\times\nn\\
&&\qquad\times\sbfunc{\pm(x-z_2)+N\tau}\prod_{a=1}^2\sbfunc{i\frac{Q}{2}\pm(x-x_a)-\tau}=\nn
\ee
\be
&&\qquad=\prod_{a=1}^2\sbfunc{i\frac{Q}{2}-2(N+a-2)\tau}\prod_{j=1}^{N-2}\sbfunc{i\frac{Q}{2}-2\mu-2(j-1)\tau}\times\nn\\
&&\qquad\times\sbfunc{-\zeta+\mu+2(N-j-1)\tau}\prod_{j=3}^N\sbfunc{\zeta+\mu+2(2j-2)\tau}\times\nn\\ 
&&\qquad\times\mathcal{Z}_{G[U(2)]}(z_a,\zeta,\mu+(N-2)\tau,i\frac{Q}{2}-\tau,i\frac{Q}{2}-(N-1)\tau)=\mathcal{Z}_{\mathcal{T}_B}\, ,
\ee
which precisely corresponds to \eqref{rankstabid} in the case $k=2$.

\section*{Acknowledgements}
We would like to thank  A.~Amariti, S.~Benvenuti, C.~Hwang, E.~Pomoni, A.~Zaffaroni  for helpful comments and discussions and in particular F.~Aprile for collaboration on the early stages of this project as well as for many enlightening discussions.
S.P. is partially supported by the ERC-STG grant 637844-HBQFTNCER, by the Fondazione Cariplo and Regione  Lombardia,  grant  n. 2015-1253  and  by  the  INFN.   M.S.  is  partially  supported by the ERC-STG grant 637844-HBQFTNCER, by the University of Milano-Bicocca grant 2016-ATESP-0586 and by the INFN.

\appendix

\section{Basic \boldmath$3d$ dualities}
\label{funddualities}

We recall here some important $3d$ dualities that are used in the derivations presented in the main text. The most fundamental of these dualities was first proposed in \cite{Benini:2017dud}:

\medskip
\noindent \textbf{Theory 1}: $U(N_c)$ with $N_f$ fundamental flavors and superpotential
\be
\mathcal{W}=\mathfrak{M}^++\mathfrak{M}^-\, .
\ee

\medskip
\noindent \textbf{Theory 2}: $U(N_f-N_c-2)$ with $N_f$ fundamental flavors, $N_f^2$ singlets (collected in a matrix $M_{ij}$) and superpotential
\be
\hat{\mathcal{W}}=\sum_{i,j=1}^{N_f}M_{ij}\tilde{q}_iq_j+\hat{\mathfrak{M}}^++\hat{\mathfrak{M}}^-\, .
\ee

\medskip
\noindent The monopole superpotential completely breaks both the axial and the topological symmetry, so that the global symmetry group of the two theories is $SU(N_f)\times SU(N_f)$. Moreover, it has the effect of fixing the R-charges of all the chiral fields to $\frac{N_f-N_c-1}{N_f}$. At the level of three-sphere partition functions, this duality is represented by the following integral identity:
\be
\mathcal{Z}_{\mathcal{T}_1}&=&\frac{1}{N_c!}\int\prod_{i=1}^{N_c}\udl{x_i}\frac{\prod_{i=1}^{N_c}\prod_{a=1}^{N_f}s_b\left(i\frac{Q}{2}\pm(x_i+M_a)-\mu_a\right)}{\prod_{i<j}^{N_c}s_b\left(i\frac{Q}{2}\pm(x_i-x_j)\right)}=\nn\\
&=&\frac{1}{(N_f-N_c-2)!}\prod_{a,b=1}^{N_f}s_b\left(i\frac{Q}{2}-(\mu_a+\mu_b-M_a+M_b)\right)\times\nn\\
&\times&\int\prod_{i=1}^{N_f-N_c-2}\udl{x_i}\frac{\prod_{i=1}^{N_f-N_c-2}\prod_{a=1}^{N_f}s_b\left(\pm(x_i-M_a)+\mu_a\right)}{\prod_{i<j}^{N_f-N_c-2}s_b\left(i\frac{Q}{2}\pm(x_i-x_j)\right)}=\mathcal{Z}_{\mathcal{T}_2}\, ,
\label{twomonopoles}
\ee
where $M_a$, $\mu_a$ are real masses corresponding to the Cartan subalgebra of the diagonal and the anti-diagonal combinations of the two $SU(N_f)$ flavor symmetries. Hence, the vector masses sum to zero $\sum M_a=0$, while the axial masses have to satisfy the constraint
\be
2\sum_{a=1}^{N_f}\mu_a=iQ(N_f-N_c-1)\, ,
\label{constraintCONF2}
\ee
which is often referred to in the mathematical literature as ``balancing condition".

From this duality, we can derive two others by performing suitable real mass deformations. The first one involves theories with only one monopole linearly turned on in the superpotential \cite{Benini:2017dud}:

\medskip
\noindent \textbf{Theory 1}: $U(N_c)$ with $N_f$ fundamental flavors and superpotential
\be
\mathcal{W}=\mathfrak{M}^-\, .
\ee

\medskip
\noindent \textbf{Theory 2}: $U(N_f-N_c-1)$ with $N_f$ fundamental flavors, $N_f^2$ singlets (collected in a matrix $M_{ij}$), an extra singlet $S^+$ and superpotential
\be
\hat{\mathcal{W}}=\sum_{i,j=1}^{N_f}M_{ij}\tilde{q}_iq_j+\hat{\mathfrak{M}}^++S^+\hat{\mathfrak{M}}^-\, .
\ee

\medskip
\noindent Implementing the real mass deformation on the partition functions, we get the following identity:
\be
\mathcal{Z}_{\mathcal{T}_1}&=&\frac{1}{N_c!}\int\prod_{i=1}^{N_c}\udl{x_i}\e^{i\pi\left(\sum_{i=1}^{N_c}x_i\right)(\eta-iQ)}\frac{\prod_{i=1}^{N_c}\prod_{a=1}^{N_f}s_b\left(i\frac{Q}{2}\pm(x_i+M_a)-\mu_a\right)}{\prod_{i<j}^{N_c}s_b\left(i\frac{Q}{2}\pm(x_j-x_i)\right)}=\nn\\
&=&\frac{1}{(N_f-N_c-1)!}\e^{-i\pi\left(2\sum_{a=1}^{N_f}M_a\mu_a+(\eta-iQ)\sum_{a=1}^{N_f}M_a\right)}s_b\left(i\frac{Q}{2}-\eta\right)\times\nn\\
&&\quad\times\prod_{a,b=1}^{N_f}s_b\left(i\frac{Q}{2}-(\mu_a+\mu_b-M_a+M_b)\right)\times\nn\\
&&\quad\times\int\prod_{i=1}^{N_f-N_c-1}\udl{x_i}\e^{i\pi\eta\sum_{i=1}^{N_c}x_i}\frac{\prod_{i=1}^{N_f-N_c-1}\prod_{a=1}^{N_f}s_b\left(\pm(x_i-M_a)+\mu_a\right)}{\prod_{i<j}^{N_f-N_c-1}s_b\left(i\frac{Q}{2}\pm(x_j-x_i)\right)}=\mathcal{Z}_{\mathcal{T}_2}\, ,\nn\\
\label{onemonopole}
\ee
where $\eta$ is the holomorphic combination between the real mass for the restored combination of the topological and the axial symmetry and the mixing coefficient of this abelian symmetry with the R-symmetry. The balancing condition is in this case
\be
\eta+2\sum_{a=1}^{N_f}\mu_a=iQ(N_f-N_c)\, .
\label{constraintM-}
\ee

Finally, with a different real mass deformation we can flow to Aharony duality \cite{Aharony:1997gp}:

\medskip
\noindent \textbf{Theory 1}: $U(N_c)$ with $N_f$ flavors and superpotential $\mathcal{W}=0$.

\medskip
\noindent \textbf{Theory 2}: $U(N_f-N_c)$ with $N_f$ flavors, $N_f^2$ singlets (collected in a matrix $M_{ij}$), two extra singlets $S^{\pm}$ and superpotential $\hat{\mathcal{W}}=\sum_{i,j=1}^{N_f}M_{ij}\tilde{q}_iq_j+S^-\hat{\mathfrak{M}}^++S^+\hat{\mathfrak{M}}^-$.

\medskip
\noindent At the level of partition functions, the result of the real mass deformation is
\be
\mathcal{Z}_{\mathcal{T}_1}&=&\frac{1}{N_c!}\int\prod_{i=1}^{N_c}\udl{x_i}\e^{i\pi\xi\left(\sum_{i=1}^{N_c}x_i\right)}\frac{\prod_{i=1}^{N_c}\prod_{a=1}^{N_f}s_b\left(i\frac{Q}{2}\pm(x_i+M_a)-\mu_a\right)}{\prod_{i<j}^{N_c}s_b\left(i\frac{Q}{2}\pm(x_j-x_i)\right)}=\nn\\
&=&\e^{-i\pi\xi\sum_{a=1}^{N_f}M_a}s_b\left(i\frac{Q}{2}-\frac{iQ(N_f-N_c+1)-2\sum_{a=1}^{N_f}\mu_a\pm\xi}{2}\right)\times\nn\\
&\times&\prod_{a,b=1}^{N_f}s_b\left(i\frac{Q}{2}-(\mu_a+\mu_b-M_a+M_b)\right)\times\nn\\
&\times&\frac{1}{(N_f-N_c)!}\int\prod_{i=1}^{N_f-N_c}\udl{x_i}\e^{i\pi\xi\sum_{i=1}^{N_c}x_i}\frac{\prod_{i=1}^{N_f-N_c}\prod_{a=1}^{N_f}s_b\left(\pm(x_i-M_a)+\mu_a\right)}{\prod_{i<j}^{N_f-N_c}s_b\left(i\frac{Q}{2}\pm(x_j-x_i)\right)}=\mathcal{Z}_{\mathcal{T}_2}\, ,\nn\\
\label{aha}
\ee
where $\xi$ is the FI parameter for the restored topological symmetry, while $\sum_a\mu_a=\mu$ with $\mu$ being the holomorphic combination between the real mass for the axial symmetry and the mixing coefficient of this abelian symmetry with the R-symmetry.

\section{Partition function computations}

\subsection{Piecewise proof of the self-duality of $FM[SU(2)]$}
\label{piecwisefm2}

The equality of the partition functions \eqref{fmsunself} implied by the self-duality of $FM[SU(N)]$ can be proven analytically in the abelian case $N=2$ using a piecewise procedure similar to the one used to prove abelian Mirror Symmetry and the self-duality of $T[SU(2)]$ \cite{Kapustin:2010xq}. The difference is that, rather than applying sequentially the penthagon identity, we need to apply the \emph{ultimate penthagon identity} \cite{Benvenuti:2016wet, 2005CMaPh.258..257V}
\be 
\int\udl{s}\prod_{i=1}^3s_b\left(i\frac{Q}{2}+a_i+s\right)s_b\left(i\frac{Q}{2}+b_i-s\right)=\prod_{i,j=1}^3s_b\left(i\frac{Q}{2}+a_i+b_j\right)\, ,
\label{ultimatepentha}
\ee
where the parameters have to satisfy the following constraint
\be
\sum_{i=1}^3(a_i+b_i)=-iQ\, .
\label{ultimateconstraint}
\ee
This identity corresponds to the two-monopole duality \eqref{twomonopoles} in the particular case $N_c=1$ and $N_f=3$ and the constraint \eqref{ultimateconstraint} to the balancing condition \eqref{constraintCONF2}. It can also be considered as a hyperbolic uplift of the well-known star-triangle identity. For our derivation, it is useful to rewrite it as
\be
&&
\int \udl{s} D_{p_1}(s-z_1)D_{p_2}(s-z_2)D_{p_3}(s-z_3)=  \nn\\
&&
\qquad\qquad
=\prod_{i=1}^3 s_b( p_i-p_i')\,D_{p_3'}(z_1-z_2)D_{p_2'}(z_1-z_3)D_{p_1'}(z_2-z_3)
\label{ultimatepentha2}\, ,
\ee
where we defined
\be
D_{\alpha}(x)=s_b(i\tfrac{Q}{2}+\alpha+x)s_b(i\tfrac{Q}{2}+\alpha-x)
\ee
and
\be
p_i'=-i\frac{Q}{2}-p_i\, .
\ee
The map between the parameters in \eqref{ultimatepentha} and those in \eqref{ultimatepentha2} is
\be
z_{i}= \frac{b_i-a_i}{2}, \qquad p_i=\frac{a_i+b_i}{2}\, . 
\ee
The constraint on the parameters then reads
\be
\sum_{i} p_i=-iQ/2\quad\Leftrightarrow\quad\sum_i p_i'=-iQ\, .
\label{ultimatepenthaconstraint}
\ee

\begin{figure}[t]
	\centering
	\includegraphics[scale=0.42]{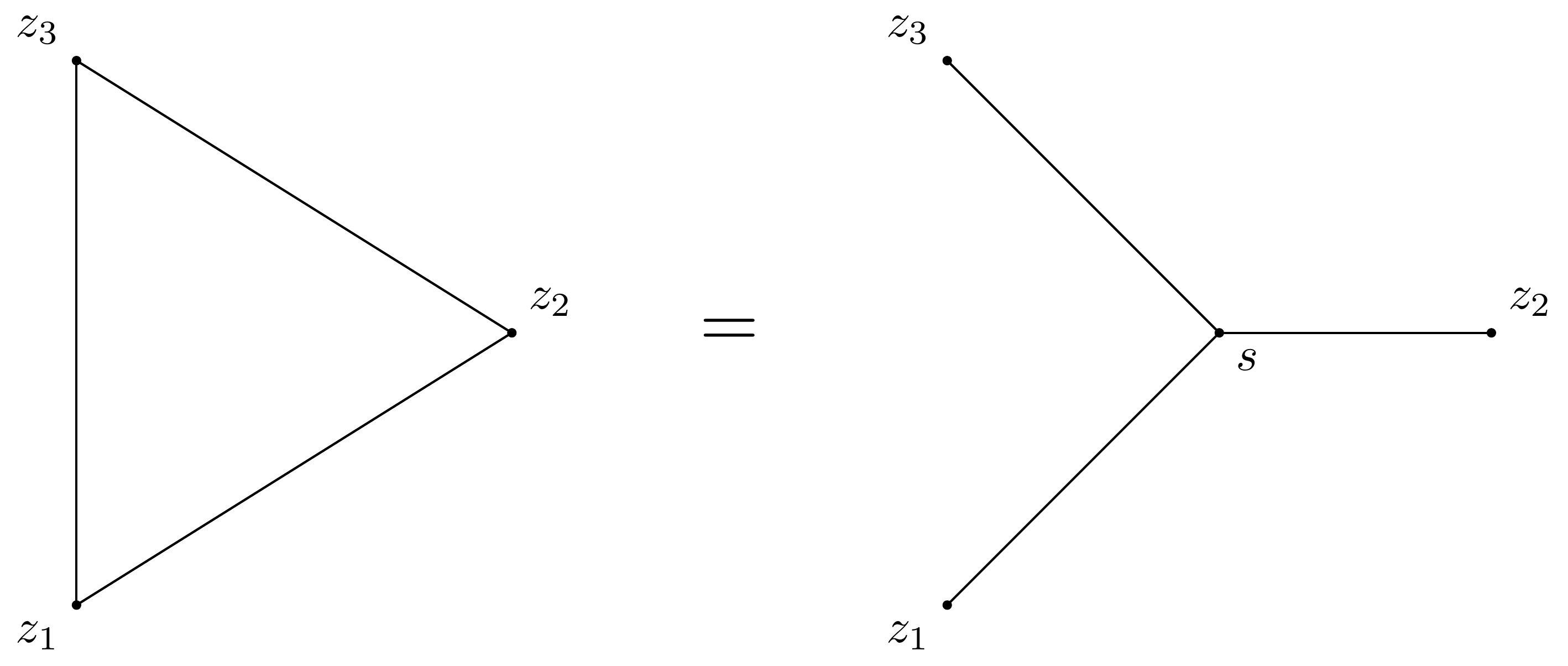}
	\caption{Diagrammatic representation of the ultimate penthagon identity. Internal points are the Coulomb branch coordinates over which we integrate in the partition function, while external points correspond to fugacities for the flavor symmetries. Each line represent a pair of a chiral and an anti-chiral charged under the symmetries corresponding to the points they link.}
	\label{ultimatepenthapic}
\end{figure}

In order to better understand the computations involving this identity, it is useful to visually represent it as in Figure \ref{ultimatepenthapic}. In the diagram, each point corresponds to a real mass parameter for a generic symmetry. In particular, internal points represent fugacities for the gauge symmetry over which we integrate, while external points correspond to the anti-diagonal combination of the two flavor symmetries. A line connecting two points represents a pair of chirals with opposite charges with respect to the corresponding symmetries. Gauge singlets that are not charged under the anti-diagonal combination of the flavor symmetries are not represented.

The starting point of the proof is the $FM[SU(2)]$ partition function
\be
&&\mathcal{Z}_2\equiv\mathcal{Z}_{FM[SU(2)]}(M_1,M_2,T_1,T_2,m_A,\Gd)= \prod_{a,b=1}^2s_b\left(i\frac{Q}{2}+(M_a-M_b)-2m_A\right)\times\nn\\
&&\times \prod_{a=1}^2s_b\left(i\frac{Q}{2}\pm(M_a-T_2)-\Gd\right)s_b\left(i\frac{Q}{2}-2m_A\right)\int\udl{x}s_b\left(iQ\pm(x-T_1)-\Gd-m_A\right)\times\nn\\
&&\times s_b\left(\pm(x-T_2)+\Gd-m_A\right)\prod_{a=1}^2s_b\left(\pm(x-M_a)+m_A\right)\, .\nn\\
\ee
We first consider the following block of double-sine functions
\be
\mathcal{B}=\sbfunc{i\frac{Q}{2}\pm(M_1-M_2)-2m_A}\prod_{a=1}^2\sbfunc{\pm(x-M_a)+m_A}
\ee
and rewrite it using \eqref{ultimatepentha2} from right to left, at the price of introducing an auxiliary integral. One can indeed check that the constraint \eqref{ultimatepenthaconstraint} is satisfied. Thus, we find
\be
&&\mathcal{Z}_2=\sbfunc{i\frac{Q}{2}-2m_A}\sbfunc{i\frac{Q}{2}-4m_A}\prod_{a=1}^2\sbfunc{i\frac{Q}{2}\pm(M_a-T_2)-\Gd}\times\nn\\
&&\qquad\times\int\udl{s}\prod_{a=1}^2\sbfunc{i\frac{Q}{2}\pm(s-M_a)-m_A}\int\udl{x}\sbfunc{iQ\pm(x-T_1)-\Gd-m_A}\times\nn\\
&&\qquad\times\sbfunc{\pm(x-T_2)+\Gd-m_A}\sbfunc{\pm(x-s)+2m_A}\, .
\ee
\begin{figure}[t]
	\centering
	\makebox[\linewidth][c]{
	\includegraphics[scale=0.35]{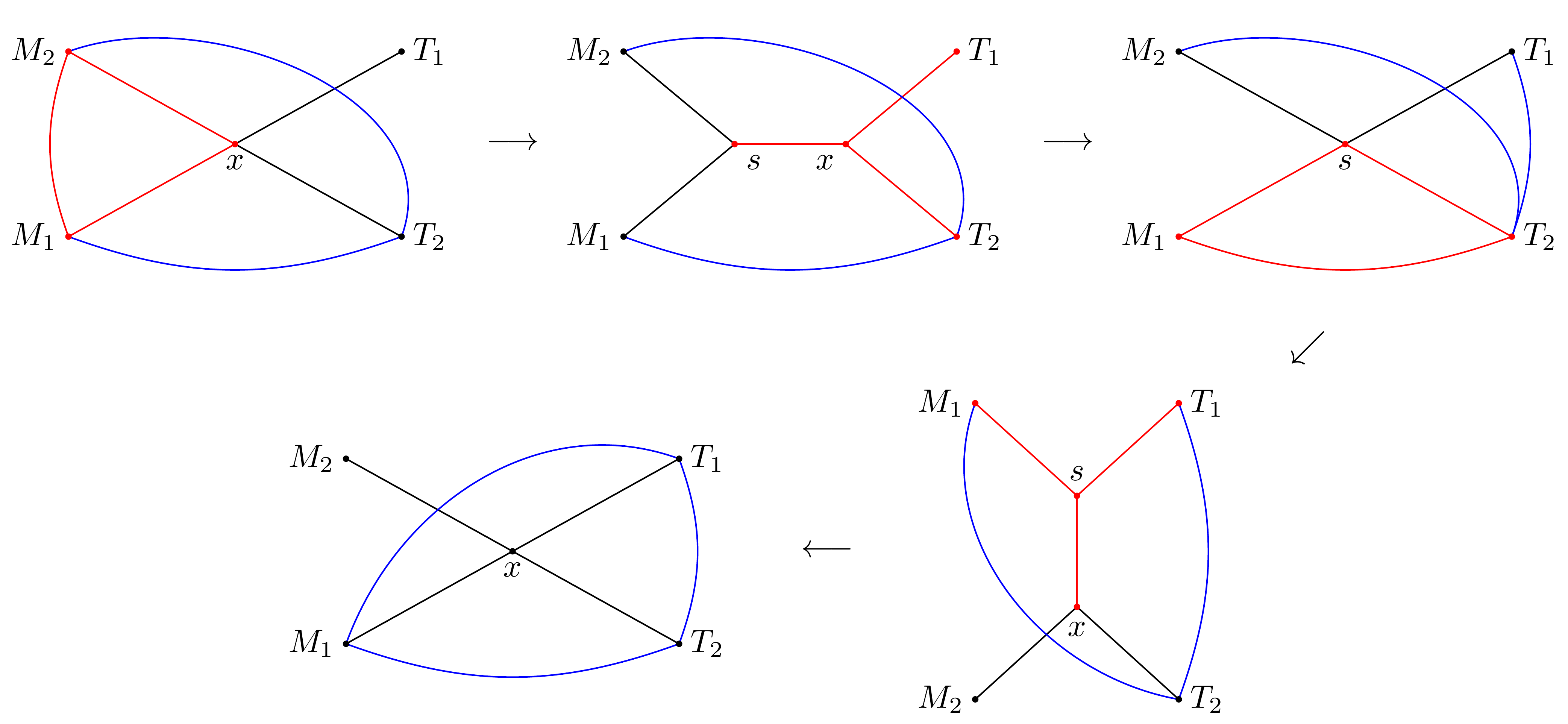}}
	\caption{Diagrammatic representations of the sequential application of the ultimate penthagon identity. We use black lines for chirals charged under the gauge symmetry, while blue lines for gauge singlets. At each step, we apply the ultimate penthagon identity on the block highlighted in red.}
	\label{startriangle}
\end{figure}

\noindent The original integral can now be evaluated using again \eqref{ultimatepentha2}. Hence, we have
\be
&&\mathcal{Z}_2=\sbfunc{i\frac{Q}{2}-2m_A}\sbfunc{-i\frac{Q}{2}+2\Gd-2m_A}\sbfunc{\frac{3}{2}iQ-2\Gd-2m_A}\times\nn\\
&&\qquad\qquad\times\sbfunc{i\frac{Q}{2}\pm(T_1-T_2)-2m_A}\prod_{a=1}^2\sbfunc{i\frac{Q}{2}\pm(M_a-T_2)-\Gd}\times\nn
\ee
\be
&&\qquad\qquad\times\int\udl{s}\sbfunc{i\frac{Q}{2}\pm(s-T_1)-\Gd+m_A}\sbfunc{-i\frac{Q}{2}\pm(s-T_2)+\Gd+m_A}\times\nn\\
&&\qquad\qquad\times \prod_{a=1}^2\sbfunc{i\frac{Q}{2}\pm(s-M_a)-m_A}\, .
\ee
At this point, we see that we have obtained the same structure of the original integral, but with the parameters re-shuffled. The manipulations we have performed so far can be represented diagrammatically as in the first line of Figure \ref{startriangle}. In order to get the desired result, we need to repeat the same moves but starting from a different block of double-sine functions, as depicted in the second line of Figure \ref{startriangle}:
\be
&&\mathcal{B}=\sbfunc{i\frac{Q}{2}\pm(M_1-T_2)-\Gd}\sbfunc{-i\frac{Q}{2}\pm(s-T_2)+\Gd+m_A}\times\qquad\qquad\qquad\qquad\nn\\
&&\qquad\qquad\qquad\qquad\qquad\qquad\qquad\qquad\qquad\qquad\times\sbfunc{i\frac{Q}{2}\pm(s-M_1)-m_A}\, .
\ee
As before, using \eqref{ultimatepentha2} from right to left we get to an intermediate step with two one-dimensional integrals
\be
&&\mathcal{Z}_2=\left[\sbfunc{i\frac{Q}{2}-2m_A}\right]^2\sbfunc{-i\frac{Q}{2}+2\Gd-2m_A}\sbfunc{i\frac{Q}{2}-2\Gd}\times\nn\\
&&\qquad\qquad\times\sbfunc{i\frac{Q}{2}\pm(T_1-T_2)-2m_A}\sbfunc{i\frac{Q}{2}\pm(M_2-T_2)-\Gd}\times\nn\\
&&\qquad\qquad\times\int\udl{x}\sbfunc{\pm(x-T_2)+m_A}\sbfunc{iQ\pm(x-M_1)-\Gd-m_A}\times\nn\\
&&\qquad\qquad\times\int\udl{s}\sbfunc{\pm(s-x)+\Gd}\sbfunc{i\frac{Q}{2}\pm(s-T_1)-\Gd+m_A}\times\nn\\
&&\qquad\qquad\times\sbfunc{i\frac{Q}{2}\pm(s-M_2)-m_A}\, .
\ee
Finally, if we remove the second integral using \eqref{ultimatepentha2} once again, we get the desired identity
\be
&&\mathcal{Z}_2= \prod_{a,b=1}^2s_b\left(i\frac{Q}{2}+(T_a-T_b)-2m_A\right)\prod_{a=1}^2s_b\left(i\frac{Q}{2}\mp(T_a-M_2)-\Gd\right)\sbfunc{i\frac{Q}{2}-2m_A}\times\nn\\
&&\qquad\qquad\times \int\udl{x}s_b\left(iQ\pm(x-M_1)-\Gd-m_A\right) s_b\left(\pm(x-M_2)+\Gd-m_A\right)\times\nn\\
&&\qquad\qquad\times \prod_{a=1}^2s_b\left(\mp(x-T_a)+m_A\right)=\mathcal{Z}_{MT[SU(2)]}(T_1,T_2,M_1,M_2,m_A,\Gd)\, .\nn\\
\ee

\subsection{Partition function for the recombination duality}
\label{partfuncrecomb}

The identity for the partition functions of the recombination duality \eqref{recombination} can be proven by applying iteratively Aharony duality \eqref{aha}, following the same procedure described in Sec.~\ref{piecewisefieldtheory} from the field theory point of view. It is useful to repeat it using partition functions to better understand the subtleties of the derivation. Let us consider also here the case $N=3$, where the partition function of Theory A is explicitly
\be
\mathcal{Z}&=&\mathcal{Z}_{\mathcal{T}_A}=\int\frac{\udl{x^{(3)}_1}\udl{x^{(3)}_2}\udl{x^{(3)}_3}}{3!}\e^{2\pi i\zeta\sum_ax^{(3)}_a}\frac{\prod_{a=1}^3\sbfunc{i\frac{Q}{2}\pm x^{(3)}_a-\mu}}{\prod_{a<b}^3\sbfunc{i\frac{Q}{2}\pm(x^{(3)}_a-x^{(3)}_b)}}\times\nn\\
&\times&\sbfunc{i\frac{Q}{2}\pm(x^{(3)}-z_3)-\Gd}\int\frac{\udl{x^{(2)}_1}\udl{x^{(2)}_2}}{2}\frac{\prod_{i,j=1}^2\sbfunc{i\frac{Q}{2}+(x^{(2)}_i-x^{(2)}_j)-2m_A}}{\sbfunc{i\frac{Q}{2}\pm(x^{(2)}_1-x^{(2)}_2)}}\times\nn\\
&\times&\prod_{i=1}^2\sbfunc{\pm(x^{(2)}_i-z_3)+\Gd-m_A}\sbfunc{iQ\pm(x^{(2)}_i-z_2)-\Gd-m_A}\times\nn\\
&\times&\prod_{a=1}^3\sbfunc{\pm(x^{(2)}_i-x^{(3)}_a)+m_A}\sbfunc{i\frac{Q}{2}-2m_A}\int\udl{x^{(1)}}\sbfunc{-i\frac{Q}{2}\pm(x^{(1)}-z_2)+\Gd}\times\nn\\
&\times&\sbfunc{\frac{3}{2}iQ\pm(x^{(1)}-z_1)-\Gd-2m_A}\prod_{i=1}^2\sbfunc{\pm(x^{(1)}-x^{(2)}_i)+m_A}\, .
\label{recombpartfunc}
\ee
We first want to apply Aharony duality to the $U(3)$ integral
\be
I_3&=&\int\frac{\udl{x^{(3)}_1}\udl{x^{(3)}_2}\udl{x^{(3)}_3}}{3!}\e^{2\pi i\zeta\sum_ax^{(3)}_a}\frac{\prod_{a=1}^3\sbfunc{i\frac{Q}{2}\pm x^{(3)}_a-\mu}}{\prod_{a<b}^3\sbfunc{i\frac{Q}{2}\pm(x^{(3)}_a-x^{(3)}_b)}}\times\nn\\
&&\qquad\qquad\times\sbfunc{i\frac{Q}{2}\pm(x^{(3)}-z_3)-\Gd}\prod_{i=1}^2\sbfunc{\pm(x^{(2)}_i-x^{(3)}_a)+m_A}\, .\nn\\
\label{recombpartfuncstep1}
\ee
Using \eqref{aha}, we can rewrite it as a one-dimensional integral
\be
I_3&=&\e^{2\pi i\zeta(\sum_ix^{(2)}_i+z_3)}\sbfunc{i\frac{Q}{2}\pm\zeta+\mu+\Gd-2m_A}\sbfunc{i\frac{Q}{2}-2\mu}\sbfunc{i\frac{Q}{2}-2\Gd}\times\nn\\
&\times&\sbfunc{i\frac{Q}{2}\pm z_3-\mu-\Gd}\prod_{i=1}^2\sbfunc{\pm(x^{(2)}_i-z_3)-\Gd+m_A}\sbfunc{\pm x^{(2)}_i-\mu+m_A}\times\nn\\
&\times&\prod_{i,j=1}^2\sbfunc{-i\frac{Q}{2}+(x^{(2)}_i-x^{(2)}_j)+2m_A}\int\udl{y^{(1)}}\e^{-2\pi i\zeta y^{(1)}}\sbfunc{\pm y^{(1)}+\mu}\times\nn\\
&\times&\sbfunc{\pm(y^{(1)}-z_3)+\Gd}\prod_{i=1}^2\sbfunc{i\frac{Q}{2}\pm(y^{(1)}-x^{(2)}_i)-m_A}\, .
\ee
Notice the contact term between the topological fugacity $\zeta$ and the real masses $x^{(2)}_i$ for the $U(2)$ gauge symmetry. When we plug this back into the partition function \eqref{recombpartfuncstep1}, this has the effect of introducing an FI contribution in the $U(2)$ integral that was not present before because of the monopole superpotential term $\mathfrak{M}^{(0,\pm1,0)}$ that breaks the topological symmetry at this node. This means that applying Aharony duality we restored the topological symmetry of the $U(2)$ node and, since the corresponding monopole operators are charged under this symmetry, they can't be in the superpotential anymore. Moreover, the FI parameters of the $\udl{x^{(2)}}$ and the $\udl{y^{(1)}}$ integral are opposite, which is compatible with the monopole superpotential term $\mathfrak{M}^{(0,\pm1,\pm1)}$ that breaks the two topological symmetries of the corresponding gauge nodes to the anti-diagonal combination $U(1)_\zeta$. If we also use the property of the double-sine functions
\be
\sbfunc{x}\sbfunc{-x}=1\, ,
\ee
which is the analogue from the point of view of partition functions of the fact that some fields have become massive and are integrated out, we see that plugging \eqref{recombpartfuncstep1} into \eqref{recombpartfunc} many of the contributions cancel and we get exactly \eqref{recombination} in the case $N=3$ and $k=1$
\be
\mathcal{Z}&=&\e^{2\pi i\zeta z_3}\sbfunc{i\frac{Q}{2}\pm z_3-\mu-\Gd}\sbfunc{i\frac{Q}{2}\pm\zeta+\mu+\Gd-2m_A}\sbfunc{i\frac{Q}{2}-2\mu}\times\nn\\
&\times&\sbfunc{i\frac{Q}{2}-2\Gd}\int\frac{\udl{x^{(2)}_1}\udl{x^{(2)}_2}}{2}\e^{2\pi i\zeta\sum_ix^{(2)}_i}\frac{\prod_{i=1}^2\sbfunc{\pm x^{(2)}_i-\mu+m_A}}{\sbfunc{i\frac{Q}{2}\pm(x^{(2)}_1-x^{(2)}_2)}}\times\nn\\
&\times&\sbfunc{iQ\pm(x^{(2)}_i-z_2)-\Gd-m_A}\sbfunc{i\frac{Q}{2}-2m_A}\times\nn\\
&\times&\int\udl{x^{(1)}}\sbfunc{-i\frac{Q}{2}\pm(x^{(1)}-z_2)+\Gd}\sbfunc{\frac{3}{2}iQ\pm(x^{(1)}-z_1)-\Gd-2m_A}\times\nn\\
&\times&\prod_{i=1}^2\sbfunc{\pm(x^{(1)}-x^{(2)}_i)+m_A}\int\udl{y^{(1)}}\e^{-2\pi i\zeta y^{(1)}}\prod_{i=1}^2\sbfunc{i\frac{Q}{2}\pm(y^{(1)}-x^{(2)}_i)}\times\nn\\
&\times&\sbfunc{\pm y^{(1)}+\mu}\sbfunc{\pm(y^{(1)}-z_3)+\Gd}\, .
\label{recombpartfunc2}
\ee
Indeed, we can recognize the prefactor $\Gl^3_1$ as well as the partition functions of the $G[U(2)]$ and $G[U(1)]$ glued together.

Since the contribution of the adjoint chiral canceled and since we have restored the FI contribution, we are allowed to apply \eqref{aha} on the $U(2)$ integral
\be
I_2&=&\int\frac{\udl{x^{(2)}_1}\udl{x^{(2)}_2}}{2}\e^{2\pi i\zeta\sum_ix^{(2)}_i}\frac{\prod_{i=1}^2\sbfunc{\pm x^{(2)}_i-\mu+m_A}}{\sbfunc{i\frac{Q}{2}\pm(x^{(2)}_1-x^{(2)}_2)}}\times\nn\\
&\times&\sbfunc{iQ\pm(x^{(2)}_i-z_2)-\Gd-m_A}\sbfunc{\pm(x^{(1)}-x^{(2)}_i)+m_A}\sbfunc{i\frac{Q}{2}\pm(y^{(1)}-x^{(2)}_i)}\, .\nn\\
\ee
Doing so, we don't replace it with a lower dimensional one as in the previous iteration, but with another two-dimensional integral. This is due to the fact that we reached the configuration with minimal rank and that $N$ is odd in this case
\be
I_2&=&\e^{2\pi i\zeta(z_2+x^{(1)}+y^{(1)})}\sbfunc{-i\frac{Q}{2}\pm\zeta+\mu+\Gd}\sbfunc{-i\frac{Q}{2}-2\mu+2m_A}\times\nn\\
&\times&\sbfunc{\frac{3}{2}iQ-2\Gd-2m_A}\sbfunc{i\frac{Q}{2}\pm z_2-\mu-\Gd}\sbfunc{-i\frac{Q}{2}\pm x^{(1)}-\mu+2m_A}\times\nn\\
&\times&\sbfunc{i\frac{Q}{2}\pm(x^{(1)}-z_2)-\Gd}\sbfunc{\pm y^{(1)}-\mu}\sbfunc{iQ\pm(y^{(1)}-z_2)-\Gd-2m_A}\times\nn\\
&\times&\int\frac{\udl{y^{(2)}_1}\udl{y^{(2)}_2}}{2}\e^{-2\pi i\zeta\sum_i y^{(2)}_i}\frac{\prod_{i=1}^2\sbfunc{i\frac{Q}{2}\pm y^{(2)}+\mu-m_A}\sbfunc{i\frac{Q}{2}\pm(y^{(2)}_i-x^{(1)})-m_A}}{\sbfunc{i\frac{Q}{2}\pm(y^{(2)}_1-y^{(2)}_2)}}\times\nn\\
&\times&\sbfunc{-i\frac{Q}{2}\pm(y^{(2)}_i-z_2)+\Gd+m_A}\sbfunc{\pm(y^{(2)}_i-y^{(1)}+m_A)}\, .
\label{recombpartfuncstep2}
\ee
The contact term has the effect of removing the FI contribution from the $y^{(1)}$ integral and of producing one in the $x^{(1)}$ integral. This means that we broke the topological symmetry on the right $U(1)$ node and turned on a monopole superpotential for it, while we did the opposite on the left $U(1)$ node. Moreover, the FI parameters of the $U(2)$ node and of the left $U(1)$ node are opposite, meaning that the monopole superpotential $\mathfrak{M}^{(\pm1,\pm1,0)}$ is turned on. Plugging \eqref{recombpartfuncstep2} into \eqref{recombpartfunc2} and simplifying the contributions of the massive fields, we get
\be
\mathcal{Z}_3&=&\Gl^3_2(m_A,\Gd,\zeta,\mu)\e^{2\pi i\zeta(z_2+z_3)}\prod_{n=2}^3\sbfunc{i\frac{Q}{2}\pm z_n-\mu-\Gd}\times\nn\\
&\times&\int\udl{x^{(1)}}\e^{2\pi i\zeta x^{(1)}}\sbfunc{-i\frac{Q}{2}\pm x^{(1)}-\mu+2m_A}\sbfunc{\frac{3}{2}iQ\pm(x^{(1)}-z_1)-\Gd-2m_A}\times\nn
\ee
\be
&\times&\int\frac{\udl{y^{(2)}_1}\udl{y^{(2)}_2}}{2}\e^{-2\pi i\zeta\sum_i y^{(2)}_i}\frac{\prod_{i=1}^2\sbfunc{i\frac{Q}{2}\pm y^{(2)}+\mu-m_A}}{\sbfunc{i\frac{Q}{2}\pm(y^{(2)}_1-y^{(2)}_2)}}\times\nn\\
&\times&\sbfunc{i\frac{Q}{2}\pm(y^{(2)}_i-x^{(1)})-m_A}\sbfunc{-i\frac{Q}{2}\pm(y^{(2)}_i-z_2)+\Gd+m_A}\times\nn\\
&\times&\sbfunc{i\frac{Q}{2}-2m_A}\int\udl{y^{(1)}}\sbfunc{iQ\pm(y^{(1)}-z_2)-\Gd-2m_A}\times\nn\\
&\times&\sbfunc{\pm(y^{(1)}-z_3)+\Gd}\prod_{i=1}^2\sbfunc{\pm(y^{(2)}_i-y^{(1)}+m_A)}\, ,
\label{recombpartfunc3}
\ee
where
\be
\Gl^3_2(m_A,\Gd,\zeta,\mu)&=&\sbfunc{i\frac{Q}{2}\pm\zeta+\mu+\Gd-2m_A}\sbfunc{-i\frac{Q}{2}\pm\zeta+\mu+\Gd}\times\nn\\
&\times&\sbfunc{i\frac{Q}{2}-2\mu}\sbfunc{-i\frac{Q}{2}-2\mu+2m_A}\times\nn\\
&\times&\sbfunc{i\frac{Q}{2}-2\Gd}\sbfunc{\frac{3}{2}iQ-2\Gd-2m_A}\, .
\ee
This coincides with \eqref{recombination} in the case $N=3$ and $k=2$.

Finally, we can apply \eqref{aha} on the $x^{(1)}$ integral
\be
I_1&=&\int\udl{x^{(1)}}\e^{2\pi i\zeta x^{(1)}}\sbfunc{-i\frac{Q}{2}\pm x^{(1)}-\mu+2m_A}\times\nn\\
&\times&\sbfunc{\frac{3}{2}iQ\pm(x^{(1)}-z_1)-\Gd-2m_A}\prod_{i=1}^2\sbfunc{i\frac{Q}{2}\pm(y^{(2)}_i-x^{(1)})-m_A}\, .
\ee
Since we passed the configuration of minimal rank, we get a three-dimensional integral
\be
I_1&=&\e^{2\pi i\zeta(z_3+\sum_iy^{(2)}_i)}\sbfunc{-\frac{3}{2}iQ\pm\zeta+\mu+\Gd+2m_A}\sbfunc{-\frac{3}{2}iQ-2\mu+4m_A}\times\nn\\
&\times&\sbfunc{\frac{5}{2}iQ-2\Gd-4m_A}\sbfunc{i\frac{Q}{2}\pm z_1-\mu-\Gd}\prod_{i,j=1}^2\sbfunc{i\frac{Q}{2}+(y^{(2)}_i-y^{(2)}_j)-2m_A}\times\nn\\
&\times&\prod_{i=1}^2\sbfunc{-i\frac{Q}{2}\pm y^{(2)}_i-\mu+m_A}\sbfunc{\frac{3}{2}iQ\pm(y^{(2)}_i-z_1)-\Gd-3m_A}\times\nn\\
&\times&\int\frac{\udl{y^{(3)}_1}\udl{y^{(3)}_2}\udl{y^{(3)}_3}}{3!}\e^{-2\pi i\zeta\sum_ay^{(3)}_a}\frac{\prod_{a=1}^3\sbfunc{iQ\pm y^{(3)}_a+\mu-2m_A}}{\prod_{a<b}^3\sbfunc{i\frac{Q}{2}\pm(y^{(3)}_a-y^{(3)}_b)}}\times\nn\\
&\times&\prod_{i=1}^2\sbfunc{\pm(y^{(3)}_a-y^{(2)}_i)+m_A}\sbfunc{-iQ\pm(y^{(3)}_a-z_1)+\Gd+2m_A}\, .
\ee
If we substitute this into \eqref{recombpartfunc3}, we finally arrive at
\be
\mathcal{Z}_3&=&\Gl^3_3(m_A,\Gd,\zeta,\mu)\prod_{n=1}^3\e^{2\pi i\zeta z_n}\sbfunc{i\frac{Q}{2}\pm z_n-\mu-\Gd}\times\nn\\
&\times&\int\frac{\udl{y^{(3)}_1}\udl{y^{(3)}_2}\udl{y^{(3)}_3}}{3!}\e^{-2\pi i\zeta\sum_ay^{(3)}_a}\frac{\prod_{a=1}^3\sbfunc{iQ\pm y^{(3)}_a+\mu-2m_A}}{\prod_{a<b}^3\sbfunc{i\frac{Q}{2}\pm(y^{(3)}_a-y^{(3)}_b)}}\times\nn\\
&\times&\sbfunc{-iQ\pm(y^{(3)}_a-z_1)+\Gd+2m_A}\times\nn\\
&\times&\int\frac{\udl{y^{(2)}_1}\udl{y^{(2)}_2}}{2}\frac{\prod_{i,j=1}^2\sbfunc{i\frac{Q}{2}+(y^{(2)}_i-y^{(2)}_j)-2m_A}}{\sbfunc{i\frac{Q}{2}\pm(y^{(2)}_1-y^{(2)}_2)}}\times\nn\\
&\times&\prod_{i=1}^2\sbfunc{\frac{3}{2}iQ\pm(y^{(2)}_i-z_1)-\Gd-3m_A}\sbfunc{-i\frac{Q}{2}\pm(y^{(2)}_i-z_2)+\Gd+m_A}\times\nn\\
&\times&\prod_{a=1}^3\sbfunc{\pm(y^{(3)}_a-y^{(2)}_i)+m_A}\times\nn\\
&\times&\sbfunc{i\frac{Q}{2}-2m_A}\int\udl{y^{(1)}}\sbfunc{iQ\pm(y^{(1)}-z_2)-\Gd-2m_A}\times\nn\\
&\times&\sbfunc{\pm(y^{(1)}-z_3)+\Gd}\prod_{i=1}^2\sbfunc{\pm(y^{(1)}-y^{(2)}_i)+m_A}\, ,
\ee
where
\be
\Gl^3_3(m_A,\Gd,\zeta,\mu)&=&\prod_{n=1}^3\sbfunc{\pm\zeta+\mu+\Gd-m_A+(3-2n)\left(i\frac{Q}{2}-m_A\right)}\times\nn\\
&\times&\sbfunc{i\frac{Q}{2}-2\mu-2(n-1)\left(i\frac{Q}{2}-m_A\right)}\times\nn\\
&\times&\sbfunc{i\frac{Q}{2}-2\Gd+2(n-1)\left(i\frac{Q}{2}-m_A\right)}\, ,
\ee
which corresponds to \eqref{recombination} in the case $N=3$ and $k=3$.

\subsection{A useful integral identity}

In Sec.~\ref{rankstabderivation}, in order to write the partition function of the theory with $k=2$ in a stable form, we used the following integral identity:\footnote{It would be interesting to interpret this identity as well as similar ones, whose $2d$ version appears in the CFT literature, as dualities for theories with monopole superpotential and both adjoint and fundamental matter. We leave this for future investigations.}
\be
Z&=&\int\frac{\udl{x_1}\udl{x_2}}{2}\e^{-4\pi i\tau(x_1+x_2)}\frac{\prod_{\ga,\gb=1}^2\sbfunc{-i\frac{Q}{2}+(x_\ga-x_\gb)+2\tau}}{\sbfunc{i\frac{Q}{2}\pm(x_1-x_2)}}\times\nn\\
&&\quad\qquad\times\prod_{\ga,\gb=1}^2\sbfunc{i\frac{Q}{2}\pm(x_\ga-y_\gb)-\tau}\sbfunc{i\frac{Q}{2}\pm(x_\ga-z_\gb)-\tau}=\nn\\
&=&\e^{-2\pi i\tau(z_1+z_2+y_1+y_2)}\sbfunc{-i\frac{Q}{2}+4\tau}\sbfunc{-\frac{3}{2}iQ+6\tau}\sbfunc{i\frac{Q}{2}-2\tau}\times\nn\\
&&\quad\qquad\times\prod_{\ga,\gb=1}^2\sbfunc{i\frac{Q}{2}\pm(y_\ga-y_\gb)-2\tau}\prod_{\ga=1}^2\sbfunc{i\frac{Q}{2}\pm(y_\ga-z_1)-2\tau}\times\nn\\
&&\quad\qquad\times\int\udl{x}\sbfunc{\pm(x+z_1)+\tau}\sbfunc{iQ\pm(x+z_2)-3\tau}\prod_{\ga=1}^2\sbfunc{\pm(x+y_\ga)+\tau}\, .\nn\\
\label{intermezzo}
\ee
This identity can be proven with a piecewise procedure similar to the one used in Appendix \ref{piecwisefm2} to prove the self-duality of $FM[SU(2)]$. More precisely, we apply the ultimate penthagon identity \eqref{ultimatepentha2} from right to left to the following block of double-sine functions
\be
\mathcal{B}=\sbfunc{-i\frac{Q}{2}\pm(x_1-x_2)+2\tau}\sbfunc{i\frac{Q}{2}\pm(x_1-z_2)-\tau}\sbfunc{i\frac{Q}{2}\pm(x_2-z_2)-\tau}\, .\nn\\
\ee
One can indeed verify that the constraint \eqref{ultimatepenthaconstraint} is satisfied for this choice. In this way, the contribution of the adjoint chiral $\Gp$ disappears, but at the price of introducing an additional $U(1)$ integral
\be
Z&=&\sbfunc{-\frac{3}{2}iQ+4\tau}\int\udl{s}\sbfunc{iQ\pm(s-z_2)-2\tau}\int\frac{\udl{x_1}\udl{x_2}}{2}\e^{-4\pi i\tau(x_1+x_2)}\times\nn\\
&\times&\frac{\prod_{\ga=1}^2\sbfunc{i\frac{Q}{2}\pm(x_\ga-z_1)-\tau}\sbfunc{\pm(x_\ga-s)+\tau}\prod_{\gb=1}^2\sbfunc{i\frac{Q}{2}\pm(x_\ga-y_\gb)-\tau}}{\sbfunc{i\frac{Q}{2}\pm(x_1-x_2)}}\, .\nn\\
\ee
Now we can replace the original integral with a lower dimensional one applying the one-monopole duality \eqref{onemonopole}. This gives
\be
Z&=&\e^{-2\pi i\tau(z_1+y_1+y_2)}\sbfunc{-\frac{3}{2}iQ+4\tau}\sbfunc{-i\frac{Q}{2}+4\tau}\times\nn\\
&\times&\prod_{\ga,\gb=1}^2\sbfunc{i\frac{Q}{2}+(y_\ga-y_\gb)-2\tau}\prod_{\ga=1}^2\sbfunc{i\frac{Q}{2}\pm(y_\ga-z_1)-2\tau}\times\nn\\
&\times&\int\udl{x}\e^{i\pi(iQ-4\tau)x}\sbfunc{\pm(x+z_1)+\tau}\prod_{\ga=1}^2\sbfunc{\pm(x+y_\ga)+\tau}\times\nn\\
&\times&\int\udl{s}\e^{i\pi(iQ-6\tau)s}\sbfunc{iQ\pm(s-z_2)-2\tau}\sbfunc{i\frac{Q}{2}\pm(x+s)-\tau}\, .
\ee
Finally, we can use again the one-monopole duality \eqref{onemonopole} to get rid of the auxiliary $\udl{s}$ integral since in this case it becomes an evaluation formula and obtain the desired result.

\section{Superconformal index computations}

We present here the results of the computations of the superconformal index we performed to test some of the dualities presented in the main text. We refer the reader to \cite{US1} for the conventions we use, which are mainly based on \cite{Imamura:2011su,Kapustin:2011jm}.

\subsection{Self-duality of $FM[SU(N)]$}
\label{scifmsun}

As an additional test of the self-duality of $FM[SU(N)]$, we compute the index perturbatively in the R-symmetry fugacity and check that all the coefficients of the power series in $x$ are symmetric under the exchange of the fugacities $m_a$ and $t_a$ for the flavor symmetries $SU(N)_M\times SU(N)_T$. We also turn on fugacities $s,p$ for the axial symmetries $U(1)_{m_A}\times U(1)_\Gd$ and we denote with $R_A$ and $R_\Gd$ respectively the mixing parameters of these axial symmetries with the R-symmetry. The test has been performed for the cases $N=2$ and $N=3$.

\subsubsection{$FM[SU(2)]$}

The superconformal index of $FM[SU(2)]$ is
\be
&&\mathcal{I}_{FM[SU(2)]}=\prod_{a,b=1}^2\frac{\xfac{\frac{m_a}{m_b} s^{-2}x^{2(1-R_A)}}}{\xfac{\frac{m_b}{m_a} s^{2}x^{2R_A}}}\prod_{a=1}^2\frac{\xfac{m_a^{\pm1} t_2^{\mp1} p^{-1} x^{2-R_\Gd}}}{\xfac{m_a^{\mp1} t_2^{\pm1} p\, x^{R_\Gd}}}\frac{\xfac{s^{-2}x^{2(1-R_A)}}}{\xfac{s^2x^{2R_A}}}\times\nn\\
&&\quad\times\sum_{m\in\mathbb{Z}}\oint\frac{\udl{u}}{2\pi iu}x^{2|m|}\frac{\xfac{u^{\pm1}t_1^{\mp1}s^{-1}p^{-1}x^{3-R_A-R_\Gd+|m|}}}{\xfac{u^{\mp1}t_1^{\pm1}s\,p\,x^{-1+R_A+R_\Gd+|m|}}}\frac{\xfac{u^{\pm1}t_2^{\mp1}s^{-1}p\,x^{1-R_A+R_\Gd+|m|}}}{\xfac{u^{\mp1}t_2^{\pm1}s\,p^{-1}x^{1+R_A-R_\Gd+|m|}}}\times\nn\\
&&\quad\times\prod_{a=1}^2\frac{\xfac{u^{\pm1}m_a^{\mp1}s\,x^{1+R_A+|m|}}}{\xfac{u^{\mp1}m_a^{\pm1}s^{-1}x^{1-R_A+|m|}}}\, .
\ee
From this expression we immediately see that all the monopole operators are uncharged under the global symmetries and have R-charge
\be
\epsilon=2|m|\, ,
\ee
which is compatible with the monopole superpotential. In order to compute the index as a power series in $x$, we have to fix the parameters $R_A$ and $R_\Gd$ such that all the chiral fields have R-charge between 0 and 2:
\be
0<R_A<1,\qquad 0<R_\Gd<2,\qquad 1-R_A<R_\Gd<1+R_A\, .
\ee
We chose the values $R_A=\frac{1}{2}$ and $R_\Gd=\frac{3}{4}$ and verified the invariance of the index under $m_a\leftrightarrow t_a$ up to order $\mathcal{O}(x^6)$. The first few terms of the expansion are
\be
\mathcal{I}_{FM[SU(2)]}&=&1+p^2 s^2x^{1/2}+\left(p^4 s^4+\frac{s^2 m_1}{m_2}+\frac{s^2 m_2}{m_1}+\frac{s^2 t_1}{t_2}+\frac{s^2 t_2}{t_1}+3 s^2-\frac{1}{s^2}\right)x+\nn\\
&+&\left(p^6 s^6+\frac{p^2 s^4 m_1}{m_2}+\frac{p^2 s^4 m_2}{m_1}+\frac{p^2 s^4 t_1}{t_2}+\frac{p^2 s^4 t_2}{t_1}+3 p^2 s^4+\frac{s^2}{p^2}+\frac{p^2 t_2^2}{m_2^2}+\right.\nn\\
&+&\frac{p^2 t_2^2}{m_2 m_1}+\frac{p^2 t_2^2}{m_1^2}+\frac{p^2 t_1^2}{m_2^2}+\frac{p^2 t_1^2}{m_2 m_1}+\frac{p^2 t_1^2}{m_1^2}+\frac{p^2 t_2 t_1}{m_2^2}+\frac{2 p^2 t_2 t_1}{m_2 m_1}+\frac{p^2 t_2 t_1}{m_1^2}+\nn\\
&+&\frac{p^2 m_1 t_1}{m_2 t_2}+\frac{p^2 m_2 t_1}{m_1 t_2}+\frac{p^2 m_2^2}{t_2^2}+\frac{p^2 m_1^2}{t_2^2}+\frac{p^2 m_2 m_1}{t_2^2}+\frac{p^2 m_1 t_2}{m_2 t_1}+\frac{p^2 m_2 t_2}{m_1 t_1}+\nn\\
&+&\frac{p^2 m_2^2}{t_2 t_1}+\frac{p^2 m_1^2}{t_2 t_1}+\frac{2 p^2 m_2 m_1}{t_2 t_1}+\frac{p^2 m_2^2}{t_1^2}+\frac{p^2 m_1^2}{t_1^2}+\frac{p^2 m_2 m_1}{t_1^2}+\frac{p^2 m_1}{m_2}+\nn\\
&+&\left.\frac{p^2 m_2}{m_1}+\frac{p^2 t_1}{t_2}+\frac{p^2 t_2}{t_1}+p^2\right)x^{3/2}+\mathcal{O}(x^2)\, .
\ee

\subsubsection{$FM[SU(3)]$}

The superconformal index of $FM[SU(3)]$ is
\be
&&\mathcal{I}_{FM[SU(3)]}=\prod_{a,b=1}^3\frac{\xfac{\frac{m_a}{m_b} s^{-2}x^{2(1-R_A)}}}{\xfac{\frac{m_b}{m_a} s^{2}x^{2R_A}}}\prod_{a=1}^3\frac{\xfac{m_a^{\pm1} t_2^{\mp1} p^{-1} x^{2-R_\Gd}}}{\xfac{m_a^{\mp1} t_2^{\pm1} p\, x^{R_\Gd}}}\times\nn\\
&&\quad\times\frac{1}{2}\sum_{\vec{m}^{(2)}\in\mathbb{Z}^2}\oint\prod_{i=1}^2\frac{\udl{u_i}}{2\pi iu_i}x^{-2R_A\left|m^{(2)}_1-m^{(2)}_2\right|+(R_A+2)\sum_{i=1}^2\left|m^{(2)}_i\right|}\left(1-\left(\frac{u^{(2)}_1}{u^{(2)}_2}\right)^{\pm1}x^{\left|m^{(2)}_1-m^{(2)}_2\right|}\right)\times\nn\\
&&\quad\times\prod_{i,j=1}^2\frac{\xfac{\frac{u^{(2)}_i}{u^{(2)}_j} s^{-2}x^{2(1-R_A)+\left|m^{(2)}_i-m^{(2)}_j\right|}}}{\xfac{\frac{u^{(2)}_j}{u^{(2)}_i} s^{2}x^{2R_A+\left|m^{(2)}_i-m^{(2)}_j\right|}}}\prod_{i=1}^2\frac{\xfac{u^{(2)\pm1}_it_1^{\mp1}s^{-1}p^{-1}x^{3-R_A-R_\Gd+|m^{(2)}_i|}}}{\xfac{u^{(2)\mp1}_it_1^{\pm1}s\,p\,x^{-1+R_A+R_\Gd+|m^{(2)}_i|}}}\times\nn\\
&&\quad\times\frac{\xfac{u^{(2)\pm1}_it_2^{\mp1}s^{-1}p\,x^{1-R_A+R_\Gd+|m^{(2)}_i|}}}{\xfac{u^{(2)\mp1}_it_2^{\pm1}s\,p^{-1}x^{1+R_A-R_\Gd+|m^{(2)}_i|}}}\prod_{a=1}^3\frac{\xfac{u^{(2)\pm1}_im_a^{\mp1}s\,x^{1+R_A+|m^{(2)}_i|}}}{\xfac{u^{(2)\mp1}_im_a^{\pm1}s^{-1}x^{1-R_A+|m^{(2)}_i|}}}\times\nn
\ee
\be
&&\quad\times\frac{\xfac{s^{-2}x^{2(1-R_A)}}}{\xfac{s^2x^{2R_A}}}\sum_{m^{(1)}\in\mathbb{Z}}\oint\frac{\udl{u^{(1)}}}{2\pi iu^{(1)}}x^{2(1-R_A)|m^{(1)}|+R_A\sum_{i=1}^2\left|m^{(1)}-m^{(2)}_i\right|}\times\nn\\
&&\quad\times\frac{\xfac{u^{(1)\pm1}t_2^{\mp1}p\,x^{R_\Gd+\left|m^{(1)}\right|}}}{\xfac{u^{(1)\mp1}t_2^{\pm1}p^{-1}x^{2-R_\Gd+\left|m^{(1)}\right|}}}\frac{\xfac{u^{(1)\pm1}t_1^{\mp1}s^{-2}p^{-1}x^{4-2R_A-R_\Gd+\left|m^{(1)}\right|}}}{\xfac{u^{(1)\mp1}t_1^{\pm1}s^2p\,x^{-2+2R_A+R_\Gd+\left|m^{(1)}\right|}}}\times\nn\\
&&\quad\times\prod_{i=1}^2\frac{\xfac{u^{(1)^\pm1}u^{(2)\mp1}_is\,x^{1+R_A+\left|m^{(1)}-m^{(2)}_i\right|}}}{\xfac{u^{(1)^\mp1}u^{(2)\pm1}_is^{-1}x^{1-R_A+\left|m^{(1)}-m^{(2)}_i\right|}}}\, .
\ee
From the overall factor of $x$ we can extract the R-charge of the monopoles
\be
\epsilon(\vec{m})=-2R_A\left|m^{(2)}_1-m^{(2)}_2\right|+(R_A+2)\sum_{i=1}^2\left|m^{(2)}_i\right|+2(1-R_A)\left|m^{(1)}\right|+R_A\sum_{i=1}^2\left|m^{(1)}-m^{(2)}_i\right|\, ,\nn\\
\ee
from which we see that all the basic monopoles have R-charge 2, as expected because of the monopole superpotential. In order for the index to have a well-defined expansion in $x$, we have to choose $R_A$ and $R_\Gd$ such that
\be
0<R_A<1,\qquad 0<R_\Gd<2,\qquad 2(1-R_A)<R_\Gd<R_A+1\, .
\ee
We computed the superconformal index choosing $R_A=\frac{1}{2}$ and $R_\Gd=\frac{4}{3}$ up to order $\mathcal{O}(x^3)$ and found perfect agreement with the self-duality of $FM[SU(3)]$. The first few terms of the expansion are
\be
\mathcal{I}_{FM[SU(3)]}&=&1+\frac{s^2}{p^2}x^{1/3}+\left(\frac{s^4}{p^4}+p^2 s^4\right)x^{2/3}+\left(\frac{s^6}{p^6}+s^6+\frac{s^2 m_2}{m_3}+\frac{s^2 m_1}{m_3}+\frac{s^2 m_1}{m_2}\right.+\nn\\
&+&\left.\frac{s^2 m_3}{m_2}+\frac{s^2 m_3}{m_1}+\frac{s^2 m_2}{m_1}+\frac{s^2 t_2}{t_3}+\frac{s^2 t_1}{t_3}+\frac{s^2 t_1}{t_2}+\frac{s^2 t_3}{t_2}+\frac{s^2 t_3}{t_1}+\frac{s^2 t_2}{t_1}\right.+\nn\\
&+&\left.5 s^2-\frac{1}{s^2}\right)x+\left(\frac{s^8}{p^8}+p^4 s^8+\frac{s^8}{p^2}+\frac{s^4 m_2}{p^2 m_3}+\frac{s^4 m_1}{p^2 m_3}+\frac{s^4 m_1}{p^2 m_2}+\frac{s^4 m_3}{p^2 m_2}+\right.\nn\\
&+&\frac{s^4 m_3}{p^2 m_1}+\frac{s^4 m_2}{p^2 m_1}+\frac{s^4 t_2}{p^2 t_3}+\frac{s^4 t_1}{p^2 t_3}+\frac{s^4 t_1}{p^2 t_2}+\frac{s^4 t_3}{p^2 t_2}+\frac{s^4 t_3}{p^2 t_1}+\frac{s^4 t_2}{p^2 t_1}+\frac{6 s^4}{p^2}+\frac{p t_3}{m_3}+\nn\\
&+&\frac{p t_3}{m_2}+\frac{p t_3}{m_1}+\frac{p t_2}{m_3}+\frac{p t_2}{m_2}+\frac{p t_2}{m_1}+\frac{p t_1}{m_3}+\frac{p t_1}{m_2}+\frac{p t_1}{m_1}+\frac{p m_3}{t_3}+\frac{p m_2}{t_3}+\frac{p m_1}{t_3}+\nn\\
&+&\left.\frac{p m_3}{t_2}+\frac{p m_2}{t_2}+\frac{p m_1}{t_2}+\frac{p m_3}{t_1}+\frac{p m_2}{t_1}+\frac{p m_1}{t_1}\right)x^{4/3}+\mathcal{O}(x^{5/3})\, .
\ee

\subsection{Rank stabilization duality}
\label{scirankstab}

We can also use the superconformal index to test the rank stabilization duality for different values of $N$ and $k$. This provides an additional consistency check for the cases where we have a derivation at the level of partition functions, but also a strong test of the duality for those cases where this was not possible. For this purpose, we turn on fugacities $z_a$ for the $U(k)_z$ flavor symmetry, $\omega$ for the topological symmetry $U(1)_\zeta$ and $s$, $p$ for the $U(1)_\tau\times U(1)_\mu$ axial symmetry. The mixing of these symmetries with the R-symmetry is parametrized by $1-R$ and $r$ respectively. We tested the duality for $k=1,2,3$ and for small values of $N$, depending on $k$.

\subsubsection{Two flavors}

The superconformal index of Theory A in the case $k=1$ takes the form
\be
&&\mathcal{I}_{\mathcal{T}_A}=\prod_{j=1}^{N-1}\frac{\xfac{s^{2j}x^{2j(1-R)}}}{\xfac{s^{-2j}x^{2-2j(1-R)}}}\sum_{\vec{m}\in\mathbb{Z}^N}\frac{\prod_{i=1}^N\omega^{m_i}}{N!}\oint\prod_{i=1}^N\frac{\udl{u_i}}{2\pi i\,u_i}s^{-2\sum_{i<j}^N|m_i-m_j|+\sum_{i=1}^N|m_i|}\times\nn\\
&&\qquad\times p^{-\sum_{i=1}^N|m_i|} x^{2(R-1)\sum_{i<j}^N|m_i-m_j|-(R+r-2)\sum_{i=1}^N|m_i|}\prod_{i<j}^N\left(1-\left(1-\frac{u_i}{u_j}\right)^{\pm1}x^{|m_i-m_j|}\right)\times\nn\\
&&\qquad\times\prod_{i,j=1}^N\frac{\xfac{\frac{u_i}{u_j}s^{-2}x^{2R+|m_i-m_j|}}}{\xfac{\frac{u_j}{u_i}s^2x^{2(1-R)+|m_i-m_j|}}}\prod_{i=1}^N\frac{u_i^{\pm1}p^{-1}x^{2-r+|m_i|}}{u_i^{\mp1}px^{r+|m_i|}}\frac{\xfac{u_i^{\pm1}z^{\pm1}s\,x^{2-R+|m_i|}}}{\xfac{u^{\mp1}z^{\mp}s^{-1}x^{R+|m_i|}}}\, ,\nn\\
\ee
while the index of Theory B is
\be
&&\mathcal{I}_{\mathcal{T}_B}=\frac{\xfac{s^{-2N}x^{2-2N(1-R)}}}{\xfac{s^{2N}x^{2N(1-R)}}}\prod_{j=1}^{N-1}\frac{\xfac{s^{-2(j-1)}p^{-2}x^{2-2(j-1)(1-R)-2r}}}{\xfac{s^{2(j-1)}p^2x^{2(j-1)(1-R)+2r}}}\times\nn\\
&&\quad\times\frac{\xfac{s^{2N-2j-1)}p\,\omega^{-1}x^{1+(2N-2j-1)(1-R)+r}}}{\xfac{s^{1-2N+2j}p^{-1}\omega\,x^{1-(2N-2j-1)(1-R)-r}}}\prod_{j=2}^N\frac{\xfac{s^{2j-3}p\,\omega\,x^{1+(2j-3)(1-R)+r}}}{\xfac{s^{3-2j}p^{-1}\omega^{-1}x^{1-(2j-3)(1-R-r)}}}\times\nn\\
&&\quad\times\sum_{m\in\mathbb{Z}}\omega^m\oint\frac{\udl{u}}{2\pi i\, u}s^{|m|}p^{-|m|}x^{(2-R-r)|m|}\frac{\xfac{u^{\pm1}p^{-1}s^{1-N}x^{2-(N-1)(1-R)-r+|m|}}}{\xfac{u^{\mp1}p\,s^{N-1}x^{(N-1)(1-R)+r+|m|}}}\times\nn\\
&&\quad\times\frac{\xfac{u^{\pm1}x^{\pm1}s^Nx^{1+N(1-R)+|m|}}}{\xfac{u^{\mp1}z^{\mp1}s^{-N}x^{1-N(1-R)+|m|}}}\, .
\ee
In order for the two indices to both have a well-defined expansion in $x$, we need to choose the R-symmetry parameters such that
\be
\frac{2N-4}{2N-3}<R<1,\qquad 0<r<(2N-3)R+4-2N\, .
\ee
We verified the matching of the superconformal indices for $N=2,3,4$. In Table \ref{sci2flav} we summarize the results of our computations.

\begin{table}[t]
\centering

\scalebox{0.93}{
\setlength{\extrarowheight}{2pt}
\makebox[\linewidth][c]{
\begin{tabular}{c|c|c|c|c}
$N$ & $R$ & $r$ & $h$ & $\mathcal{I}$ \\ \hline
2 & $\frac{2}{3}$ & $\frac{1}{3}$ & 8 & $1+\frac{\omega}{p s}x^{1/3}+\frac{1}{p s \omega}x^{1/3}+\frac{\omega^2 }{p^2 s^2}x^{2/3}+\frac{1}{p^2 s^2 \omega^2}x^{2/3}+\frac{1}{p^2 s^2}x^{2/3}+p^2 x^{2/3}+\cdots$ \\
3 & $\frac{4}{5}$ & $\frac{1}{5}$ & 5 & $1+\frac{\omega}{p s^3}x^{1/5}+\frac{1}{p s^3 \omega}x^{1/5}+\frac{\omega^2}{p^2 s^6}x^{2/5}+\frac{1}{p^2 s^6 \omega^2}x^{2/5}+\frac{1}{p^2 s^6}x^{2/5}+p^2 x^{2/5}+\cdots$ \\
4 & $\frac{6}{7}$ & $\frac{1}{7}$ & 2 & $1+\frac{\omega}{p s^5}x^{1/7}+\frac{1}{p s^5 \omega}x^{1/7}+\frac{\omega^2}{p^2 s^{10}}x^{2/7}+\frac{1}{p^2 s^{10} \omega^2}x^{2/7}+\frac{1}{p^2 s^{10}}x^{2/7}+p^2 x^{2/7}+\cdots$
\end{tabular}}}
\caption{Computation of the superconformal index for different values of $N$ up to order $\mathcal{O}\left(x^h\right)$. In the last column we report the first terms of the expansion.}
\label{sci2flav}
\end{table}

\subsubsection{Three flavors}

The superconformal index of Theory A in the case $k=2$ takes the form
\be
&&\mathcal{I}_{\mathcal{T}_A}=\prod_{j=1}^{N-2}\frac{\xfac{s^{2j}x^{2j(1-R)}}}{\xfac{s^{-2j}x^{2-2j(1-R)}}}\sum_{\vec{m}\in\mathbb{Z}^N}\frac{\prod_{i=1}^N\omega^{m_i}}{N!}\oint\prod_{i=1}^N\frac{\udl{u_i}}{2\pi i\,u_i}s^{-2\sum_{i<j}^N|m_i-m_j|+2\sum_{i=1}^N|m_i|}\times\nn\\
&&\qquad\times p^{-\sum_{i=1}^N|m_i|} x^{2(R-1)\sum_{i<j}^N|m_i-m_j|-(2R+r-3)\sum_{i=1}^N|m_i|}\prod_{i<j}^N\left(1-\left(1-\frac{u_i}{u_j}\right)^{\pm1}x^{|m_i-m_j|}\right)\times\nn\\
&&\qquad\times\prod_{i,j=1}^N\frac{\xfac{\frac{u_i}{u_j}s^{-2}x^{2R+|m_i-m_j|}}}{\xfac{\frac{u_j}{u_i}s^2x^{2(1-R)+|m_i-m_j|}}}\prod_{i=1}^N\frac{u_i^{\pm1}p^{-1}x^{2-r+|m_i|}}{u_i^{\mp1}px^{r+|m_i|}}\prod_{a=1}^2\frac{\xfac{u_i^{\pm1}z_a^{\pm1}s\,x^{2-R+|m_i|}}}{\xfac{u^{\mp1}z_a^{\mp}s^{-1}x^{R+|m_i|}}}\, ,\nn\\
\ee
while the index of Theory B is
\be
&&\mathcal{I}_{\mathcal{T}_B}=\prod_{a=1}^2\frac{\xfac{s^{-2(N-a+1)}x^{2-2(N-a+1)(1-R)}}}{\xfac{s^{2(N-a+1)}x^{2(N-a+1)(1-R)}}}\prod_{j=1}^{N-2}\frac{\xfac{s^{-2(j-1)}p^{-2}x^{2-2(j-1)(1-R)-2r}}}{\xfac{s^{2(j-1)}p^2x^{2(j-1)(1-R)+2r}}}\times\nn\\
&&\times\frac{\xfac{s^{2(N-j-1))}p\,\omega^{-1}x^{1+2(N-j-1)(1-R)+r}}}{\xfac{s^{-2(N-j-1)}p^{-1}\omega\,x^{1-2(N-j-1)(1-R)-r}}}\prod_{j=3}^N\frac{\xfac{s^{2(j-2)}p\,\omega\,x^{1+2(j-2)(1-R)+r}}}{\xfac{s^{-2(j-2)}p^{-1}\omega^{-1}x^{1-2(j-2)(1-R-r)}}}\times\nn\\
&&\times\sum_{\vec{m}^{(2)}\in\mathbb{Z}^2}\frac{\prod_{\ga=1}^2\omega^{m^{(2)}_\ga}}{2}\oint\prod_{a=1}^2\frac{\udl{u^{(2)}_a}}{2\pi i\,u^{(2)}_a}\left(1-\left(\frac{u^{(2)}_1}{u^{(2)}_2}\right)^{\pm1}x^{|m^{(2)}_1-m^{(2)}_2|}\right)\times\nn\\
&&\times\prod_{a=1}^2\frac{\xfac{u^{(2)^\pm1}_a p^{-1}s^{2-N}x^{2-(N-2)(1-R)-r+|m^{(2)}_a|}}}{\xfac{u^{(2)\mp1}_ap\,s^{N-2}x^{(N-2)(1-R)+r+|m^{(2)}_a|}}}\frac{\xfac{u^{(2)\pm1}_a z^{\pm1}_1s^{N-1}x^{1+(N-1)(1-R)+|m^{(2)}_a|}}}{\xfac{u^{(2)\mp1}_a z^{\mp1}_1s^{1-N}x^{1-(N-1)(1-R)+|m^{(2)}_a|}}}\times\nn
\ee
\be
&&\times\frac{\xfac{s^2x^{2(1-R)}}}{\xfac{s^{-2}x^{2R}}}\sum_{m^{(1)}\in\mathbb{Z}}\oint\frac{\udl{u^{(1)}}}{2\pi i\, u^{(1)}}s^{\sum_a|m^{(2)}_a|+2|m^{(1)}-\sum_a|m^{(1)}-m^{(2)}_a|}\times\nn\\
&&\times p^{-\sum_a|m^{(2)}_a|}x^{-|m^{(2)}_1-m^{(2)}_2|-(R+r-2)\sum_a|m^{(2)}_a|+2(1-R)|m^{(1)}|+R\sum_a|m^{(1)}-m^{(2)}_a|}\times\nn\\
&&\times\frac{\xfac{u^{(1)\pm1}z^{\pm1}_1s^{2-N}x^{1-(N-2)(1-R)+|m^{(1)}|}}}{\xfac{u^{(1)\mp1}z^{\mp1}_1s^{N-2}x^{1+(N-2)(1-R)+|m^{(1)}|}}}\frac{\xfac{u^{(1)\pm1}z^{\pm1}_2s^Nx^{1+N(1-R)+|m^{(1)}|}}}{u^{(1)\mp1}z^{\mp1}_2s^{-N}x^{1-N(1-R)+|m^{(1)}|}}\times\nn\\
&&\times\prod_{a=1}^2\frac{\xfac{u^{(1)\pm1}u^{(2)\mp1}_a s^{-1}x^{1+R+|m^{(1)}-m^{(2)}_a|}}}{\xfac{u^{(1)\mp1}u^{(2)\pm1}_a s\, x^{1-R+|m^{(1)}-m^{(2)}_a|}}}\, .
\ee
In order for the two indices to both have a well-defined expansion in $x$, we need to choose the R-symmetry parameters such that
\be
\frac{2N-5}{2(N-2)}<R<1,\qquad 0<r<2(N-2)R+5-2N\, .
\ee
We verified the matching of the superconformal indices for $N=3,4,5$. In Table \ref{sci3flav} we summarize the results of our computations.

\begin{table}[t]
\centering

\scalebox{0.93}{
\setlength{\extrarowheight}{2pt}
\makebox[\linewidth][c]{
\begin{tabular}{c|c|c|c|c}
$N$ & $R$ & $r$ & $h$ & $\mathcal{I}$ \\ \hline
3 & $\frac{3}{4}$ & $\frac{1}{4}$ & 3 & $1+\frac{\omega}{p s^2}x^{1/4}+\frac{1}{p s^2 \omega}x^{1/4}+\frac{\omega^2}{p^2 s^4}x^{1/2}+\frac{1}{p^2 s^4 \omega^2}x^{1/2}+\frac{1}{p^2 s^4}x^{1/2}+p^2x^{1/2}+\cdots$ \\
4 & $\frac{5}{6}$ & $\frac{1}{6}$ & 2 & $1+\frac{\omega}{p s^4}x^{1/6}+\frac{1}{p s^4 \omega}x^{1/6}+\frac{\omega^2}{p^2 s^8}x^{1/3}+\frac{1}{p^2 s^8 \omega^2}x^{1/3}+\frac{1}{p^2 s^8}x^{1/3}+p^2x^{1/3}+\cdots$ \\
5 & $\frac{7}{8}$ & $\frac{1}{8}$ & 1 & $1+\frac{\omega1}{p s^6}x^{1/8}+\frac{1}{p s^6 \omega}x^{1/8}+\frac{\omega^2}{p^2 s^{12}}x^{1/4}+\frac{1}{p^2 s^{12} \omega^2}x^{1/4}+\frac{1}{p^2 s^{12}}x^{1/4}+p^2 x^{1/4}+\cdots$
\end{tabular}}}
\caption{Computation of the superconformal index for different values of $N$ up to order $\mathcal{O}\left(x^h\right)$. In the last column we report the first terms of the expansion.}
\label{sci3flav}
\end{table}

\subsubsection{Four flavors}

The superconformal index of Theory A in the case $k=3$ takes the form
\be
&&\mathcal{I}_{\mathcal{T}_A}=\prod_{j=1}^{N-3}\frac{\xfac{s^{2j}x^{2j(1-R)}}}{\xfac{s^{-2j}x^{2-2j(1-R)}}}\sum_{\vec{m}\in\mathbb{Z}^N}\frac{\prod_{i=1}^N\omega^{m_i}}{N!}\oint\prod_{i=1}^N\frac{\udl{u_i}}{2\pi i\,u_i}s^{-2\sum_{i<j}^N|m_i-m_j|+3\sum_{i=1}^N|m_i|}\times\nn\\
&&\qquad\times p^{-\sum_{i=1}^N|m_i|} x^{2(R-1)\sum_{i<j}^N|m_i-m_j|-(3R+r-4)\sum_{i=1}^N|m_i|}\prod_{i<j}^N\left(1-\left(1-\frac{u_i}{u_j}\right)^{\pm1}x^{|m_i-m_j|}\right)\times\nn\\
&&\qquad\times\prod_{i,j=1}^N\frac{\xfac{\frac{u_i}{u_j}s^{-2}x^{2R+|m_i-m_j|}}}{\xfac{\frac{u_j}{u_i}s^2x^{2(1-R)+|m_i-m_j|}}}\prod_{i=1}^N\frac{u_i^{\pm1}p^{-1}x^{2-r+|m_i|}}{u_i^{\mp1}px^{r+|m_i|}}\prod_{a=1}^3\frac{\xfac{u_i^{\pm1}z_a^{\pm1}s\,x^{2-R+|m_i|}}}{\xfac{u^{\mp1}z_a^{\mp}s^{-1}x^{R+|m_i|}}}\, ,\nn\\
\ee
while the index of Theory B is
\be
&&\mathcal{I}_{\mathcal{T}_B}=\prod_{a=1}^3\frac{\xfac{s^{-2(N-a+1)}x^{2-2(N-a+1)(1-R)}}}{\xfac{s^{2(N-a+1)}x^{2(N-a+1)(1-R)}}}\prod_{j=1}^{N-3}\frac{\xfac{s^{-2(j-1)}p^{-2}x^{2-2(j-1)(1-R)-2r}}}{\xfac{s^{2(j-1)}p^2x^{2(j-1)(1-R)+2r}}}\times\nn\\
&&\times\frac{\xfac{s^{2N-2j-3}p\,\omega^{-1}x^{1+(2N-2j-3)(1-R)+r}}}{\xfac{s^{-2N+2j+3}p^{-1}\omega\,x^{1-(2N-2j-3)(1-R)-r}}}\prod_{j=4}^N\frac{\xfac{s^{2j-5}p\,\omega\,x^{1+(2j-5)(1-R)+r}}}{\xfac{s^{-2j+5}p^{-1}\omega^{-1}x^{1-(2j-5)(1-R-r)}}}\times\nn\\
&&\times\sum_{\vec{m}^{(3)}\in\mathbb{Z}^3}\frac{\prod_{a=1}^3\omega^{m^{(3)}_a}}{3!}\oint\prod_{a=1}^3\frac{\udl{u^{(3)}_a}}{2\pi i\,u^{(3)}_a}\prod_{a<b}^3\left(1-\left(\frac{u^{(3)}_a}{u^{(3)}_b}\right)^{\pm1}x^{|m^{(3)}_a-m^{(3)}_b|}\right)\times\nn\\
&&\times\prod_{a=1}^3\frac{\xfac{u^{(3)^\pm1}_a p^{-1}s^{3-N}x^{2-(N-3)(1-R)-r+|m^{(3)}_a|}}}{\xfac{u^{(3)\mp1}_ap\,s^{N-3}x^{(N-3)(1-R)+r+|m^{(3)}_a|}}}\frac{\xfac{u^{(3)\pm1}_a z^{\pm1}_1s^{N-2}x^{1+(N-2)(1-R)+|m^{(3)}_a|}}}{\xfac{u^{(3)\mp1}_a z^{\mp1}_1s^{2-N}x^{1-(N-2)(1-R)+|m^{(3)}_a|}}}\times\nn
\ee
\be
&&\times\frac{1}{2}\sum_{m^{(2)}\in\mathbb{Z}^2}\oint\prod_{\ga=1}^2\frac{\udl{u^{(2)}_\ga}}{2\pi i\, u^{(2)}_\ga}\left(1-\left(\frac{u^{(2)}_1}{u^{(2)}_2}\right)^{\pm1}x^{|m^{(2)}_1-m^{(2)}_2|}\right)\prod_{\ga,\gb=1}^2\frac{\xfac{\frac{u^{(2)}_\ga}{u^{(2)}_\gb}s^2x^{2(1-R)+|m^{(2)}_\ga-m^{(2)}_\gb|}}}{\xfac{\frac{u^{(2)}_\gb}{u^{(2)}_\ga}s^{-2}x^{2R+|m^{(2)}_\ga-m^{(2)}_\gb|}}}\times\nn\\
&&\times\prod_{\ga=1}^2\frac{\xfac{u^{(2)\pm1}_\ga z^{\pm1}_1s^{3-N}x^{1-(N-3)(1-R)+|m^{(2)}_\ga|}}}{\xfac{u^{(2)\mp1}_\ga z^{\mp1}_1s^{N-3}x^{1+(N-3)(1-R)+|m^{(2)}_\ga|}}}\frac{\xfac{u^{(2)\pm1}_\ga z^{\pm1}_2}s^{N-1}x^{1+(N-1)(1-R)+|m^{(2)}_\ga|}}{u^{(2)\mp1}_\ga z^{\mp1}_2s^{1-N}x^{1-(N-1)(1-R)+|m^{(2)}_\ga|}}\times\nn\\
&&\times\prod_{a=1}^3\frac{\xfac{u^{(2)\pm1}_\ga u^{(3)\mp1}_a s^{-1}x^{1+R+|m^{(2)}_\ga-m^{(3)}_a|}}}{\xfac{u^{(3)\mp1}_\ga u^{(3)\pm1}_a s\, x^{1-R+|m^{(2)}_\ga-m^{(3)}_a|}}}\frac{\xfac{s^2x^{2(1-R)}}}{\xfac{s^{-2}x^{2R}}}\sum_{m^{(1)}\in\mathbb{Z}}\oint\frac{\udl{u^{(1)}}}{2\pi i\, u^{(1)}}\times\nn\\
&&\times s^{2|m^{(2)}_1-m^{(2)}_2|+\sum_a|m^{(3)}_a|+2\sum_\ga|m^{(2)}_\ga|+2|m^{(1)}-\sum_a\sum_{\ga=1}^2|m^{(2)}_\ga-m^{(3)}_a|-\sum_\ga|m^{(1)}-m^{(2)}_\ga|}p^{-\sum_a|m^{(3)}_a|}\times\nn\\
&&\times x^{-\sum_{a<b}|m^{(3)}_a-m^{(2)}_b|-2R|m^{(2)}_1-m^{(2)}_1|-(R+r-2)\sum_a|m^{(3)}_a|+2(1-R)\sum_\ga|m^{(2)}_\ga|+2(1-R)|m^{(1)}|}\times\nn\\
&&\times x^{R\sum_a\sum_{\ga}|m^{(2)}_\ga-m^{(3)}_a|+R\sum_\ga|m^{(1)}-m^{(2)}_\ga|}\frac{\xfac{u^{(1)\pm1}z^{\pm1}_2s^{2-N}x^{1-(N-2)(1-R)+|m^{(1)}|}}}{\xfac{u^{(1)\mp1}z^{\mp1}_2s^{N-2}x^{1+(N-2)(1-R)+|m^{(1)}|}}}\times\nn\\
&&\times\frac{\xfac{u^{(1)\pm1}z^{\pm1}_3s^Nx^{1+N(1-R)+|m^{(1)}|}}}{u^{(1)\mp1}z^{\mp1}_3s^{-N}x^{1-N(1-R)+|m^{(1)}|}}\prod_{\ga=1}^2\frac{\xfac{u^{(1)\pm1}u^{(2)\mp1}_\ga s^{-1}x^{1+R+|m^{(1)}-m^{(2)}_\ga|}}}{\xfac{u^{(1)\mp1}u^{(2)\pm1}_\ga s\, x^{1-R+|m^{(1)}-m^{(2)}_\ga|}}}\, .
\ee
In order for the two indices to both have a well-defined expansion in $x$, we need to choose the R-symmetry parameters such that
\be
\frac{2(N-3)}{2N-5}<R<1,\qquad 0<r<(2N-5)R+6-2N\, .
\ee
We verified the matching of the superconformal indices for $N=4,5$. In Table \ref{sci4flav} we summarize the results of our computations.

\begin{table}[t]
\centering

\scalebox{0.93}{
\setlength{\extrarowheight}{2pt}
\makebox[\linewidth][c]{
\begin{tabular}{c|c|c|c|c}
$N$ & $R$ & $r$ & $h$ & $\mathcal{I}$ \\ \hline
4 & $\frac{4}{5}$ & $\frac{1}{5}$ & 2 & $1+\frac{\omega}{p s^3}x^{1/5}+\frac{1}{p s^3 \omega}x^{1/5}+\frac{\omega^2}{p^2 s^6} x^{2/5}+\frac{1}{p^2 s^6 \omega^2} x^{2/5}+\frac{1}{p^2 s^6} x^{2/5}+p^2 x^{2/5}+\cdots$ \\
5 & $\frac{6}{7}$ & $\frac{1}{7}$ & 1 & $1+\frac{\omega}{p s^5}x^{1/7}+\frac{1}{p s^5 \omega}x^{1/7}+\frac{\omega^2}{p^2 s^{10}} x^{2/7}+\frac{1}{p^2 s^{10} \omega^2} x^{2/7}+\frac{1}{p^2 s^{10}} x^{2/7}+p^2 x^{2/7}+\cdots$
\end{tabular}}}
\caption{Computation of the superconformal index for different values of $N$ up to order $\mathcal{O}\left(x^h\right)$. In the last column we report the first terms of the expansion.}
\label{sci4flav}
\end{table}

\bibliographystyle{JHEP}

\end{document}